\documentclass[12pt]{JHEP}
\usepackage{amsmath,epsfig}
\usepackage{amssymb,amsfonts}
\usepackage{latexsym}
\usepackage{epsfig}
\relax
\renewcommand{\theequation}{\arabic{section}.\arabic{equation}}
\def\be{\begin{equation}}
\def\ee{\end{equation}}
\def\bea{\begin{eqnarray}}
\def\eea{\end{eqnarray}}
\def\simleq{\; \raise0.3ex\hbox{$<$\kern-0.75em\raise-1.1ex\hbox{$\sim$}}\; }
\def\simgeq{\; \raise0.3ex\hbox{$>$\kern-0.75em\raise-1.1ex\hbox{$\sim$}}\; }

\newcommand\fverb{\setbox\pippobox=\hbox\bgroup\verb}
\newcommand\fverbdo{\egroup\medskip\noindent%
                        \fbox{\unhbox\pippobox}\ }
\newcommand\fverbit{\egroup\item[\fbox{\unhbox\pippobox}]}

\newcommand{\bear}{\begin{eqnarray}}

\newcommand{\eear}{\end{eqnarray}}
\def\hri#1#2{\href{http://arxiv.org/abs/#1}{[ArXiv:#1][#2]}}
\def\hre#1#2{\href{http://arxiv.org/abs/#1/#2}{[ArXiv:#1/#2]}}
\def\hspi#1#2{\href{http://www.slac.stanford.edu/spires/find/hep/www?irn=#1}{#2}}
\newbox\pippobox

\def\ls{\ell_s}
\def\ie{{\it i.e.~}}
\def\lab{\label}
\def\6{\partial}
\def\f{\Phi}
\def\a{\alpha}
\def\nn{\nonumber}

\def\half{\frac12}

\def\le{\left}
\def\ri{\right}
\def\cO{{\cal O}}

\def\pa{\partial}

\def\e{\epsilon}
\def\m{\mu}
\def\n{\nu}
\def\r{\rho}
\def\s{\sigma}

\def\sp{\;\;\;,\;\;\;}

\def\de{\partial}

\def\sq
\def\a{\alpha}
\def\b{\beta}
\def\l{\lambda}

\def\tr{{\rm Tr}}

\def\II{\relax{\rm I\kern-.18em I}}

\title{Exploring improved holographic theories for QCD: Part II}

\author{U. G{\"u}rsoy$^{1,2}$,
\href{http://hep.physics.uoc.gr/~kiritsis/}{E. Kiritsis}$^{1,3}$, F. Nitti$^1$\\
$^1$\href{http://cpht.polytechnique.fr/cpht/cordes/}{CPHT, Ecole Polytechnique, CNRS,
 91128, Palaiseau, France}\\
 ( UMR du CNRS 7644).\\
~\\
$^2$\href{http://www.lpt.ens.fr/}{Laboratoire de Physique Th\'eorique},\\
Ecole Normale Sup\'erieure,
24, Rue Lhomond, Paris 75005, France.\\
~\\
$^3$\href{http://hep.physics.uoc.gr/}{Department of Physics, University of Crete
71003 Heraklion, Greece}\\
~\\}

\preprint{ArXiv:0707.1349 \\ CPHT-RR028.0507}      

\abstract{This paper is a continuation of \href{http://arxiv.org/abs/0707.1324}{ArXiv:0707.1324}
where improved holographic theories for QCD were set up and explored.
Here, the IR confining geometries are classified and analyzed.
They all end in a ``good" (repulsive) singularity in the IR.
The glueball spectra are gapped and discrete, and they favorably compare to the lattice data.
Quite generally, confinement and discrete spectra imply each other.
Asymptotically linear glueball masses can also be achieved.
Asymptotic mass ratios of various glueballs with different spin also turn out to be universal.
Meson dynamics is implemented via space filling $D_4-\bar D_4$ brane pairs. The associated tachyon dynamics
is analyzed and chiral symmetry breaking is shown.
The dynamics of the RR axion is analyzed, and the non-perturbative running of the QCD $\theta$-angle is obtained.
It is shown to always vanish in the IR.
}


\begin{document}

\def\g{\gamma}
\def\go{\g_{00}}
\def\gi{\g_{ii}}

\maketitle 

\section{Introduction and summary}

This paper is a direct sequel of \cite{part1}, and the reader is guided
there for a comprehensive introduction and summary of results of both papers.
In the first part of this work, two of the authors establish and
motivate a general 5D holographic setup
to describe 4D gauge theories with a large
number of colors (large $N_c$). The setup described there constitutes
 a bottom-up
approach, motivated in part from known features of 5D non-critical
string theory and in part by what we expect from QCD.

The pure gauge dynamics is  encoded holographically
in the solution of a two-derivative action for the expected 5D fields: the 5D
metric (dual to the YM stress tensor), a scalar (the dilaton, dual to $Tr[F^2]$),
 and a pseudoscalar (the axion, dual to $Tr[F\wedge F]$). The
dilaton potential is expected to be non-trivial  and is expected to  obtain non-trivial contributions
from the non-propagating four-form. In
practice,
  the potential (and the associated  superpotential) is  in one-to-one correspondence with the QCD $\beta$-function
and is chosen in such a way as to reproduce known features
(e.g. UV asymptotic freedom and IR confinement) of the gauge theory. This is
what makes our approach  phenomenological.

The resulting backgrounds present an improvement over pre-existing
models of ``phenomenological holography'', e.g.
\cite{erlich,pomarol}: among other advantages, the backgrounds we
present incorporate the running of the coupling and asymptotic
freedom; establish a one-to-one correspondence between the 5D
geometry and the gauge theory parameters, namely the $\beta$-function $\beta(\l)$,
and the dynamically generated IR  scale $\Lambda_{QCD}$ and do
not require specifying the boundary conditions in the IR. Moreover,
they provide a natural environment  to study non-perturbative
dynamical phenomena such as confinement, generation of the mass gap
and chiral-symmetry breaking.

Part I is devoted to establish and motivate the setup, and to analyze
the perturbative UV regime  of the correspondence.
 The present work, on the other hand,  focuses on  the analysis of the
non-perturbative regime.  As one of our main results, we
establish a relation between color confinement
(i.e. an area law for the Wilson loop)  and the properties
of the geometry in the IR, and we show that confining
backgrounds always exhibit a mass gap and generically a discrete
spectrum. This is a nontrivial statement, as in our models
there is no IR boundary (which would automatically guarantee
both confinement {\emph{and}} a mass gap). In most of this
 work we  focus
on the pure Yang-Mills sector, which we describe holographically
by a 5D Einstein-Dilaton system. We discuss the addition of $N_f$ matter
flavors in the quenched approximation $N_f\ll N_c$, so that we can
neglect the backreaction of the 5D fields dual to the operators containing
quarks.

The structure of the present paper is as follows. In Section
\ref{hqcd} we give an overview of the setup discussed in
\cite{part1}. We recall how  asymptotic freedom demands the UV geometry
to be asymptotically $AdS_5$ with logarithmic corrections. We then
review the holographic dictionary, mapping field theory quantities
to their geometrical counterparts. In particular, there is a
one-to-one correspondence between the superpotential associated to
the geometry and the exact field theory $\beta$-function.
In the last subsection we analyze
the IR behavior of space-times  that have AdS$_5$ UV asymptotics, showing
that backgrounds that are not conformal in the IR necessarily
exhibit an IR singularity.

In Section \ref{conf} we provide  a complete characterization of 5D
asymptotically $AdS_5$ backgrounds that exhibit confinement in the IR.
Here, as a definition of confinement, we require  that the
Wilson loop exhibits an area law behavior. We compute the Wilson loop
holographically, using the prescription of \cite{maldarey}, as the
action of a classical string world-sheet with fixed UV boundary
conditions. We show that  confinement requires the scale factor to
vanish sufficiently fast in the IR. We formulate this requirement in
terms of the superpotential and the 4D $\beta$-function. We then
discuss  the holographic computation of the 't Hooft loop, relevant
for the potential between two color-magnetic charges, in order to investigate
whether screening behavior is present.

In Section 4 we perform a holographic computation of  the (regularized)
QCD vacuum energy.

In Section \ref{axion} we discuss the dynamics of
 the 5D  axion. This field does not backreact
on the geometry in the large $N_c$ limit, however its normal modes
give the spectrum of 4D pseudoscalar glueballs. In addition, its UV source is associated
to the $theta$-angle of YM and its IR
properties are relevant for the discussion of the effective QCD
$\theta$-parameter. We argue  that this field  must vanish in the
IR. In this way a pseudo-scalar glueball condensate is
dynamically generated. This suggests that the effects of the $\theta$-parameter
are screened in the IR.

In Section \ref{spectrum} we discuss the general features of the
low-energy particle spectrum in our model, obtained from the
fluctuations of the bulk fields around the background. For scalar
and tensor  glueballs, in all confining potentials the spectrum is
gapped and discrete. With the exception of a certain class of
``pathological'' geometries in which the singularity is not
screened,  the particle  spectrum can be computed unambiguously
imposing normalizability of the eigenfunctions. We find potentials where
the glueballs have a linear asymptotic spectrum, i.e. $m_n^2 \sim
n$. We discuss the addition of flavor branes  along the lines of
\cite{ckp}, where it was proposed that chiral symmetry breaking is
correctly described by open string tachyon condensation. We show
that this idea can be naturally implemented in our setup. We discuss
the asymptotics of the tachyon background and the qualitative
features of the spectrum of mesons.

In Section 7 we discuss the parameters of the gravity backgrounds
and their relations with the gauge theory  parameters. We show that,
once confining asymptotics in the IR are imposed, the 5D backgrounds are
completely specified by the dilaton potential plus the arbitrary
choice of a single integration constant, that  only affects the
overall energy scale. It is in one-to-one correspondence  to a
choice of $\Lambda_{QCD}$ in the dual field theory.

In Section \ref{numerics} we give some concrete examples, in which
we specify the exact $\beta$-function, solve numerically for the
corresponding geometry and compute, again numerically, the glueball
spectrum. We compare our results with the available lattice data,
and with similar computations in other models. In particular, we show
that the ``linear confinement'' background fits particularly well the
lattice data. On the other hand, in hard-wall models like
\cite{erlich,pomarol}, or generically in models with  a
``quadratic'' mass spectrum, the agreement is not as good.

Some of the technical details can be found in the appendices. In
particular, the reader interested in the details of the characterization
of confining  backgrounds, including their various geometric properties,
is referred to Appendix \ref{confinement}.

\section{Building blocks of holographic QCD} \label{hqcd}

In this section we review some properties of the 5d backgrounds and
their holographic interpretation. Some of these points where
extensively discussed in \cite{part1}.

\subsection{The 5D backgrounds}

As detailed in \cite{part1}, we take a ``minimal'' non-critical
approach to holographic large $N_c$ QCD type theories, in which the
5D string theory dual has, as low energy excitations, the duals of
the lowest-dimension gauge invariant operators. In the pure
glue sector these are: the five-dimensional
metric $g_{\mu\nu}$ (dual to the YM stress tensor);  a scalar field
$\phi$ which we call the dilaton (dual to the YM operator $\tr F^2$)
 and an axion, dual to $\tr F\Tilde F$. We may  ignore  the axion
when searching for the QCD vacuum solution as its contribution
is subleading in $1/N_c$ \cite{wit3}. It can be included in the sequel  (this is
discussed in section \ref{axion}) as it does affect some important
physics, in particular that of flavor singlet mesons. The scalar $\phi$
encodes the running of the YM  coupling, and it is naturally
identified with  the 5D string dilaton.

We should emphasize that we think of the 5D bulk theory as a (non-critical)
string theory, not just gravity.
However, we restrict ourselves to the two-derivative effective action,
including a general dilaton potential,
that contains also a subclass of higher $\alpha'$ terms
as argued in \cite{part1}.

Therefore, the string-frame action describing the low-lying
excitations is: \be\lab{action-string} S_S = {M^3 N_c^2}\int d^5 x
\sqrt{-g_S}{1\over \l^2} \left[R + 4 g_S^{\mu\nu}{\de_\mu \l\over
\l} {\de_\nu\l \over \l}+ {V_S}(\l)\right] \ee where we have
introduced the 't Hooft coupling \be \l\equiv N_c e^{\phi}\;.
\label{sss}\ee It is related to the 't Hooft coupling of the gauge
theory, up to a multiplicative constant. $V_S(\l)$ is an ``effective
potential'' originating from integrating-out the non-dynamical
4-form, \cite{part1} and other higher curvature corrections. We do
not attempt here to derive ${V_S}(\l)$ from first principles. We
determine certain of its  properties   by requiring that the
geometry that follows from $V_S$ reproduces some known features of
the Yang-Mills dynamics. In particular, requiring UV asymptotic
freedom  constraints the asymptotics of the potential for  small
values of $\l$. Requiring an area law for the Wilson loop on the
other hand constrains  the asymptotics of the potential for large
$\l$. {}From now on we also define for convenience a renormalized
dilaton $\Phi$ as \be \l=e^{\Phi}\;. \ee

We mostly work with the Einstein frame metric,
\be\lab{stringtoeinst}
g_{\m\n} = e^{-{4\over 3}\Phi} (g_S)_{\m\n},
\ee
for which the action reads:
\be\lab{action-einst}
S = {M^3 N_c^2}\int d^5 x \sqrt{-g}\left[ R - {4\over 3}
 g^{\mu\nu}\de_\mu \Phi \de_\nu\Phi + V(\Phi)\right]\sp V(\Phi) = e^{4\Phi/3} {V_S}(\Phi)\,.
\ee In the large $N_c$ limit we assume, $g_{\m\n}$ and $\Phi$ are
independent of $N_c$.

We search  backgrounds of the form:
\be
\lab{sol} g_{\mu\nu} =
du^2 + e^{2A(u)}\eta_{ij} dx^i dx^j = e^{2A(r)}\left(dr^2 +
\eta_{ij} dx^i dx^j\right), \quad \Phi = \Phi(u),
\ee
 where
$x^i$ are the 4D  space-time coordinates, and $\eta_{ij}=diag(-,+,+,+)$.
We write
the metric in two different coordinate systems, related by:
\be\lab{rtou} {d r \over d u} = e^{-A(u)}. \ee We name the first set
of coordinate system  the {\em domain-wall
  coordinates}.
The second set involving $r$ will be called
 the {\em conformal coordinates} as the metric is explicitly
conformally flat in this coordinate system. Throughout the paper, we will use a prime for $d/du$ and a dot for $d/dr$.

The independent Einstein's equations take the following form (in
domain-wall coordinates):
\be\label{einsteq}
\Phi^{'2}(u) = -{9\over
4} A''(u), \quad V(\Phi) = 3 A''(u) + 12 A^{'2}(u).
\ee
These
equations can  be written in first-order form in terms of a
superpotential $W(\Phi)$ \cite{sken1}:
\bea
&&\Phi' = {d W\over d\Phi}, \qquad  A' = -{4\over 9} W, \label{einsteinsuper}\\
 && V(\Phi) =
-{4\over 3}\left({d W\over d\Phi}\right)^2 + {64\over 27} W^2.\label{VtoW}
\eea
Given any scale factor $A(u)$ such that $A''(u)\leq 0$,
one can invert the relation between $\Phi$ and $u$ using the first equation
 (\ref{einsteq}) and find a  superpotential $W(\Phi)=-4/9
\, A'(u(\Phi))$. This determines  a potential, such that the given
$A(u)$ is a solution. This useful property allows to investigate the
backgrounds under consideration starting directly from a
parametrization of the metric, rather than the dilaton potential.

In conformal coordinates, Einstein's equations (\ref{einsteq}) read:
\be \lab{einsteqconf}
\dot{\Phi}^{2}(r) = -{9\over 4} \left(\ddot{A}(r)-\dot{A}^{2}(r)\right) ,
 \quad V(\Phi) =
 e^{-2A(r)}\left(3 \ddot{A}(r) + 9 \dot{A}^{2}(r)\right),
\ee
or, in terms of the superpotential:
\be\label{conf1st}
\dot{\Phi} = {dW \over d\Phi}e^A , \quad \dot{A} =    -{4\over 9} W e^A.
\ee

As shown in  \cite{part1}, asymptotic freedom in the gauge theory
 translates into an asymptotic $AdS_5$
region\footnote{We should note that ``asymptotically AdS$_5$" here
has a different meaning that the similar term in the mathematics
literature (see for example \cite{Skenderis} and the references
therein). Here the corrections to the AdS$_5$ metric are also
logarithmic, while there they are powers of the conformal coordinate
$r$.}, where the dilaton asymptotes to $-\infty$, and $W\to W_0 >0$:
\be\lab{UVas} A(u) \sim - u/\ell + O\left(\log u\right), \quad
\Phi(u) \sim -\log\left[-{u\over \ell}-\log(\ell \Lambda)\right] +
O(1), \qquad u \to -\infty, \ee or, in conformal coordinates: \be
\lab{UVasr} A(r) \sim -\log r/\ell + O\left({1\over \log r}\right),
\quad \Phi \sim -\log (-\log (r\Lambda)) + O(1), \qquad r\sim \ell
e^{u/\ell} \to 0. \ee The $AdS$ scale $\ell$ is fixed by the  value
of $W(\l)$ for $\l=0$: $\ell =(9/4)W_0^{-1}$;  $\Lambda$ is an
integration constant whose meaning will be clarified in section
\ref{pars}. The subleading terms are also fixed, order by order by
matching the $\beta$-function coefficients. This corresponds to a
dilaton potential of the form $V(\Phi) \sim V_0 + V_1 e^{\Phi} + V_2
e^{2\Phi}+O(e^{3\Phi})$. Since $\l\sim e^{\Phi}$ is small in this
region, we expect this potential to be generated by the full
resummation of the $\a'$ expansion, and is to be interpreted as an
``effective potential.'' The information of its weak coupling
expansion maps on the field theory side, to the perturbative
$\b$-function.

In this paper we are mostly concerned with the opposite regime, i.e.
the large $\l$ IR region. One of our goals is to find
 which solutions to  (\ref{einsteq}), satisfying the UV asymptotics (\ref{UVas}),
provide an area-law for the Wilson loop, and what kind of potentials
are necessary to realize those solutions. Before addressing this
problem, we give some preliminary discussion about the holographic
dictionary and the infra-red properties of 5D geometries.

\subsection{Holographic dictionary}

In order to exploit the gauge-gravity duality, we must first
establish a dictionary between the 5D and 4D quantities. In
particular we must identify the Yang Mills coupling and energy scale
on the gravity side. For this we use the dictionary established in
\cite{part1}.

At a given position in the fifth dimension, the four-dimensional
energy scale is set by the scale factor $e^A(u)$, as seen from eq
(\ref{sol}). This, we argue, leads to the identification:
\be\label{energy} \log E \leftrightarrow A(u). \ee Close to the
$AdS_5$ boundary, this reduces to the familiar identification $E =
1/r$. The correspondence (\ref{energy}) does not fix the absolute
units of the 4D energy scale with respect to the scale factor. This
is consistent with the observation that a constant shift in $A(u)$
leaves Einstein's equations (\ref{einsteq}) invariant, and can be
used to change the unit energy in a given background.

Notice that the overall scale factor in (\ref{energy}) is in the {\em
Einstein frame}. In pure $AdS_5$ with a constant dilaton this
distinction does not matter, but in our backgrounds the dilaton does
not asymptote to a constant in the UV, therefore this clarification
is needed. In particular, the Einstein frame scale factor has the
important property of being monotonically decreasing with $u$
(see Section 2.3). This property is not shared by the string
frame metric.
 Monotonicity is crucial if we want our geometry to be dual to a single RG flow
from the UV to the IR (and not, e.g, to two different UV theories
that flow to  the same IR). For a discussion on the string
corrections to (\ref{energy}), see \cite{part1}.

\subsubsection*{The $\beta$-function}

We identify\footnote{There are some ambiguities in this identification that are discussed in \cite{part1}.}
the  4D 't Hooft coupling $\l=g_{YM}^2 N_c$ as\footnote{As stated in \ref{sss},
the  string dilaton, $\phi\equiv \log g_s$,  is related
to $\Phi$ by $\phi\equiv \Phi - \log N_c$.  $\Phi$
is the appropriate variable to use in the large $N_c$ limit.}
\be\label{coupling}
\l=  e^{\Phi}.
\ee

With the identification (\ref{energy}), it follows that the $\b$-function
of the 't Hooft coupling is related to 5D fields as:
\be\label{beta}
\b(\l) \equiv {d \l \over d \log E} = \l {d \Phi \over d A},
\ee
or, in terms of the phase-space variable $X$, introduced in \cite{part1},
\be\label{x}
 X \equiv {\Phi'\over 3 A'} \sp \b = 3 \l X
\ee
The above definition is independent of
reparametrizations of the radial coordinate, and  can be expressed either in the $r$, $u$
coordinates, or by using  $\Phi$ as a radial coordinate.

Einstein's equations can be rewritten in terms of
$X(\Phi)$ as:
\begin{eqnarray}\lab{fprime}
\f'  &=& -\frac43 W_0 X  e^{-\frac43\int_{-\infty}^{\f} X d\f},\\
\lab{Aprime}
A' &=& -\frac49 {W_0} e^{-\frac43\int_{-\infty}^{\f} X d\f},
\end{eqnarray}
Here, $W_0>0$ is the asymptotic value of the superpotential as
$\f\to -\infty$. It is related to the asymptotic $AdS_5$ scale $\ell$ by
\be\label{ell}
W_0 = {9\over 4\ell}.
\ee
{}From these equations, the superpotential is related to $X$ as:
\be\label{spotx}
 X(\Phi)   = -{3\over 4} {d \log W(\Phi) \over d\Phi}\,.
\ee
We deduce that  fixing the function $X(\l)$ (hence
the $\b$-function) specifies the superpotential
(up to an overall multiplicative constant). Then, the equations of motion
 (\ref{fprime}) and (\ref{Aprime}) determine  the geometry up to
two integration constants and an overall length scale $\ell$. As we
show in Section 7, only one of the two integration constants is
physical, the second one being an artifact of reparametrization
invariance.

\subsection{Infrared properties of asymptotically $AdS_5$ backgrounds} \label{IRasympt}

In the holographic approach to strongly coupled gauge theories,
confinement at low energies is typically related to the termination of space-time
at a certain point in the radial coordinate. In five-dimensional
holography, with asymptotic $AdS_5$ in the UV, this often implies the presence
of a singularity in the bulk. We show here that, unless the IR is conformal,
a curvature
singularity is always present
 when we restrict ourselves to a two-derivative effective action. Specifically, we prove the
following statement:

{\bf Proposition:} {\em Consider any solution of (\ref{einsteq})
such that $\exp A(r) \sim \ell/r$ as $r\to 0$ (with $r>0$). Then,}
\begin{itemize}
\item{\em The scale factor $e^{A(r)}$ is  monotonically decreasing}
\item  There are only three  possible, mutually exclusive
 IR behaviors:
\begin{enumerate}
\item {\em there is  another asymptotic $AdS_5$ region, at $r\to \infty$,
where  $\exp A(r) \sim \ell'/r$, and  $\ell' \leq \ell$}
(equality holds if and only if the space is exactly $AdS_5$ everywhere);
\item {\em  there is a curvature singularity at some finite value
of the radial coordinate, $r=r_0$;}
\item {\em  there is a curvature singularity at $r \to \infty$, where
the scale factor vanishes and the space-time shrinks to zero size.}
\end{enumerate}
\end{itemize}

That the scale factor must  be monotonic in this context is well
known, and it is most clear in the $u$ coordinates: the first
equation in (\ref{einsteq}) implies that $A''(u)< 0$, therefore
$A'(u)$ must be monotonically decreasing. In the UV, $A(u) \sim
-u/\ell$ so for any $u$   we must have \be\lab{derbound} A'(u) \leq
-1/\ell <0, \quad \forall u. \ee As a consequence, $A(u)$ itself
must be monotonically decreasing from the UV to the IR. This is a
version of the holographic $c$-theorem \cite{cth}.

We now investigate possible IR
behaviors. In conformal coordinates, the bound (\ref{derbound}) translates to :
\be\lab{derboundr}
{d\over dr} e^{-A(r)} \geq {1\over \ell}.
\ee
Suppose that the $r$ coordinate extends to $+\infty$. Then, either
the l.h.s. of (\ref{derboundr}) asymptotes
to constant $\ell^{'-1} > \ell^{-1}$, or it asymptotes to infinity. In the former case,
we obtain:
\be
 e^{A(r)} \sim  {\ell'\over r}, \quad r\to \infty.
\ee
This implies  that the space-time is asymptotically $AdS_5$ in the
IR, with a smaller $AdS_5$ radius $\ell'$. The gauge theory flows to an IR
conformal fixed point, and is therefore not
confining.\footnote{Five-dimensional holographic duals of the
Bank-Zaks fixed points are in this class, \cite{BCCKP}.}

If instead  ${d\over dr}~e^{-A(r)} \to +\infty$ as $r\to +\infty$,
then the curvature scalar diverges, as can be seen from its
expression: \be\lab{curvature} R(r) =
-e^{-2A}\left(12 \dot{A}^{2} + 8\ddot{A}\right). \ee
In this case, $e^{-2A(r)}$ diverges faster than $r^2$, and
$\dot{A}^{2}$ and $\ddot{A}$ do not vanish faster than $r^{-2}$,
forcing eq. (\ref{curvature}) to diverge as $r\to \infty$. Moreover, the
scale factor $e^{A(r)}$ vanishes for large $r$, as claimed.

There is another possibility, i.e. that the space-time ends at a finite
value $r_0$. This can happen because
the scale factor $e^{A}$ shrinks to zero, or some of its derivatives
diverge\footnote{We are always assuming that the space-time terminates due
to some
non-trivial dynamics, rather than because of some  boundary at some
otherwise regular point $r=r_0$. This is in contrast with the original
AdS/QCD approach
which advocates an AdS space with an IR boundary.} at $r_0$.
In either case eq. (\ref{curvature}) indicates that
 we are in the presence of a curvature singularity at $r_0$.

These considerations were derived in the context of 5D
Einstein-Dilaton gravity, but they are more general, because they
follow only from the condition $A''(u)<0$. This can be shown to be
equivalent to the Null Energy Condition (NEC) (see e.g.
\cite{nitti}). Therefore the arguments of this subsection can be
extended to any bulk field content, provided its stress tensor
satisfies the NEC.

\section{Confining backgrounds} \label{conf}

Here we would like to characterize the backgrounds that exhibit
confinement. By ``confinement'' we understand an area law behavior for the Wilson
loop. Our analysis allows a simple  classification of confining background
in terms of the metric, superpotential, or $\b$-function IR  asymptotics.

\subsection{The Wilson loop test\label{wilsonloop}}

In this subsection we review the holographic computation of the
Wilson Loop, \cite{maldarey,cobi}. The potential energy $E(L)$ of an
external quark-antiquark pair separated by a distance $L$ and
evolved in time $T$, can be computed holographically by the action
of a classical string embedded in the 5D space-time, with a single
boundary which is a rectangular loop with sides $L$ and $T$ on the
$AdS_5$ boundary. We have,
\be
T E(L) =  S_{NG}[X^\mu_{min}(\s, \tau)]\,.
\ee
Here  $S_{NG}$  is
the Nambu-Goto action evaluated on the world-sheet embedding
$X^\mu_{min}(\s,\tau)$ with minimum area:
\be
S_{NG} = T_f \int d\tau
d\s \sqrt{-\det g_S},  \qquad (g_S)_{\a\b} = (g_S)_{\mu\nu} \de_\a
X^\mu\de_\b X^\nu, \quad \a,\b = 1,2
 \ee
 with $T_f={1\over 2\pi\ell_s^2}$ the fundamental
string tension and $(g_S)_{\mu\nu}$ the bulk {\em string frame} metric.
For a generic 5D metric of the form:
\be
(g_S)_{\mu\nu}dx^\mu dx^\nu =
g_{ss}(s) ds^2 - g_{00}(s)dt^2 +  g_{\parallel}(s)d\vec{x}^2,
\ee
\cite{cobi} showed that, for differentiable world-sheets, one can
write:
\be
\label{wilson} E(L) = T_f f(s_F) L - 2 T_f\int^{s_F}_{s_B}
ds {g(s)\over f(s)}\sqrt{f^2(s) - f^2(s_F)}
\ee
 where the functions
$f(s), g(s)$ are defined as:
\be
 f^2(s) = g_{00}(s)g_{\parallel}(s),
\quad g^2(s) = g_{00}(s)g_{ss}(s)
\ee
 and $s_F$ is the turning point
of the world-sheet in the bulk. Implicitly, $s_F$ depends on $L$
through the relation:
\be
\label{length} L = 2\int_{s_B}^{s_F} ds {
g(s)\over f(s)} {1\over \sqrt{f^2(s)/f^2(s_F) -1}},
 \ee
 where $s_B$
denotes the boundary. For large $L$, the second term in eq.
(\ref{wilson}) is subleading.

Expressions (\ref{wilson})  and (\ref{length}) drastically simplify
 if we use conformal coordinates,
$s=r$,
\be
(g_S)_{\mu\nu}(r) = e^{2A_S(r)} \eta_{\mu\nu}, \quad A_S(r) = A(r) + {2\over 3}\Phi(r),   \quad f(r) = g(r) = e^{2A_S(r)},
\ee
to obtain:
\be\lab{Lofr}
L = 2\int_{0}^{r_F} dr  {1\over \sqrt{e^{4A_S(r)-4A_S(r_F)} -1}}.
\ee
In the neighborhood of $r=0$ the integral is finite, because the integrand behaves  as $e^{-2A_S(r)}
\sim r^2$, and $r_F \sim L^3$ for small $L$.  Around $r_F$ we may expand the
 denominator as:
\be
  {1\over \sqrt{e^{4A_S(r)-4A_S(r_F)} -1}} \simeq {1\over \sqrt{4A_S'(r_F)(r_F-r)
+ 8A_S''(r_F)(r_F-r)^2 + \ldots }}\,. \ee The integral is  finite for
generic $r_F$ and grows indefinitely as $r_F$ approaches a
stationary point $r_*$ of $A_S(r)$, where $A_S'(r_*)=0$.  This must
correspond to
 a minimum since $A_S$ starts decreasing at $r=0$. In other words,
if there exists  such a stationary point $r_*$, then
\be
 r_F \to r_* \quad as \quad L\to \infty.
\ee

The large $L$ behavior of the quark-antiquark potential energy  is
thus (from (\ref{wilson}) \be\lab{tension} E(L) \sim T_f
e^{2A_S(r_*)} L \ee which exhibits an area law if and only if $A_S$
is finite at the minimum. From eq. (\ref{tension}) we read off the
confining string tension as, \be\label{qcdstring} T_s = T_f
e^{2A_S(r_*)} \ee

Notice that the finiteness of the string tension is not directly
related to the value of the metric at the end of space, as sometimes
assumed. Even if the space-time shrinks to zero-size at the
singularity, which is the generic behavior of the {\em Einstein's
frame} metric, this does not impede an area law: the string frame
scale factor has a global minimum at a regular point in the bulk,
and classical string world-sheets  {\em never probe} the region of
space beyond that point and never reach the singularity.

Equation (\ref{tension}) captures the leading behavior of the
quark-antiquark potential. In QCD the first subleading correction is
the L\"uscher term, $\sim 1/L$. As  shown in \cite{cobi2}, this term
arises in some
 confining backgrounds (e.g. \cite{wittenmodel})
from the first quantum corrections to the classical Wilson loop
in \cite{cobi}. It would be interesting to see if this
is also the case in the models we are considering.

\subsection{Confining IR asymptotics\label{conf-sec}}

We are now ready to answer the question: {\em which IR asymptotics
give rise to confinement}.

Here we discuss a special class of metrics,  that demonstrate
 particularly  interesting features: namely the space-times with
 infinite range of the conformal coordinate, $r\in (0,\infty)$. In Appendix
 \ref{confinement} we  give a complete discussion including other types of
 backgrounds. There,  we also present the asymptotic values of some of the interesting
quantities. The reader can find a summary of the classification in
Table \ref{summarytable} at the end of this section.

Consider a class of space-times whose Einstein frame metric has the
form (\ref{sol}), with the asymptotics:
\be\label{metric-asympt0}
A(r) \to -C r^\a+\ldots,  \qquad r\to \infty,\quad \a,C>0
\ee
up to
\emph{generic} subleading terms. Here, $C^{-1/\a} \equiv R$ is a
length scale controlling the IR dynamics.

The singularity is at $r\to \infty$,
and the space-time shrinks to zero-size there\footnote{The singularity is at
a \emph{finite} value $u_{IR}$ of the  $u$ coordinate. See appendix \ref{confinement}.}.
 To check
whether the fundamental string is confining  we need the string-frame scale factor,
\be\label{strscale0}
A_S(r) = A(r) + {2\over 3}\Phi(r).
\ee
As we have discussed in the previous subsection, confinement  is
equivalent to the existence  of a minimum of the expression (\ref{strscale0}), where
$e^{A_S}$ is non-zero. Due to  the $AdS$ UV asymptotics,
$A_S \to +\infty$ as $r\to 0$. Therefore a necessary and sufficient condition
for confinement is that $A_S$ {\it does not asymptote to $-\infty$ } at the IR singularity,
$r\to +\infty$\footnote{Since we are assuming that the  singularity
is at $r\to \infty$,
and $\Phi$ is monotonically increasing from $\Phi=-\infty$
at $r=0$, $A_S$ cannot diverge to $-\infty$ at some finite $r$. Therefore, if there
is a minimum for $A_S$, the string tension is certainly finite.}.

The asymptotics of the dilaton can be obtained using the first equation in (\ref{einsteqconf}):
\be\label{dilr1}
\Phi(r) \sim -{3\over 2} A(r) + {3\over 4} \log |\dot A(r)| + \Phi_0 .
\ee
Indeed, (\ref{dilr1}) solves eq. (\ref{einsteqconf}) up to
a term proportional to $(\ddot A/\dot A)^2 \sim r^{-2}$,
regardless of the subleading behavior in (\ref{metric-asympt0}).

Using (\ref{dilr1}) we obtain the asymptotic form of the string frame scale factor (\ref{strscale0}):
 \be\label{strscale}
A_S  \sim {1\over 2}\log |\dot A(r)| \sim  {(\a-1)\over 2}\log r/R,
\qquad ds^2_S \sim \left({r\over R}\right)^{\a-1}
\left(dr^2 + \eta_{ij}dx^i dx^j\right)
\ee
Notice that the leading power-law term has  canceled! Moreover the first
surviving term is completely determined only by the leading power divergence
of the Einstein frame scale factor.

With the  simple result (\ref{strscale}), we can immediately
determine which backgrounds lead to confinement:
\begin{itemize}

  \item  $\a \geq 1 \Longrightarrow$ confinement: \\
the string frame scale factor approaches
 $+\infty$ in the IR, thus it has a minimum at finite $r$.
The special case $\a=1$ also leads to
confinement.  The minimum is reached as $r\to \infty$, and  the
confining string tension is $T_f \lim_{r\to \infty}  \exp[2
A_S(r)]$.\footnote{One could think of a situation where the
string frame scale factor has multiple minima at $r_i$, with nonzero
values for $\exp[A(r_i)]$ (otherwise there would be a singularity at
finite $r$.) In this case, the classical analysis implies that the
string world-sheet has to stop at the minimum closest to the $AdS$
boundary, and never knows about the existence of the others.
However, quantum corrections may plausibly trigger the decay into
other minima with lower tension. We don't consider this possibility
any further, and we will always treat backgrounds with a single
minimum of $A_S$.} Notice that when $\a=1$ the asymptotic geometry
(in the string frame) is 5D Minkowski space-time with linear
dilaton.

\item  $\a<1 \Longrightarrow$ no confinement:\\
 $A_S$ asymptotes
to $-\infty$ for large $r$, hence the confining string tension vanishes. It is
easy to show that the same result applies if $\a=0$, and the scale
factor $A(r)$ goes to $-\infty$ slower than any power-law (e.g.
logarithmically).\footnote{As discussed in Section \ref{IRasympt}
$A(r)$ cannot asymptote to  a finite constant.}

\end{itemize}

We can relate the asymptotics (\ref{metric-asympt0}) to the
$\b$-function and to the superpotential, as follows: first we
compute the $X$-variable, defined in eq (\ref{x}), as a function of
$r$, then, using eq. (\ref{dilr1}) we can invert asymptotically the
relation between $\Phi$ and $r$ and substitute it in the expression
above. This gives: \be\lab{xconf2} X(\l) = -{1\over 2} \left[1 +
{3\over 4}{\a-1 \over \a}{1\over \log \lambda} + \ldots\right],
\qquad \l\to \infty. \ee We note that, generically, the point $r_*$ where
$A_S'=0$ corresponds to $X = -1/2$. In
(\ref{xconf2}), the point $X=-1/2$ is first reached at $r_*$,
and then at the singularity $r=+\infty$ where $\l$ diverges.

The asymptotic form
 of the superpotential is, from eq. (\ref{spotx}):
\be\lab{spotconf}
 W(\Phi) \sim \Phi^{\a-1\over 2\a} e^{2\Phi/3}, \qquad \Phi\to +\infty.
\ee

Notice that in the leading asymptotics of the superpotential or of
 $X(\l)$  there is no trace of the dimensionfull constant $C$ that controls the
``steepness''  of the warp  factor in eq. (\ref{metric-asympt0}).
The appearance of the parameter $R=C^{-1/\a}$ in the metric is the
manifestation, in conformal coordinates, of the  dynamical generation
of the IR scale, as we will show explicitly in Section 7.
It is fixed by the integration constants of
Einstein's eqs, rather than by  fundamental parameters appearing in
$W(\l)$.

The idea that some aspect of the geometry, which determines the IR scale, can
be related to the integration constants rather than some \emph{a priori} chosen
parameter, was already present in the ``braneless approach'' to $AdS$/QCD of
\cite{cr}.  As we will discuss  in section \ref{beware} however, the spectral
properties of the
background analyzed in \cite{cr} suffer from some pathologies,
that make it conceptually equivalent to  models with a hard IR cutoff,
in which some additional, arbitrary  boundary conditions in the IR must be supplied.

We can also relax the requirement that $A(r)$ grows as a simple
power-law, since from eq. (\ref{strscale}) we see that all that is
needed for confinement is the condition $\log |\dot{A}| > 0 $
asymptotically. This is true for any function $A(r)$ whose
asymptotics is bounded above and below as: \be C_1 r^{\a_1} < -A(r)
< C_2 r^{\a_2}, \qquad \a_{1,2}\geq 1, \qquad C_{1,2} \geq 0 \ee

\subsection{General confinement criteria} \label{criterion}

In Appendix \ref{confinement} we analyze also the backgrounds where
the singularity is at finite $r=r_0$. They always  exhibit area law.
The analysis in  the previous section, together with Appendix
\ref{confinement}, allows us to formulate a general criterion for
confinement in 5D holographic models: \vspace{0.5cm}
\\
\centerline {\bf General criterion for confinement (geometric version)}\\
{\em A geometry that shrinks to zero size in the IR
 is dual to a confining 4D theory if and only
if the Einstein metric in conformal coordinates vanishes as (or faster than)
$e^{-C r}$ as $r\to \infty$, for some $C> 0$. }
\\
(It is understood here that a metric vanishing at finite $r=r_0$
also satisfies the above condition.) \vspace{0.5cm}

Comparing the superpotentials found in all the examples studied in
Appendix \ref{confinement}, eqs. (\ref{pq1}),  (\ref{pq2}),
(\ref{pq3}), and (\ref{pq4}), we see that one can treat
simultaneously all  cases by using the following parametrization
for large $\l$ :
\be
\label{pq} W(\l) \sim (\log\l)^{P/2} \l^Q, \quad
\beta(\l)= 3\l X(\l) \sim -{9\over 4}\l \left(Q + {P\over 2} {1\over
\log\l}\right),
\ee
 where $P$ and $Q$ are real numbers. This implies
for the Einstein and string frame dilaton potentials: \be V(\Phi)
\sim  (\log\l)^{P} \l^{2Q}, \qquad V_S(\Phi) \sim (\log\l)^{P}
\l^{(2Q-4/3)} \ee

An equivalent characterization of the
confining backgrounds is: \vspace{0.5cm}
\\
\centerline{\bf General criterion for confinement (superpotential)}\\
{\em A 5D background is dual to a confining theory if the
superpotential grows as (or faster than) $(\log\l)^{P/2}\l^{2/3}$ as
$\l\to\infty$ for some $P\geq 0$.} \vspace{0.5cm}

The relation between parameters $P$ and $\a$ (appearing in (\ref{metric-asympt0}),
is given  in Table \ref{summarytable}.

One can also relate the IR properties directly to the large $\l$
asymptotics of the $\beta$-function. Computing $X(\l) =
\b(\l)/(3\l)$ from the superpotential via eq. (\ref{spotx}), one
obtains the following form of the same criterion: \vspace{0.5cm}
\\
\centerline{\bf General criterion for confinement ($\b$-function)}\\
{\em A 5D background is dual to a confining theory if and only if
\be\lab{general1}
\lim_{\l \to \infty} \left(X(\l) +{1\over 2}\right) \log \l  = K, \qquad -\infty\leq K \leq 0
\ee
}
In the above form\footnote{We are assuming that the limit exists,
and that the $\beta$-function does not oscillate infinitely many
times across $-3/(2 \l)$ as $\l\to \infty$. This possibility seems
remote from a physical point of view.} the condition for confinement
does not make any explicit reference to any coordinate system. Yet,
we can determine the geometry if we specify $K$. In particular:
\begin{enumerate}
\item $K = -\infty$: the scale factor vanishes at some finite
  $r_0$, not faster than a power-law.
\item  $-\infty< K< -3/8$: the scale factor vanishes at some finite $r_0$ faster than any power-law.
\item  $-3/8<K<0$:   the scale factor vanishes as $r\to \infty$ faster
than $e^{-C r^{1+\e}}$ for some  $\e>0$.
\item  $K=0$:  the scale factor vanishes as $r\to \infty$ as
$e^{-C r}$ (or faster), but slower than  $e^{-C r^{1+\e}}$ for any
$\e>0$.
\end{enumerate}
The borderline case, $K=3/8$, is certainly confining (by continuity), but
whether or not the singularity is at finite $r$ depends on the
subleading terms. When $K$ is finite,  we can relate it to  the
parameters $Q$ and $P$ appearing in the superpotential: if $K<\infty$,
then $Q=2/3$ and $P=-8K/3$.
The classification of the various possible IR asymptotics in terms
of their confining properties is summarized in Table \ref{summarytable}.
\begin{table}[h]
\hspace{-1cm}
\begin{tabular}{|c||c|c|c|c|c|}
\hline
\multicolumn{1}{|c||}{ }
&
\multicolumn{3}{|c|}{$r\in (0,\infty)$}
&
\multicolumn{2}{|c|}{$r\in (0,r_0)$}\\
\hline
\hline
\multicolumn{1}{|c||}{}&\multicolumn{1}{|c|}{} &\multicolumn{2}{|c|}{}
&\multicolumn{1}{|c|}{} &  \multicolumn{1}{|c|}{}\\
\multicolumn{1}{|c||}{$A(r)\sim$}&
\multicolumn{1}{|c|}{$-\gamma \log r$}&
\multicolumn{2}{|c|}{$-C r^\a$}&
\multicolumn{1}{|c|}{$ -C(r_0-r)^{-\tilde{\a}}$}&
\multicolumn{1}{|c|}{$\delta \log(r_0-r)$}
\\
\cline{3-4}
 & & $0<\a<1$ & $\a\geq 1$ & & \\
\hline
& & & & & \\
{\bf Confining} & {\bf No} & {\bf No} & {\bf Yes} & {\bf Yes} & {\bf Yes} \\
& & & & & \\
\hline
& & & & & \\
Q & ${2\over3}\sqrt{1-{1\over \gamma}}< {2\over3}$ &${2\over 3}$
&${2\over 3}$ &
${2\over 3}$ & ${2\over3}\sqrt{1+{1\over \delta}}> {2\over3}$  \\
& & & & & \\
\hline
& & & & & \\
P & arbitrary  & ${\a-1\over \a}<0$ & ${\a-1\over \a}\in [0,1)$  &
${\tilde{\a}+1\over \tilde{\a}} >1$ & arbitrary \\
& & & & & \\
\hline
& & & & & \\
K & $>0$ & $>0$ &$-{3\over 8}P \in \left(-{3\over 8}, 0\right.\left.\right]$ &
 $-{3\over 8}P \in \left(-\infty,{3\over 8} \right)$ & $-\infty$  \\
& & & & & \\
\hline
\end{tabular}
\caption{ \
Summary of confining asymptotics.
 As required by the  NEC, the parameters
$\a,\tilde{\a},\gamma, \delta, C$ are all assumed positive,
  and
$\g\geq 1$
 }
 \label{summarytable}
\end{table}

We note that, if we classify the backgrounds in terms of $P$ and
$Q$, our analysis covers the entire range of these parameters. As a
result, our classification is not limited to superpotentials
that behave asymptotically as  (\ref{pq}), but it also applies
to any superpotential that for large $\l$  is bounded between
two functions of the form (\ref{pq}), for two appropriate pairs
$(Q_1,P_1)$ and $(Q_2, P_2)$.

For most of the confining backgrounds, although the space-time is
singular in the Einstein frame, the string frame geometry is regular
for large $r$ (see Appendix \ref{confinement}). In fact, in these
situations, all curvature invariants vanish for large $r$. The
dilaton however diverges. Therefore, in the string frame
the singularity manifests itself as a strong coupling
region in a weakly curved space-time.

Interestingly, as discussed in the previous subsection, string
world-sheets do not probe the strong coupling region, at least
classically. This is because the geodesic surfaces ending on the AdS
boundary do not stretch beyond the minimum of the scale factor. At
that point, the t'Hooft coupling may be of order one, but the {\em
string} coupling $g_s = \l/N_c$ is still small. This can be
intuitively attributed to the fact that the string tries to stay
away from the region where the metric becomes large, since this
would generate a larger world-sheet area.

Therefore,  singular confining backgrounds have generically the
property that  the singularity is {\em repulsive}, i.e. only highly
excited states can probe it.  This will also be reflected in the
analysis of the particle spectrum, in the next subsection. This
consideration makes our conclusions more robust, since they are
insensitive to the region near the singularity, where quantum
effects may become important. As the classical string world-sheet never
probes the strong curvature region, a semiclassical analysis is
reliable.

One
could also worry that a direct coupling of the dilaton to the world-sheet curvature
scalar could spoil this analysis. This is not so, as shown in Appendix C.

\subsection{Magnetic charge screening} \lab{monopole}

In confining theories, one expects the dual magnetic gauge group to
be Higgsed, leading to a screening of the magnetic charges. In our
setup, magnetic monopoles can be described as the endpoints of
D1-branes. Therefore the calculation of the monopole-antimonopole
potential proceeds exactly like the one for the quark-antiquark
potential, with a D-string replacing the fundamental string. In this
section we discuss the case of infinite range backgrounds, leaving
the finite case to Appendix \ref{monopoleapp}.

The D-string action is\footnote{One might expect that the magnetic and electric quark potentials
are related by $\lambda \to 1/\lambda$ duality, in the Einstein frame. This is the case for gauge theories that are deformations
of ${\cal N}=4$ sYM and others that descend from ten dimensions but not in general. In the Einstein frame
the electric string action is proportional to $\l^{4\over D-2}$ while the magnetic one to $\l^{6-D\over D-2}$.
For $D=10$, these factors are inverses of each other, but not for $D=5$ relevant here.}
\be S_D = T_{D_1}\int d^2\xi e^{-\Phi}\sqrt{
- det \, g_{\a\b}},
\ee
where $g_{\a\b}$ is the induced metric on
the world-sheet and the target space metric is in the string frame.
We work in the conformal coordinates, \be ds^2 = e^{2A_S} \left(dr^2
+ \eta_{ij}dx^i dx^j\right), \ee and reabsorb the factor of the dilaton in
the conformal factor of  the target space metric, reducing the
problem to a string with Nambu-Goto action propagating in a target
space with an effective metric: \be\label{Dframe} ds^2 = e^{2A_D(r)}
\left(dr^2 +  \eta_{ij}dx^i dx^j \right), \quad A_D = A_S-{\Phi\over 2} = A
+ {\Phi\over 6}. \ee The properties of the string embedding can then
be deduced using the same techniques as in the previous subsections.

 For large $ L $ the energy of this a configuration is simply given by
\be\label{poten} E = e^{2A_D(r_F(L))} L + \cdots\, , \ee
where $r_F(L) $ is the bulk position of turning point of the worldsheed with length $L$
on the boundary. The relation between $r_F$ and $L$ is given by a formula
similar to eq. (\ref{Lofr}), with the substitution $A_S \to A_D$.

To avoid the magnetic charge confinement, it must be that the scale
factor $e^{A_D}$ of the ``D-string frame'' metric,
eq. (\ref{Dframe}), vanishes at the IR singularity.

In the confining backgrounds of section \ref{conf-sec}, with large
$r$ asymptotics (\ref{metric-asympt0})-(\ref{dilr1}) and $\a\geq 1$,
 the scale factor  $e^{A_D}$ in eq. (\ref{Dframe}) does indeed vanish
as $r\to\infty$; the magnetic string tension is  zero and
the magnetic charges are not confined. The question remains, whether
they feel an inverse
 power-law potential or they are truly screened in which case
the potential falls-off  exponentially or faster. Below, we show
that the latter holds for the backgrounds under consideration.

 In order to answer this question, one has to study the potential energy
(\ref{poten}) for large $L$: to do this, one has to
 invert asymptotically the relation between
$r_F$ and $L$ from the D-string analog of eq. (\ref{Lofr}), and insert it into eq. (\ref{poten}).

The asymptotic form of the D-string metric is \be\label{Dasyinf}
A_D^{(\a)}(r) \sim -{3C\over 4} r^\a + \ldots
 \qquad  \a \geq 1,
\ee where we are restrict to the confining case. We must evaluate
\be L^{(\a)}(r_F) = \int_0^{r_F} {dr \over
\left[e^{4\left(A_D^{(\a)}(r) - A_D^{(\a)}(r_F)\right)}
-1\right]^{1/2}}. \ee By assumption, there are no other
singularities of $\exp[A_D]$ for any finite $r$, and we assume that
there are no other local extrema. Thus,
the only region in which $L(r_F)$ could diverge is $r_F \to \infty$.

We show below that, for $\a\geq 1$, $L^{(\a)}(r_F)$ is  finite in this
limit. We first divide the integration range in two regions,
$0<r<r_1$, $r_1<r<r_F$, such that in the second region the
asymptotic form of the scale factor (\ref{Dasyinf}) holds. Consider
the integral in the first region:

\bea
&& \!\!\!\!\!\!\!\!\int_0^{r_1}
{dr \over \left[e^{4\left(A_D(r_F) -  A_D(r)\right)} -1\right]^{1/2}}
=  \int_0^{r_1} dr { e^{4\left(A_D(r_F)-A_D(r_1)\right)}\over
\left[e^{4\left(A_D(r) -  A_D(r_1)\right)} - e^{4\left(A_D(r_F) -  A_D(r_1)\right)}\right]^{1/2}} \nonumber \\
 && < \,\,{e^{4A_D(r_F)}\over e^{4A_D(r_1)}}\int_0^{r_1} {dr \over
\left[e^{4\left(A_D(r) -  A_D(r_1)\right)} - 1 \right]^{1/2}}
 =  {e^{4A_D(r_F)}\over e^{4A_D(r_1)}}\,\, L(r_1).
\eea The inequality follows from our  assumption that
$A_D$ is monotonically decreasing. Since   $L(r_1)$ is finite for
finite $r_1$, and $\exp[4A_D(r_F)]\to 0$ as $r_F\to \infty$, the
r.h.s vanishes in this limit. Therefore, for large $r_F$ the
dominant contribution to $L(r_F)$ comes from the region $r>r_1$.

To analyze the behavior of the integral
over the asymptotic region, consider first the case  $\a=1$. We have:

\be
L^{(1)}(r_F) \sim \int_{r_1}^{r_F} {dr \over \left[e^{3C(r_F - r)} - 1\right]^{1/2}}
= {1\over 3C} \int^{3C(r_F-r_1)}_0 {dy \over  \sqrt{e^{y}-1}},
\ee
 and
\be
\lim_{r_F\to +\infty} L^{(1)}(r_F)
= {1\over 3C} \int^{+\infty}_0{ d y \over  \sqrt{e^{y}-1}} = L_{max} < +\infty.
\ee

Next consider $\a>1$. For large $r<r_F$, the following
inequality holds: \be r_F^\a - r^\a > r_F^{\a-1}(r_F - r). \ee It follows
that \be L^{(\a)}(r_F) \sim \int_{r_1}^{r_F} {dr \over
\left[e^{3C(r_F^\a - r^\a)} - 1\right]^{1/2}} < \int_{r_1}^{r_F} {dr
\over \left[e^{3C r_F^{\a-1}(r_F - r)} - 1\right]^{1/2}}
 \sim
{1\over r_F^{\a-1}} L^{(1)}(r_F),
\ee
which implies that for $\a>1$
\be
\lim_{r_F\to +\infty}L^{(\a)}(r_F) = 0.
\ee

We showed that $L^{(\a)}(r_F)$ cannot be larger than a maximum value
$L_{max}^{(\a)}$, which is reached at $+\infty$ if $\a=1$, and at
some finite $r_{max}$ if $\a>1$. Therefore two monopoles at a
distance larger than $L_{max}$ cannot be connected by a smooth
world-sheet. In this case, the configuration that minimizes the
action consists of two straight lines separated by a distance $L$
and joined by a line at constant $r=\infty$. This configuration has the
same energy as the one with two straight lines only as the
contribution from the piece at the singularity vanishes\footnote{One
should take this argument with a grain of salt. This is because,
unlike the configuration that stretches only up to $r_F$, this
configuration falls into the singularity, hence one should worry
about various string and quantum  corrections to the classical solution. At any
rate, our final statement about the magnetic screening is valid as
existence of an $L_{max}$ is sufficient for that.}. Therefore for
$L>L_{max}$ the monopoles are non-interacting. This shows that in the
backgrounds with an infinite range of $r$, the magnetic charges are
screened.

The finite $r_0$ case is discussed in Appendix \ref{monopoleapp},
where we show that the monopole charges are screened, except in
backgrounds with power-law decay $\exp A \sim (r_0-r)^\delta$
 with $\delta <1/15$. This case falls into the
range $0<\delta<1$, which turns out to be problematic also for other
reasons as we show in the discussion of the particle spectrum in
Section \ref{spectrum}.

\subsubsection{Absence of screening in hard-wall models}

In the simplest models proposed as a holographic description of
chiral dynamics of QCD \cite{erlich,pomarol}, the space-time ends at
an IR boundary before any singularity.  According to our discussion in this section,
one  finds linear confinement both for the electric \emph{and}
the magnetic charges. This is contrary to the expectations from the
gauge theory dynamics. In fact, the computation of the magnetic
string Wilson loop is exactly the same as that of the electric one,
since the wall has the same effect on both objects. This was
computed for the cut-off AdS$_5$ background in e.g. \cite{wilsonadsqcd},
where the expected area law was found.

\section{The QCD vacuum energy }

An interesting question in YM theory concerns the value of the vacuum energy,
 and this is closely related to the so called
gluon condensate, $\langle Tr[F^2]\rangle$.
Typically they are UV divergent. Of course one would try to renormalize
them by subtracting the divergences.
We do not know of an unambiguous way to define them beyond perturbation
 theory\footnote{See however \cite{holdom} for another discussion.}.
 Indeed, once the divergences are subtracted
one might as well subtract also the finite piece. We also stress that
the gluon condensate is also defined in a semi-phenomenological fashion
via the SVZ sum rules. Therefore, without quarks, it is not obvious how to define
it\footnote{This is unlike the CP-odd condensate $\langle Tr[F\wedge F]\rangle$
that we will calculate in section \ref{axion}.
The reason is that the operator $Tr[F\wedge F]$ does not need a holographic renormalization.}.

Because of this we will calculate the divergent (holographic) full vacuum energy
to leading order in $1/N_c$.
To do this, we will introduce the usual UV cutoff near the AdS$_5$ boundary and
 will compute the
Euclidean action of the vacuum QCD  solution. The Gibbons-Hawking boundary term
is important in this calculation.

\begin{equation}
   S_5=S_E+S_{GH}
    \label{ve1}\end{equation}
\be
 S_E=-{M^3N_c^2}\int d^5x\sqrt{g}
\left[R-{4\over 3}(\partial\Phi)^2+V(\Phi) \right]
\label{ve2}\ee
\be
 S_{GH}=2M^3N_c^2\int_{\partial M}d^4x \sqrt{h}~K
\label{ve3}\ee

Evaluated on a solution the Einstein action is
\begin{equation}
   S_E={2\over 3}M^3N_c^2\int d^5x\sqrt{g}~~
V(\Phi)={2\over 3}M^3N_c^2V_4\int_{\e}^{r_{0}} dr~e^{5A}~~
V(\Phi)=
 \label{ve4}\end{equation}
 $$
 =2M^3N_c^2V_4\int_{\e}^{r_{0}}dr~\frac{d}{dr}(e^{3A}\dot{A})=2M^3N_c^2V_4\left[e^{3A}\dot{A}\right]_{\e}^{r_0}
 $$
 where $V_4$ is the space-time volume and we introduced a IR cutoff $\e$ in the bulk.
For all our confining backgrounds, the contribution from the singularity vanishes automatically. This is
a good consistency check of the procedure as only the contribution from the
 UV boundary should survive. We therefore obtain
\be
{\cal S}_E=-2M^3N_c^2V_4~e^{3A(\e)}\dot{A(\e)}
 \label{ve5}\ee

For the GH term, the trace of the extrinsic curvature is
$
K=4e^{-A}\dot{A}$
and therefore
\be
{\cal S}^{\e}_{GH}=8M^3N_c^2V_4 ~e^{3A(\e)}\dot A(\e)
\ee
Putting everything together we obtain for the vacuum energy density
(Euclidean action divided by the space-time volume)
\be
{\cal E}_{\rm QCD}=6M^3 N_c^2 ~e^{3A(\e)}\dot A(\e)
\ee
Note that this is negative as $\dot A<0$.
We can re-express the result in terms of the cutoff energy scale, $\Lambda_{UV}$
$$A(\e)=\log\Lambda_{UV}$$

Then the bare vacuum energy density satisfies
\be
{\partial \log {\cal E}_{\rm QCD}\over \partial \log \Lambda_{UV}}=4-
{4\over 9}\Phi'^2=4-{4\beta^2(\l)\over 9\l^2}\simeq
4-{4b_0^2\l^2\over 9}+{\cal O}(\l^3)
\ee
Another (related) equation determines the coupling dependence
\be
{\partial \log {\cal E}_{\rm QCD}\over \partial\l(\Lambda_{UV})}={4\over \beta(\l)}-{\beta(\l)\over 4\l^2}
\ee

\section{The axion background\lab{axion}}

The axion $a$ is dual to the instanton density $Tr[F\wedge F]$. In
particular its UV boundary value is the UV value of the QCD
$\theta$-angle. Moreover, its profile $a(r)$ in the vacuum
 solution may  be interpreted as the ``running"
$\theta$-angle in analogy with the dilaton, that we interpret  as the running coupling constant .
Such an interpretation should be qualified, as it may suggest the false
impression that UV divergences renormalize the $\theta$
angle in QCD. We will return to this later.

The question of the $\theta$ dependence of large $N_c$ QCD and the associated
 $\eta'$ problem has led to  several advances that culminated
with the Witten-Veneziano solution, \cite{wit1,wit2}.
It states that although naively the $\theta$ dependence is expected to be
non-perturbative, at large $N_c$ this is not so.
It enters at order $1/N_c^2$ in YM theory. It generates a $\theta$-depended
vacuum energy that scales as ${\theta^2 \over {N_c^2}}$
 and provides the correct mass (of order $1/N_c$) to the $\eta'$.
Such expectations have been verified in the holographic realization of a
four-dimensional confining gauge theory based on $D_4$ branes, \cite{wit3}.

Here we analyze the structure of the background solution for the
axion in five dimensions relevant for pure YM theory. The action in the Einstein
frame and the corresponding  equation of motion are:
\be
S_a = {M^3\over 2}\int d^5x \sqrt{-g} ~Z(\l)~\left(\de_\mu a\right)^2,\qquad   {1\over
\sqrt{g}}\partial_{\m}\left[Z(\l)\sqrt{g}g^{\m\n}\partial_{\n}\right]
a=0 \lab{d12a}
\ee
 where $Z(\l)$ captures a part of the $\a'$
corrections. It was shown in appendix B.1 of \cite{part1} that
$Z(\l)$ depends in general on the  't Hooft coupling $\l$.
In perturbative string theory and  to leading order in $\a'$,
$Z(\l)=\l^2$. However, as was the case for the potential,  we expect a constant leading term also here,
\be
Z(\l)=Z_a+{\cal O}(\l^2)\sp \l\to 0
\ee
 The axion field equation
is to be solved on a given  metric and dilaton background,
i.e. we neglect the backreaction of the axion \cite{part1}.

For a radially dependent axion the equation becomes
\be
\ddot a+\left(3\dot A+(\partial_{\l}\log Z)\dot\l\right)\dot a=0
\label{31s}
\ee
This equation can be integrated once as
\be
\dot a={C~e^{-3A}\over \ell~Z(\l)}
\label{d12}\ee

The equation (\ref{31s}) has two independent solutions. One is a constant, $f_0(r)=constant$.
The other $f_1(r)$ can be obtained by integrating (\ref{d12}) and choosing the initial conditions
 so that it vanishes at the boundary $r=0$:
 \be
 f_1=\int_0^r {dr\over \ell} {e^{-3A}\over Z(\l)}
 \label{d12c}\ee
where we divided by the AdS length so that the function is dimensionless.
 A first observation is that the function $f_1(r)$ is strictly increasing.

Since near the boundary, $Z=Z_a+\cdots$, $ e^{\Phi}=-{1\over b_0\log(r\Lambda)}+\cdots$ and
 $e^A={\ell\over r}+\cdots$ we obtain
\be
\lim_{r\to 0}f_1(r)={r^4\over 4Z_a\ell^4}\left[1+{\cal O}\left({1\over \log(r\Lambda)}\right)\right]
\label{d12d}\ee
where we chose an arbitrary normalization for this solution. This solution
is the one normalizable in the UV.

The constant solution should be related to the UV value of the  $\theta$-angle as
\be
f_0=\theta_{UV}+2\pi k\sp k\in Z
\ee
The different values of the integer $k$ correspond to an infinite number of vacua, \cite{wit3}.
Such vacua exhibit in general oblique confinement and they are degenerate to leading order in the $1/N_c$
expansion. As it is known and  we will also see it explicitly below, their degeneracy is lifted to the next order.
They can be separated by domain walls that are the $D_2$ branes.

The full background solution therefore reads
\be
a(r)=\theta_{UV}+2\pi k+C~f_1(r)
\label{d12e}\ee
where we take by convention $\theta_{UV}\in [0,2\pi)$.
The coefficient $C$ is proportional to
the expectation value of the QCD instanton density operator in the QCD vacuum.
Using the precise holographic formula and (\ref{d12d}) we obtain
\be
C={4Z_a\ell^4\over (2\Delta-4)}{\langle F\wedge F\rangle\over 32\pi^2}=
{Z_a\ell^4}{\langle F\wedge F\rangle\over 32\pi^2}
\ee

Substituting the solution in the effective action we obtain, for the
($\theta$-dependent) energy per unit three-volume,
 the following boundary terms
\be
{\cal E}(\theta_{UV})={M^3\over 2}\int dr\sqrt{g}Z(\Phi)(\partial a)^2=
{M^3\over 2} e^{3A}Z(\l) a\dot a\Big|_{r=0}^{r=r_0}=
{M^3\over 2\ell}C ~a(r)\Big|_{r=0}^{r=r_0}
\label{d12f}\ee
where we have used the equations of motion to write the on-shell action as a boundary term.
$r_0$ is the position of the singularity in the IR. It may be finite or infinite,
 as discussed in the previous sections.

We expect that the only contribution to the $\theta$ dependent
vacuum energy should come from the UV boundary.
The reason is that there should be only one boundary in the theory. The presence of a second boundary
would imply that the holographic dynamics of the theory is incomplete.
Therefore, we should not expect a contribution from $r=r_0$.
In order for this to be true, the axion should vanish at the singularity.\footnote{Allowing
 any other value at the singularity, $a(r_0)\equiv a_0$ we obtain for the
vacuum energy density, ${\cal E}\sim (\theta_{UV}-a_0)^2$ in contradiction
 with all large-$N_c$ expectations, \cite{wit1,wit2}.}
We
must therefore have,
\be
E(\theta_{UV})={M^3\over
2\ell}C(\theta_{UV}+2\pi k)\sp a(r_0)=\theta_{UV}+2\pi k+Cf_1(r_0)=0
\label{d12g}
\ee
Solving the IR equation assuming $f_1(r_0)\not=0$ we
obtain
 \be
 E(\theta_{UV})=-{M^3\over 2\ell}~{\rm Min}_{ k}~{(\theta_{UV}+2\pi k)^2\over
f_1(r_0)}\sp {a(r)\over  \theta_{UV}+2\pi k}=\left[1-{f_1(r)\over
f_1(r_0)}\right]={\int_r^{r_0} {dr \over e^{3A}Z(\l)}
\over \int_0^{r_0} {dr \over e^{3A}Z(\l)}} \label{d12h}
\ee
where the minimum on $k$ (that we denote by $k_0$) is obtained in order to
choose the k-vacuum that minimizes the energy.
>From (\ref{d12h}) we can extract the topological susceptibility as
\be
\chi={M^3\over \int_0^{r_0}{dr\over e^{3A}Z(\l)}}
\ee

 We have
obtained the expected quadratic and non-analytic  behavior for $E(\theta)$.
We have also determined that the instanton vacuum condensate
is non-zero
\be
{\langle F\wedge F\rangle\over 32\pi^2}=-{\theta_{UV}+2\pi k_0\over Z_a\ell^4
f_1(r_0)}=-{\theta_{UV}+2\pi k_0\over \ell^3\int_0^{r_0}
dr{Z(0)\over e^{3A}Z(\l)}}
\ee
Notice, that the constant $Z(0)$ drops out of all quantities of interest associated
 with the axion, except the topological susceptibility.
Moreover, the condensate is finite without a renormalization of the $F\wedge F$
 operator. This is in accordance with lattice results
\cite{giusti}.

We also observe a very interesting corollary: {\em  If we view $a(r)$ as an effective
 $\theta$-angle  (in analogy with the t' Hooft coupling)
it  vanishes in the IR! }.

We now study the dimensionless constant $f_1(r_0)$ that is proportional to the inverse
of the topological vacuum susceptibility
\be
 f_1(r_0)=\int_0^{r_0} {dr\over \ell} {e^{-3A}\over Z(\l)}
 \label{d12i}\ee
 The integrand is a positive function as $Z(\l)$ is multiplying the axion kinetic energy and is therefore
 expected to be non-negative.
 Moreover we do not expect the integrand to diverge at a point before the singularity $r_0$, as
 $e^{A}$ vanishes only at $r_0$, and $Z(\l)$ is also not expected to vanish.
 Therefore, the only potential pathological behavior is a divergence at $r_0$.

 To study the region around the singularity we will have to study the two cases ($r_0$ finite or infinite) separately.
\begin{itemize}

\item We first  consider the IR asymptotics in the infinite range case, namely the singularity at $r=\infty$.
{}From section  \ref{conf-sec}, for large r and in the Einstein frame:
\be
\log\l={3\over 2} C r^\a + \cdots
\sp
A=-Cr^\a + \cdots
\ee
We also assume that  for large $\l$,
\be
Z(\l)\sim \l^d+\cdots\sp \l\to\infty
\label{cd}\ee
Then:

(1) if $d\not =2$
\be
f_1(r_0=\infty)=\int^{\infty} {dr\over \ell}~ \exp[{3C\over 2}(2-d)r^\a+\cdots]
\ee
In order for this not to diverge, we ask $d>2$.
In this case the low energy asymptotics of the axion are
\be
{a(r)\over \theta_{UV}+2\pi k_0}
\simeq {1\over f_1(\infty)}\int_r^{\infty}dr \exp\left[-{3\over 2}(d-2)C r^\a\right]=
\ee
$$=
{1\over \a f_1(\infty)}\left({3(d-2)C \over 2}\right)^{1\over \a}\Gamma\left[{1\over \a},{3\over 2}(d-2)C r^\a\right]
$$
$$
\simeq
{1\over \a f_1(\infty)}\left({3(d-2)C\over 2}\right)^{{2\over \a}-1}r^{\a-1}\exp\left[-{3\over 2}(d-2)Cr^\a\right]
\sim E^{{3\over 2}(d-2)}\left(\log E\right)^{\a-1\over \a}
$$
where in the last expression we have replaced the radial variable with the holographic energy using (\ref{energy}).

(2) For $d=2$,
\be
f_1(r_0=\infty)=\int^{\infty} dr~  r^{-{3\over
2}(\a-1)}+\cdots
\ee
In order to obtain a finite result, $\a>5/3$.
This is stronger than the confinement condition $\a\geq 1$. The low
energy asymptotics of the axion are
\be
{a(r)\over \theta_{UV}+2\pi k_0}
\simeq {1\over
f_1(\infty)}\int_r^{\infty}dr ~ r^{-{3\over 2}(\a-1)}=
{2\theta_{UV}\over (3\a-5)f_1(\infty)}r^{-{(3\a-5)\over 2}}\sim
\left(-\log E\right)^{-{(3\a-5)\over 2\a}}
\ee
and the effective
$\theta$-angle grows slowly in the IR. However, as it is shown in
section \ref{pss}, in order for the $0^{+-}$ glueballs to have a
discrete spectrum, we must demand $d>2$ and therefore this case
seems not relevant for QCD.

\item Similar remarks apply to confining backgrounds with $r_0$ finite.
In particular $f_1(r_0)$ is finite if $d\geq 2$. When $d>2$ then at
low energy \be \theta(E)\sim E^{{3\over 2}(d-2)} \ee while for $d=2$,
the low energy running is by the inverse power of the logarithm of
the energy.
\end{itemize}

A comment is in order here concerning the relevance of higher-derivative corrections
 to the  IR asymptotics of the solution.
Such corrections can be obtained by substituting $Z(\l)\to Z(R,(\partial\l)^2,\l)$.
This is indeed the most general form as higher powers of $(\partial a)^2$ are suppressed by extra powers of $1/N_c$.
All arguments made above go through with this more general kinetic function.

The oblique confinement  vacua, labeled by $k$, are long-lived in the large-$N_c$ limit,
\cite{wit3} with lifetimes that scale as ${\cal O}(N_c)$.
The $D_2$ brane, discussed in \cite{part1} is a domain wall separating
two such consecutive vacua, as $k$ jumps by $\pm 1$ when crossing the $D_2$
domain walls.

\subsection{Screening of CP violation in the IR\label{cp}}

The essence of the strong CP  problem lies in the fact that  a non-zero $\theta$-parameter
in QCD breaks CP (except at $\theta=\pi$) and provides a non-trivial contribution to
 the neutron dipole moment. The stringent  experimental limits on this quantity
constrain  $\theta$ to be very small, ($\preceq 10^{-9}$) \cite{edp}.
This is known as the strong CP problem: why is $\theta$ so small in nature?

\begin{figure}[h]
 \begin{center}
\leavevmode \epsfxsize=12cm \epsffile{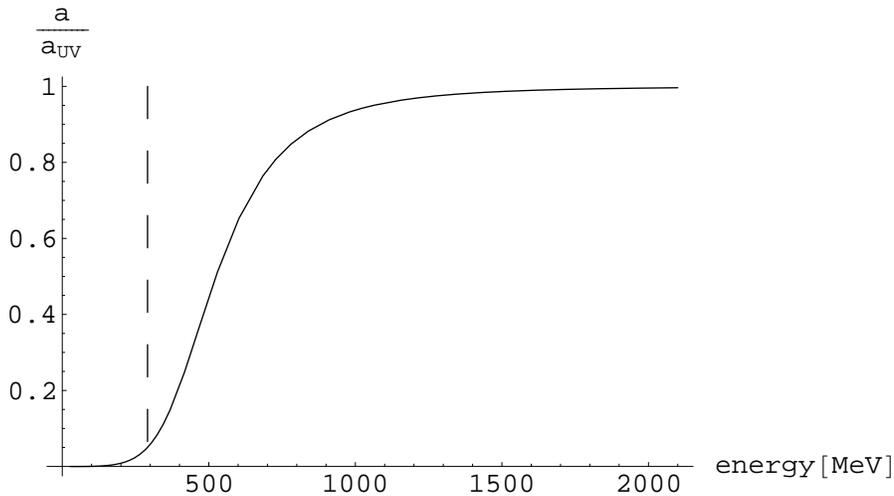}\hspace{0.5cm}
\end{center}
 \caption[] {An example of the axion profile (normalized to one in the UV) as a function of energy,
in one of the explicit cases we treat numerically in Section
\ref{numerics}. The energy scale is in MeV, and  it is
normalized to match the mass of the lowest scalar glueball from lattice
data, $m_0=1475 MeV$. The axion kinetic function is taken as $Z(\l)
=Z_a( 1 + c_a \l^{4})$, with $c_a=100$ while it does not depend on the value of $Z_a$.
 The vertical
dashed line corresponds to $\Lambda_{QCD}$ as defined in eq. (\ref{l2l2}). In this particular case
$\Lambda = 290 MeV$. } \label{c}
\end{figure}

\begin{figure}[h]
 \begin{center}
\leavevmode \epsfxsize=12cm \epsffile{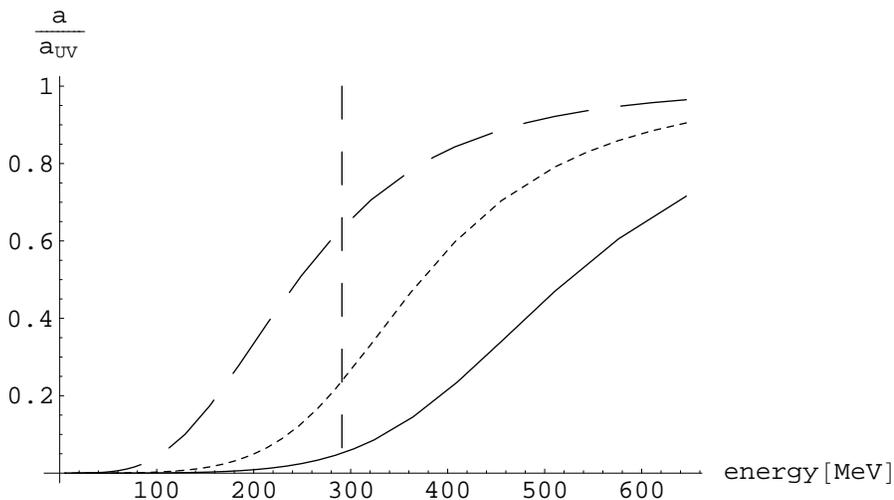}\hspace{0.5cm}
\end{center}
 \caption[]{
A detail showing the different axion profiles for different values of $c_a$.
The values  are   $c_a=0.1$ (dashed line), $c_a=10$ (dotted line)  and $c_a=100$ (solid line).}
 \label{cc}\end{figure}

In pure YM, the case  that we are studying here, the nature of the strong CP problem changes, as there are no quarks, and no baryons.
However, the issue is how strong are the CP violating effects of a nontrivial $\theta_{UV}$ in observable data.
For a pure gauge theory, a CP-violating effect can be the  decay of an (excited) $0^{+-}$ glueball to $0^{++}$ glueballs.
How important such effects are at low energy depends on the way the axion solution behaves at all radial distances.
The behavior of the radial axion profile is shown in figures \ref{c} and \ref{cc}.

It is tempting to think of the radial axion solution as a ``running $\theta$-angle" in analogy with the similar intuition
concerning the dilaton that has been justified quantitatively in numerous holographic setups.
Here however, such an interpretation needs to be qualified, as there are strong indications from lattice \cite{giusti},
that $\theta$ as a coupling in the bare Lagrangian does not receive UV-singular corrections, that would force it to renormalize.
That being said, it is direct on the other hand in the holographic context to see that parity violations at low energy
although proportional to $\theta_{UV}$, have numerical coefficients that are due to this radial change of the axion and such coefficients
can be small.

It is well known that in theories where non-perturbative corrections can be controlled, that $\theta_{eff}$ at low energy receives
finite corrections and is different from $\theta_{UV}$. A simple example of this are ${\cal N}=2$ gauge theories in four-dimensions.
There, the effective field theory in the Coulomb branch can be exactly solved, \cite{sw}, and exact effective superpotential calculated.
Its real part is the effective $\theta$-angle, which does receive instanton corrections.
 Although here the effective theory is abelian the effective
$\theta$-angle can affect low energy physics as some monopoles and dyons can be light.
In large-$N_c$ YM, instantons are apparently  suppressed exponentially, but as was argued in early works, \cite{wit1,wit2},
such non-perturbative effects can affect $\theta$-related physics at leading orders in $1/N_c$.
We therefore advocate that the radial change of the axion can be interpreted as a finite renormalization of the  $\theta$-angle in QCD.
Of course the quantitatively precise statement is that the axion solution must be used in accordance with the standard rules of the
gauge theory/gravity correspondence to calculate physical quantities.

 The discussion above as well as figures \ref{c} and \ref{cc} suggest that the CP-violating effects
 of $\theta_{UV}$ are screened at sufficiently low energy,
whatever the UV value of the $\theta$-parameter. Moreover, the IR
vanishing of the axion is power-like as we have shown above. As we
argue in section \ref{pss}, the expected value of the exponent  $d$ in (\ref{cd}) is $d=4$ from
parity independence of asymptotic glueball spectra.

There have been attempts to calculate QCD data at small values of $\theta_{UV}$,  on the lattice.
Results in particular exist on the topological susceptibility as well as the $\theta$-dependence of the glueball spectrum, \cite{thetalat}.
Such calculations can in principle be confronted with similar calculations using the present framework, but we will not do this here.

The relevance of this discussion in the case where quarks are present is more complex and we will not attempt it here
although we will show several of the relevant ramifications in section \ref{flavor}.

\section{The particle spectrum}
\lab{spectrum}

In gauge-gravity dualities, the particle spectrum of the 4D theory
is obtained from the spectrum of fluctuations of the bulk fields
around the background. Reviews of glueball spectra calculations in the holographic context can be found in \cite{glue}.
In this section we first give a general
overview of the spectra of various particle species (glueballs and
mesons). Then, in Section \ref{numerics},  we compute numerically
the glueball spectrum of some  concrete backgrounds that exhibit asymptotic
freedom in the UV and confinement in the IR. The main results of
this analysis can be summarized as follows:

\begin{enumerate}

\item In the previous section we showed that in order for the theory to confine, the Einstein frame  scale factor
must vanish at least as fast as $\exp[-C r^\a]$ with $\a\geq 1$,
$C\geq 0$. Remarkably, this is the {\em same condition} one obtains
from the requirement of mass-gap in the spectrum. Therefore, using
holography, we can directly relate the existence of a confining
string with the existence of a mass gap.

\item Among the class of
confining backgrounds we have considered, we find examples that
exhibit an asymptotic  ``linear'' mass spectrum, $m^2_n \sim n$.
\end{enumerate}

In this section we are mostly interested in confining backgrounds in
which the scale factor  exhibits  exponential decay at $r \to
\infty$;  in the last subsection we briefly discuss the backgrounds
with a singularity at finite $r$.

\subsection{General properties of the spectra}
\lab{general}

Here we discuss the spectrum from a general point of view and leave
the details and comparison with the lattice
results
to Section \ref{numerics}. We mostly work in the conformal frame,
where the properties of the spectrum are more transparent. The
spectrum of particles up to spin 2 is determined by the fluctuation
equations of the various bulk fields in the solution. Typically, one
can identify fluctuations $\xi(r,x^i)$ with a diagonal kinetic term
and a quadratic action of the form \be\lab{genac} S[\xi] \sim \int
dr d^4x ~e^{2B(r)}\left[ \left(\de_r\xi\right)^2 +
\left(\de_i\xi\right)^2 + M^2(r) \xi^2 \right], \ee where $B(r)$
and $M^2(r)$ are functions depending on the background and on the
type of fluctuation in question.

The linearized field equation reads:
\begin{equation}\lab{scaff}
\ddot{\xi}+2\dot{B}\dot{\xi}+ \Box_4 \xi - M^2(r)\xi = 0.
\end{equation}
To look for 4D mass eigenstates, the standard procedure is to write:
\be
\xi(r,x)  = \xi(r)\xi^{(4)}(x), \qquad \Box \xi^{(4)}(x) = m^2\xi^{(4)}(x).
\ee
Then, eq. (\ref{scaff}) can be put into a Schr\"odinger form
by defining  a wave-function
associated to the fluctuation $\xi$,
\be\label{wave}
\xi(r) = e^{-B(r)}\psi(r)\,.
\ee
Eq. (\ref{scaff}) becomes
\begin{equation}\lab{schro}
-\frac{d^2}{dr^2}\psi + V(r) \psi = m^2\psi,
\end{equation}
with the potential given by,
\begin{equation}
\lab{gluepot}
V(r) = \frac{d^2B}{dr^2}+\le(\frac{dB}{dr}\ri)^2 + M^2(r).
\end{equation}
The  Schr\"odinger equation (\ref{schro}) is  to be solved on the
space of square-integrable functions $\psi(r)$, as can be seen
inserting (\ref{wave}) into the quadratic action: the kinetic term
of  a given 4D mode $\xi^{(4)}(x)$ reads:
\be
\left(\int dr
e^{2B(r)} |\xi(r)|^2 \right) \int d^4x
\left(\de_\m\xi^{(4)}(x)\right)^2 = \left(\int dr |\psi(r)|^2
\right)\int d^4x  \left(\de_\m\xi^{(4)}(x)\right)^2\,.
 \ee
Requiring  finiteness of the kinetic term  leads to
\be
\int dr
|\psi(r)|^2 < \infty.
\ee

Therefore, in these coordinates, the problem of finding
the spectrum translates into a standard quantum mechanical problem.
The general features of the spectrum can be inferred from the
properties of the effective Schr\"odinger potential (\ref{gluepot}). Given the
functions $B(r)$ and $M(r)$ we can obtain useful information
without finding explicit solutions.

In the case we are mostly interested in, i.e. the infinite-range
case, a number of interesting properties of the spectrum can be
derived in full generality.

\subsection{Existence of a mass gap}

Consider first the effective potential in  the asymptotically $AdS_5$ region, $r\sim 0$.
There,  the potential behaves
universally, since  $B(r) \sim 3/2 A(r)$ in the UV
for all kinds of fluctuations:
\be\lab{peats}
V(r) \sim {15\over 4} {1\over r^2} \to +\infty,\qquad r\to 0
\ee

Next, notice that the  equation (\ref{schro}) can be written as:
\be
\left(P^\dagger P + M^2(r)\right)\psi = m^2\psi, \qquad P = (-\de_r + \dot{B}(r))
\ee
 Taking into account also the behavior near $r=0$, the following
statements hold:
\begin{enumerate}
\item {\em if $M^2(r)\geq 0$ the spectrum is non-negative}
\item {\em If the space-time extends to $r=+\infty$, and if moreover $V(r)$ does not vanish as $r\to \infty$, then there is a mass gap}.
\end{enumerate}
The second statement  follows from the fact that, if the
singularity is at $r\to +\infty$, there can be no normalizable zero-mode
solutions to eq. (\ref{schro}) with $M^2(r)=0$. Indeed, the
$m=0$ solutions can be found exactly:
\be\label{zeromodes}
\psi_0^{(1)}(r) = e^{B(r)}, \qquad \psi_0^{(2)} = e^{B(r)} \int_0^r dr' e^{-2B(r')}\ee
It is easy to see that $ \psi_0^{(1)}$ is not normalizable in the UV,
since there $B\sim -3/2 \log r$. On
the other hand,  $\psi_0^{(2)}$ is not normalizable in the IR, no  matter what
is the behavior of $B(r)$ as $r\to \infty$.
See \cite{nitti} for a detailed discussion. Since there are no normalizable
zero-modes, the only obstacle to having a mass gap would be a continuous
spectrum starting at zero. This cannot be the case if $V(r)$ does not
vanish as $r\to \infty$.

If the singularity is at a finite $r=r_0$, this argument still holds
if the scale factor vanishes at least as
 $(r_0-r)^\delta$, with $\delta>1$. Then,
again one can show that there are no zero-modes. For $0<\delta<1$
things  are more subtle, and we will leave this case for separate
discussion in Section \ref{beware}.

For the various particle types we analyze (vector mesons, and glueballs of
spins up to 2) we will see that property $1$ always holds, in fact
$M^2=0$. Moreover, for all particles we consider, it turns out that
the function $B(r)$ has the same IR asymptotics as $A(r)$. In the
backgrounds with infinite $r$ range, as $r\to \infty$: \be A(r) \sim - \left({r\over
R}\right)^{\a}, \ee therefore \be\lab{potIR} V(r) = \dot{B}^{2}(r) +
\ddot{B}(r) \sim R^{-2}\left({r\over R}\right)^{2(\a-1)}. \ee We see
that  the mass gap condition  is $\a \geq 1$. {\em This is
the same condition we found independently for quark confinement}. If
we require $\a>1$ strictly, we moreover obtain a purely discrete
spectrum, since then  $V(r) \to +\infty$ for large $r$. If $\a=1$ the
spectrum becomes continuous for $m^2 \geq V(r\to\infty)$.

\subsection{Large $n$ mass asymptotics and linear confinement\label{lco}}

In the confining backgrounds, where the potential behaves as in eq.
(\ref{potIR}) for large $r$  and as (\ref{peats}) for small $r$,
the large eigenvalue asymptotics of eq. (\ref{schro}) may be
obtained through the WKB approximation: the quantization condition
is approximately given by the quantization of the action integral:
\be
 n \pi = \int_{r_1}^{r_2} \sqrt{m^2 -V(r)} dr
\ee
where $r_1$ and $r_2$ are the turning points.
For large $m^2$, $r_1\sim 0$, and  $ (r_2/R)^{2(\a-1)}\simeq R^2 m^2$,
so we can write:
\be
 n \pi = m \int_0^{R (m R)^{1/(\a-1)}}  \sqrt{1 - {V(r)\over m^2}}dr
\ee Assuming $m^2\gg V(r)$ in the intermediate region, the second
term under the square  root becomes relevant only when $V(r)$ takes
its asymptotic form. We can therefore  write \be n\pi \simeq m
\int_0^{R (m R)^{1/(\a-1)}} \sqrt{1 - \left[\left({r\over
R}\right)^{\a-1}{1\over m R}\right]^2} dr = \left({m\over
\Lambda}\right)^{\a\over \a-1} \int_0^1 d x\sqrt{1-x^{2(\a-1)}} \ee
where $\Lambda = R^{-1}$. For large $n$: \be\lab{slope} m ~\sim
~\Lambda ~n^{\a-1\over \a}. \ee In particular we have ``linear
confinement'' ($m^2 \sim n$) if $\a =2$.\footnote{A dilaton and/or
a warp factor $A(r)$ behaving as  $r^2$ for large $r$, were advocated in
\cite{linearconfinement}, in order to obtain a linear spectrum for
mesons. In that work, the authors suggest  an $AdS_5$ space-time
together with a dilaton with   $r^2$ asymptotics. Such a background
is sometimes referred to as a  ``soft wall'' model, and has
been used to compute meson-related quantities (see e.g. \cite{radu}
for recent work). A similar approach was adopted in  \cite{forkel}
to treat both baryons and mesons. In that work it is the scale factor,
rather than the dilaton, that drives the IR asymptotics.

We should stress that those backgrounds, unlike the ones
we study here,  are not obtained as solutions of any set of field equations.
For example, from our previous discussion  it is apparent that,
if the dilaton grows as $r^2$ in the IR,
its backreaction is such that the space-time cannot be  close to $AdS_5$ for
large $r$, independently of the
form of the dilaton potential. Moreover, as we discuss in
Section 5.4, the dynamics of mesons could be described by a
different mechanism \cite{ckp} which does not necessarily require $\a=2$ for a
linear meson spectrum.}. For $\a\to \infty$ the spectrum looks similar
to the one of a ``particle in a box'' potential, $m^2 \sim n^2$,
characteristic of ``hard wall'' models and more generically of any
background with finite $r_0$ (see Section \ref{finiter0}).

\subsection{Universal asymptotic mass ratios}

Here we  derive some general properties of the glueball spectrum
that are independent of the specific potential chosen. In this
section, we consider the backgrounds where $X\to -1/2$ at the
singularity. As we have seen, this
is generic in confining backgrounds with singularity at
$r=+\infty$.
 The function $B$ in (\ref{scaff}) generally asymptotes
to
\begin{equation}\lab{asympB}
B\to B_f\log(\l),\quad as \quad \l\to\infty,
\end{equation}
where the coefficient $B_f$ depends on the type of particle.
It is essentially
determined by the normalization of the kinetic term in the effective action
of the specific type of background fluctuation that correspond to the particle
in question.

One can also write down the effective Schr\"odinger potential
(\ref{gluepot}) using $\lambda$ as a coordinate,
 \begin{equation}
\lab{schro2} V_s(\l) = \frac{3V_0}{4}X^2
e^{\frac23\int^{\l}\frac{d\l'}{\l'}\le(\frac1X -4X\ri)}
\le(\l^2\frac{d^2B}{d\l^2}+\l\frac{dB}{d\l}\le(\frac{1}{3X}+1-\frac{4X}{3}
+\l\frac{d\log|X|}{d\l}\ri)+\le(\l\frac{dB}{d\l}\ri)^2\ri) .
\end{equation}
{}From (\ref{schro2}) we observe that $V_s$ in the IR
asymptotes to,
\begin{equation}\lab{potIRlambda} V_s \to
\frac{9}{4l^2}e^{2A_0+\frac23\int^{\infty}\frac{d\l'}{\l'}\le(\frac1X-4X\ri)}B_f^2\,>\,0.
\end{equation}
The exponential depends on the specified $\beta$-function of the gauge
theory. However the constant $B_f$ is universal for a given type of
particle, \ie it is independent of the specified running of the
gauge coupling. For example $B_f$ is 1 both for the $0^{++}$
glueballs and the $2^{++}$ glueballs, i.e.
\be
\frac{V_s(0^{++})(r)}{V_s(2^{++})(r)} \to 1, \qquad r\to +\infty.
\ee

This means that  the glueballs have a spectrum
whose slope is independent of their spin for large mass:
\be
\label{unirat2} \frac{m^2_{n\to\infty}(0^{++})}{m^2_{n\to\infty}(2^{++})}=1
\ee
This
fits nicely with the semi-classical string models (see e.g.
\cite{kadaylov}) for the glueballs that predict \be\lab{glpred}
\frac{m^2}{2\pi \s_a} = 2n + J + c, \ee where $\s_a$ is the adjoint
string tension, $J$ is the angular momentum and $c$ is some number
of order 1. Our finding (\ref{unirat2}) is in accord with the
general prediction of such models that the adjoint string tension is
universal for glueballs with different spin.

Next, we move to specific analysis of the spectra of different species of glueballs.

\subsection{Glueball spectra}

At the lowest mass level the bulk theory contains the dilaton $\Phi$,
the metric $g_{\mu\nu}$, and the axion $a$. The spectrum of physical
fluctuations of these fields is dual to the spectrum of glueballs in
the gauge theory, as these fields come from the closed sting
sector. The physical massive fluctuations
of the minimal metric$+$dilaton system
consists of  one spin-2 mode (5 degrees of freedom), and one spin-0
mode.\footnote{See e.g. \cite{nitti}
for a complete discussion of the
identification of the physical fluctuations
and the corresponding field equations.
In the massless sector there are a massless spin-2
(2 polarization), one massless spin-1 (2 polarizations) and 2
massless spin-0  modes.  However we will not have massless modes in
our spectra, so we will not consider this case further. In
\cite{nitti} it was shown that in general the presence of a massless
spin-2 mode is only possible if an IR singularity appears and if
special non-local boundary conditions are put at the singularity.
This is compatible with the Weinberg-Witten theorem \cite{ww}.
} The
fluctuations of the axion field correspond to pseudoscalar
glueballs. They  do not mix with those in the scalar sector of the
metric-dilaton system, since we neglect the backreaction of the
axion on the geometry\footnote{However, they are expected to mix with $\eta'$ if we introduce flavor branes.}.

Throughout this section we consider only the IR asymptotics of the
type (\ref{metric-asympt0}),
\be\label{asymptA}
A(r) \sim - \left({r\over R}\right)^\a + \ldots, \qquad \a \geq 1
\ee
with no further assumptions on the subleading behavior. We will consider the case
with singularity at finite $r$ in Section \ref{finiter0}

\subsubsection{Scalar glueballs}

In 5D Einstein-Dilaton gravity there exists a single gauge invariant
spin-0 mode\footnote{Here ``gauge invariace'' refers to the linearized
5D diffeomorphisms. The precise definition of this field is
\be
\zeta = \psi - {1\over 3 X(r)} \delta\phi=\psi - {\l\over \beta(\l)} \delta\phi,
\ee
 where $\delta\phi$
and $\psi$ are the fluctuations in the dilaton and in the scalar part
of the $g_{ij}$ metric component. See e.g.\cite{nitti}.},
$\zeta(r,x)$, satisfying the equation (\ref{scaff}) with
\be
\label{B0} B_0(r) = {3\over 2}A(r) + {1\over 2}\log  X^2, \qquad
M(r)=0,
\ee
 The effective Schr\"odinger potential is given by eq.
(\ref{gluepot}). Notice that, both for large and small $r$, the
second term in $B_0$ (\ref{B0}) is negligible. Therefore the leading
asymptotics are,
\be
\lab{scalarpot} V_0(r) \sim \left\{\begin{array}{ll} {9\over 4} R^{-2}
\left({r\over R}\right)^{2(\a-1)}, & \;\; r\to \infty, \\
\sim {15\over 4}{1\over r^2}, & \;\; r\to 0. \end{array}\right.
\ee
 We have a mass gap and discrete spectrum if and only if $\a>1$.

In the UV, the gauge invariance of $\zeta$ indicates that it is dual to the renormalization group invariant operator
$\beta(\l)Tr[F^2]$ \cite{narison}.

\subsubsection{Tensor glueballs}

The massive spin-2 glueballs are described by transverse traceless
tensor fluctuations $h_{ij}$ of the 4D part of the metric:
\be
ds^2 = e^{2A(r)}\left( dr^2 +(\eta_{ij} + h_{ij}) dx^i
dx^j\right)
\ee
These fluctuations satisfy the equation
(\ref{scaff}) with
\be\label{B2} B_2(r) =  {3\over 2}A(r), \qquad
M(r)=0
\ee
The effective Schr\"odinger potential has the same
asymptotics as (\ref{scalarpot}):
\be\lab{tensorpot}
 V_2(r) \sim \left\{\begin{array}{ll}
{9\over 4} R^{-2} \left({r\over R}\right)^{2(\a-1)}, & \;\; r\to
\infty, \\   {15\over 4}{1\over r^2}, & \;\; r\to 0. \end{array} \right.
\ee
Together with (\ref{scalarpot}) this confirms (\ref{unirat2}).
However, due to the difference between (\ref{B0}) and (\ref{B2}),
the spin-0 and spin-2 glueball spectra are not degenerate. This is unlike the
$AdS/QCD$ models with exact $AdS$ metric and constant dilaton: there
the scalar and tensor modes are exactly degenerate.
We will
see in an explicit background that the lowest-lying spin-0 glueball
is lighter than the lowest spin-2 glueball. We expect this fact to
be generic, although we can not provide a proof in our set-up.

\subsubsection{Pseudo-scalar glueballs\label{pss}}

The Einstein frame axion action in the conformal coordinates reads:
\be
\label{axionaction}
S_a = -{M^3\over 2} \int d^5 x  Z(\l)
e^{3A} (\de a)^2.
\ee
Since the axion appears quadratically, this is
also the action for the fluctuations. We thus have:
\be\label{B0-}
B_a(r)  =  {3\over 2}A(r) + {1\over 2} \log Z(\l).
\ee
To leading
order in  string perturbation theory, $Z(\l) = \l^2$.
However, this in general is expected to receive corrections from the
5-form, similar to the dilaton potential. Indeed, if this were not
the case one would find a puzzling result: one would obtain a
continuous spectrum for the pseudo-scalar glueballs starting at
$m=0$. To see this, assume as in Section \ref{axion} that   $Z(\l)
= \l^d$ for large $\l$. Then, using eq. (\ref{dilr1}) in
(\ref{B0-}) we obtain:
\bea\label{B0-2} B_a(r) && = {3\over 2}\left(1-{d\over 2}\right)A(r)
+ {d\over 2} {3\over 4} (\a-1)\log r/R \nn\\
&& \sim
\left\{ \begin{array}{ll} {3\over 4}(d-2) \left({r/ R}\right)^\a & d\neq 2 \\
{3\over 4} (\a-1)\log r/R \quad & d=2,
\end{array}\right.
\eea where we used (\ref{asymptA}). The IR asymptotics of the
Schr\"odinger potential are (using (\ref{gluepot})),
 \be\label{axionpot}
V_a(r) \sim \left\{\begin{array}{ll} {9\over 16}\left[(d-2)^2/ R^2 \right]
\left(r/ R\right)^{2(\a-1)} & d\neq 2 \\
\left[{9\over 16}(\a-1)^2-{3\over 4}(\a-1)\right]{1\over r^2} \quad
& d=2.
\end{array}\right.
\ee Thus the potential and the spectrum have the same features as
the other glueballs, \emph{unless} the perturbative result $d=2$ is
unmodified.

The asymptotic mass ratio for large $n$ of the $0^{-+}$ to $0^{++}$
glueball states can be read-off comparing  the large $r$ asymptotics
of (\ref{axionpot}) for $d\neq2$ and (\ref{scalarpot}): \be
{V(0^{-+})\over V(0^{++})}  \to {1\over4 }(d-2)^2 \ee

Using the expected asymptotic glueball universality argument (as in
(\ref{unirat2})) \be
\frac{m^2_{n\to\infty}(0^{-+})}{m^2_{n\to\infty}(0^{++})} =
\frac{m^2_{n\to\infty}(0^{++})}{m^2_{n\to\infty}(2^{++})}=1, \ee we
can determine \be d=4. \ee This result predicts an interesting
renormalization of the bare axion
 kinetic term, (\ref{axionaction}).

It is appropriate to point out that the effective Schr\"odinger potential
 for the $0^{-+}$ trajectory of glueballs can be written in terms
of the background axion solution  (\ref{d12h}) as
\be
V_a(r)={1\over 4}{\dddot a\over \dot a}
\ee
An interesting corollary of this relation is that the potential is independent
of the UV $\theta$-angle of QCD, $\theta_{UV}$.

When $\theta\not =0$ glueballs of different parities mix as well as their energies become $\theta$-dependent
to next order in $1/N_c$.
The $\theta$-dependence of the glueball spectrum can be calculated by considering
 the first order (${\cal O}(1/N_c^2)$)
backreaction of the axion solution to the QCD vacuum. This can be an interesting
test as lattice data on this exist, \cite{thetalat}.
We will not attempt however the calculation in this paper.

\subsection{Singularity at finite $r_0$}\label{finiter0}

In the previous subsections we considered backgrounds with infinite
range in $r$. Here we discuss the case in which the IR singularity
is at some finite $r=r_0$. As discussed in Appendix
\ref{confinement} and summarized in Table \ref{summarytable}, these
backgrounds generically lead to a confining string potential. To
analyze the mass spectrum, consider the case when the IR singularity
has the following form: \be\lab{finite} A(r) \sim \delta
\log(r_0-r), \quad r\to r_0. \ee
The effective Schr\"odinger potential (\ref{gluepot}) has the same
asymptotic form in the IR both for the scalar and the tensor
glueballs. This is because the functions (\ref{B0}) and  (\ref{B2}),
differ only by a function of $X(r)$ which, as shown in  Appendix
\ref{confinement}, asymptotes to a ($\delta$-dependent) constant as $r\to
r_0$. Then, both for the spin-0 and the spin-2 glueballs, the
effective Schr\"odinger potential has the following asymptotic form:
\be\lab{potasfinite} V(r) \sim {15\over 4}{1\over r^2}\quad (r\to
0), \qquad V(r) \sim {9\over 4}{\delta(\delta-2/3)\over (r-r_0)^2},
\quad (r\to\infty). \ee For $\delta>2/3$, $V\to +\infty$ in the IR,
and by the same general argument we used in subsection 4.1 we obtain a
mass gap and a discrete spectrum. The treatment of the case
$0<\delta<2/3$ (in fact $0<\delta<1$) requires extra care, as we
discuss in the next subsection.

The large mass asymptotics of both the scalar and the tensor
glueballs in the backgrounds (\ref{finite}) are universal. They
depend neither on $\delta$ nor the details of the metric in the
bulk: due to (\ref{potasfinite}), the Schr\"odinger equation for
large eigenvalues is effectively the one for a particle in a box of
size $r_0$, so for large mass eigenstates we  obtain
\be\lab{massfinite} m_n^2 \sim {n^2\over r_0^2}. \ee This does not {\it a priori}
prevent the mesons to have a linear mass spectrum, since
this is driven  by the tachyon dynamics\footnote{This observation
avoids the arguments put forward in \cite{schreiber} regarding the
meson spectra in gravity duals.}, as in the infinite range
case.

In the case of power-law behavior for $A(r)$,
\be
A(r) \sim -{C\over (r_0 -r)^{\tilde{\a}}},\qquad \tilde{\a}, C>0
\ee
the potential in the IR always asymptotes to $+\infty$, and it is
  steeper than $(r_0-r)^{-2}$ in the IR:
\be
V(r)\sim {9\over 4}{C^2\over (r_0-r)^{2\tilde{\a}+2}}
\ee
\subsubsection{The pathologies for  $0<\delta<1$}\label{beware}

As discussed in \cite{nitti} in
a different context (see also \cite{brax} for a related discussion),
this range of parameters is somewhat pathological,
 since it requires additional
boundary conditions at the singularity, and the spectrum is
not determined by the normalization condition alone.

The Schr\"odinger equation for a generic
mass eigenstate  close  to $r_0$ is:
\be
-\ddot{\psi} +  V(r) \sim -\ddot{\psi}+  {9\over 4}{\delta(\delta-2/3)\over (r-r_0)^2} \psi =m^2 \psi.
\ee
For $r\sim r_0$ we can neglect the mass term on the r.h.s, and find
the asymptotic solution close to $r_0$:
\be\label{2norm}
\psi(r) \sim c_1 (r_0 -r)^{3\delta/2 } + c_2  (r_0 -r)^{1- 3\delta/2}.
\ee
For $\delta<1$ both solutions are square-integrable, and they
both vanish at $r_0$ if in addition  $\delta<2/3$. Therefore,
for $0<\delta<1$, normalizability alone is not enough to fix the
spectrum uniquely. One has to specify some extra boundary
conditions at the singularity, which  may be given by fixing
the ratio $c_1/c_2$.\footnote{In operator language, the Hamiltonian
of this problem is symmetric but not essentially self-adjoint, and
it admits an infinite number of self-adjoint extensions, each with
a different spectrum, parametrized by the choice of $c_1/c_2$.}
In contrast, for $\delta\geq1$ normalizability in the IR \emph{forces}
the choice $c_2=0$, and there is no ambiguity.

Ultimately it is this extra input at the singularity
that determines the  spectrum
in a background with $\delta<1$, and not the dynamics
of the theory at any finite energy. This situation
is not so different from the hard-wall models \cite{erlich,pomarol},
where one also has to specify IR boundary conditions
for the fluctuations to compute the spectrum.

We note here that the background studied of Csaki and Reece in \cite{cr}
falls in this class of examples: one can easily check that  its metric in conformal frame
behaves as in  eq. (\ref{finite}) with $\delta=1/3$. In computing
the spectrum,  IR Neumann boundary conditions are chosen in \cite{cr},
but according to the present discussion this is as good a  choice
as any other.

\subsection{Adding flavor\label{flavor}}

A small number $N_f\ll N_c$ of quark flavors can be included
in our setup by adding space-time filling ``flavor-branes''.
In this case they are pairs of space-filling $D4-\bar{D4}$ branes.
It was proposed in  \cite{ckp} that
the proper treatment of the flavor sector (including chiral symmetry breaking)
 involves the dynamics of the open string tachyon of the $D4-\bar{D4}$ system.
According to this, the  meson sector of the 4D gauge theory is
captured holographically  by the open string DBI+WZ action, which schematically reads, in the
string frame,
\be\lab{actiongauge1}
S[T,A^L,A^R]  =S_{DBI}+S_{WZ}
\ee
where the DBI action for the pair is
\be\label{DBI}
S_{DBI}=\int dr d^4 x  ~{N_c\over \l} ~
{\bf Str}\left[V(T) \bigg(\sqrt{-\det \left(g_{\mu\nu} + D_{\{\mu} T^{\dagger} D_{\nu\}} T +
F^L_{\mu\nu}\right)}+
\right.
\ee
$$
\left.+\sqrt{-\det \left(g_{\mu\nu} + D_{\{\mu} T^{\dagger} D_{\nu\}} T + F^R_{\mu\nu}\right)}\bigg)\right]
$$
Here $T$ is the tachyon, a complex $N_f\times N_f$ matrix. $A^{L,R}_{\mu}$ are the world-volume gauge fields
of the $U(N_f)_L\times U(N_f)_R$ flavor symmetry, under which the tachyon is transforming as the $(N_f,\bar N_f)$,
a fact reflected in the presence of the covariant derivatives\footnote{We are using the conventions of \cite{ckp}.}
\be
D_{\m}T\equiv \partial_{\mu}T-iTA^L_{\m}+iA^R_{\m}T\sp D_{\m}T^{\dagger}\equiv \partial_{\mu}T^{\dagger}-iA^L_{\m}T^{\dagger}+iT^{\dagger}A^R_{\m}
\ee
transforming covariantly under
\be
T\to V_RTV_L^{\dagger}\sp A^L\to V_L(A^L-iV_L^{\dagger}dV_L )V_L^{\dagger}\sp A^R\to V_R(A^R-iV_R^{\dagger}dV_R )V_R^{\dagger}
\ee
as well as the field strengths $F^{L,R}=dA_{L,R}-iA_{L,R}\wedge A_{L,R}$ of the $A^{L,R}$ gauge fields.
$\l\equiv e^{\Phi}=N_c e^{\phi}$ is as usual the 't Hooft coupling.
We have also used the symmetric trace ($\equiv Str$) prescription although higher order terms of the non-abelian DBI action
are not known. It turns out that such a prescription is not relevant for the vacuum structure in the meson sector
(as determined by the classical solution of the tachyon) neither for the mass spectrum.
The reason is that we may treat the light quark masses as equal to the first approximation and then in the vacuum,
$T=\tau {\bf 1}$ with $\tau$ real, and this is insensitive to non-abelian ramifications.
Expanding around this solution, the non-abelian ambiguities in the higher order terms do not enter at quadratic order.
Therefore, for the spectrum we might as well replace $Str\to Tr$.

The WZ action on the other hand is given by\footnote{This expression was
proposed in \cite{tach1}
and proved in \cite{KrausLarsen,TTU} using boundary string field theory}:
\be\label{WZtach}
S_{WZ}=T_4 \int_{M_5} C\wedge \mathrm{Str} ~\exp\left[{i
2\pi\a'\mathcal{F}}\right]
\ee
where $M_5$ is the world-volume of the
${\rm D}4\,$-$\overline{{\rm D}4}$ branes that coincides with the full space-time. Here,
$C$ is a formal sum of the RR potentials $C=\sum_n (-i)^{\frac{5-n}{2}}C_n$,
and $\mathcal{F}$ is the curvature of a superconnection
${\cal A}$.
In terms of the tachyon field matrix $T$
and the gauge fields  $A^L$ and  $A^R$
living respectively on the branes and
antibranes, they are
(We will set $2\pi \alpha'=1$
and use the notation of
\cite{KrausLarsen}):
\be
i\mathcal{A}=\left(\begin{array}{cc} iA_L & T^\dagger\\
T & iA_R\end{array}\right)\,,\qquad
i\mathcal{F}=\left(\begin{array}{cc} iF_L-T^\dagger T & DT^\dagger\\
DT & iF_R-TT^\dagger\end{array}\right)
\label{AFdef}
\ee
The curvature of the superconnection is defined as:
\be
{\cal F} = d{\cal A} - i {\cal A} \wedge {\cal A}\sp
d{\cal F} - i {\cal A} \wedge {\cal F} + i {\cal F} \wedge {\cal A} = 0
\ee
Note that under (flavor) gauge transformation it transforms homogeneously
\be
{\cal F}\to \left(\begin{array}{cc} V_L& 0\\
0 & V_R\end{array}\right)~{\cal F}~\left(\begin{array}{cc} V^{\dagger}_L& 0\\
0 & V^{\dagger}_R\end{array}\right)
\ee
In \cite{ckp} the relevant definitions
and properties of this {\it supermatrix}
formalism can be found.

By expanding we obtain
\be
S_{WZ}=T_4\int C_5\wedge Z_0+C_3\wedge Z_2+C_1\wedge  Z_4+C_{-1}\wedge Z_6
\ee
where $Z_{2n}$ are appropriate forms coming from the expansion of the exponential of the superconnection.
In particular, $Z_0=0$, signaling the global cancellation of 4-brane charge, which is equivalent to the cancelation
of the gauge anomaly in QCD.
Further, as was shown in \cite{ckp}
\be
Z_2=d\Omega_1\sp \Omega_1= i STr(V(T^{\dagger}T))Tr(A_L-A_R)-\log\det (T)d (Str V(T^{\dagger}T))
\ee
This terms provides the Stuckelberg mixing between $Tr[A^L_{\m}-A^R_{\m}]$ and the QCD axion that is dual to $C_3$.
Dualizing the full action we obtain
\be
S_{CP-odd}={M^3\over 2N_c^2}\int d^5x\sqrt{g}Z(\l)\left(\pa a+i\Omega_1\right)^2
\ee
$$
={M^3\over 2}\int d^5x\sqrt{g}Z(\l)\left(\pa_{\mu} a+\zeta \pa_{\m}V(\tau)-\sqrt{N_f\over 2}V(\tau)A^A_{\m}\right)^2
$$
with
\be
\zeta =\Im \log\det T\sp A_{L}-A_{R}\equiv {1\over 2N_f}A^A\II+(A^a_{L}-A^a_{R})\lambda^a
\ee
and where we have set the tachyon to it vev $T=\tau {\bf 1}$ .
This term is invariant under the $U(1)_A$ transformations
\be
\zeta\to \zeta+\e\sp A^A_{\mu}\to  A^A_{\mu}-\sqrt{2\over N_f}\partial_{\mu}\e\sp a\to a-N_f\e V(\tau)
\ee
reflecting the QCD $U(1)_A$ anomaly.
It is this Stuckelberg term together with the kinetic term of the
tachyon field that is responsible for the mixing between the QCD axion
and the $\eta'$. In terms of degrees of freedom, we have two scalars
$a,\zeta$ and an (axial)  vector, $A^A_{\m}$.
We can use gauge invariance to remove the longitudinal components of
 $A^A$. Then an appropriate linear combination of the two scalars will become
 the $0^{-+}$ glueball field
while the other will be the $\eta'$. The transverse (5d) vector will
provide the tower of $U(1)_A$ vector mesons.

The next term in the WZ expansion couples the baryon density to a one-form RR
field $C_1$. There is no known operator expected to be dual to this
bulk form.
However its presence and coupling to baryon density can be understood as follows.
Before decoupling the $N_c$ $D_3$ branes, its dual form $C_2$ couples to the U(1)
$_B$ on the $D_3$ branes via the standard $C_2\wedge F_B$ WZ coupling.
This is dual to a free field, the doubleton, living only at the boundary of the bulk.
Once we add the probe $D_4+\bar D_4$ branes the free field is now a linear combination
of $A^B$ and an $N_f/N_c$ admixture of $A^V$ originating
on the flavor branes. The orthogonal combination is the baryon number current on the
flavor branes and it naturally couples to $C_1$.
Therefore the $C_1$ field is expected to be dual to the topological baryon current at
 the boundary.

Finally the form of the last term requires some explanation.
By writing $Z_6=d\Omega_5$  we may rewrite this term as
\be
\int F_0\wedge \Omega_5\sp F_0=dC_{-1}
\ee
$F_0\sim N_c$ is nothing else but the dual of the five-form field strength.
This term then provides the correct Chern-Simons form
that reproduces the flavor anomalies of QCD. Its explicit form in terms of the gauge fields $A_{L,R}$
and the tachyon was given in equation (3.13) in \cite{ckp}.

To proceed further and analyze the vacuum solution we set $T=\tau~{\bf 1}$ and set the vectors to zero.
Then the action (\ref{actiongauge1}) collapses to
\be\lab{actiongauge}
S[\tau,A_M]  = N_cN_f\int dr d^4 x  ~e^{-\Phi}
V(\tau) \sqrt{-\det \left(g_{\mu\nu} + \de_\mu \tau \de_\nu \tau \right)}
\ee
Following \cite{ckp}
we assume the following tachyon potential, motivated/calculated in studies of tachyon condensation:
\be\lab{tachyonpot}
V(\tau) = V_0 e^{-{\mu^2\over 2} \tau^2}
\ee
where $\mu$ has dimension of mass. It is fixed by the requirement that $\tau$
has the correct bulk mass to couple to the quark bilinear operator on the boundary.

In our  minimal setup,
the brane-antibrane system fills the whole bulk. Therefore these fields
are  bulk fields. We will eventually expand the action
at most to quadratic order in the gauge fields.

Chiral symmetry breaking in the IR is described by a non-trivial tachyon profile.
For small $N_f$ we can neglect the backreaction of the tachyon on the metric-dilaton
system, and solve the equation for the tachyon profile on a given background,
e.g. one  of the confining backgrounds we discussed.
Once a solution for the tachyon is found,  the spectrum of mesons is
given by the spectrum of  fluctuations around this background. For example,
vector mesons are described by the fluctuations
of the components  $A_i$ around the $A_i=0$ configuration,
in a given background for the metric, dilaton and tachyon.

\subsubsection{Tachyon dynamics}

In the conformal frame,
the action (\ref{actiongauge}) becomes:
\be
S[\tau] = N_c N_f\int dr d^4 x   e^{4A_s(r)-\Phi(r)} V(\tau)\sqrt{e^{2A_s(r)} +\dot{\tau}(r)^2},
\ee
from which we obtain the nonlinear field equation:
\be\lab{tachyoneq}
\ddot{\tau} + \left(3\dot{A}_S -  \dot{\Phi}\right)\dot{\tau} +  e^{2A_S} \mu^2\tau +
e^{-2A_S}\left[4\dot{A}_S - \dot{\Phi}\right](\dot{\tau})^3 + \mu^2 \tau (\dot{\tau})^2=0.
\ee
The tachyon is dual to the dimension 3 quark bilinear operator. Near the boundary, $r\to 0$, we expect
$\tau=m r+ \sigma r^3 +\ldots$\footnote{ For simplicity, here we take   all quark masses to be equal.
The tachyon is therefore proportional to the identity matrix in flavor space.}.
Thus, in the UV We may ignore the non-linear terms in eq. (\ref{tachyoneq}), which then
reduces to
the equation for a free scalar field with mass $\mu$ on an asymptotically $AdS_5$ background.
In order for this to be dual to the quark bilinear operator, with naive dimension 3 (to  leading order),
we need $3 = 2 + \sqrt{4 - \mu^2 \ell^2}$, hence $\mu^2\ell^2 =3$.

It is argued in \cite{ckp} that consistency
of the bulk gauge theory (i.e. absence of extra gauge anomalies in the IR )
 requires the tachyon to diverge before or at the singularity.
In  Appendix \ref{tachyonapp} we analyze the possible singularities
of the solutions of eq. (\ref{tachyoneq}), in backgrounds with IR asymptotics
(\ref{asymptA}). We
show that   the  only consistent solution for  $r\to \infty$,
is such that the tachyon diverges exponentially:
\be\lab{tachyonIRas2}
\tau(r) \sim \tau_0 \exp\left[{2\over\a} {R\over\ell^2} \,r\right],  \qquad r\to \infty,
\ee
where $\tau_0$ is an integration
 constant determined by  UV initial conditions.

We also analyze possible  singularities of
the solutions  at finite $r$. We find that generically, the
tachyon cannot diverge at any finite $r$, where both $A_S$ and $\Phi$
are regular, except special  points where $4\dot{A_s}-\dot{\Phi}=0$. This
does not happen in our backgrounds.
Instead,  the generic solution of (\ref{tachyoneq}) has a singularity at finite $r_*$,
where   $\tau(r_*)$ stays finite but its derivatives diverges:
\be
\tau \sim \tau_* + \gamma \sqrt{r_* - r}.
\ee
Such solutions are unphysical, since around $r_*$ the backreaction on the metric
 is no longer negligible: the tachyon stress tensor diverges as $1/(r_*-r)$, and our
assumption that the tachyon does not perturb the background is invalid.
On the other hand this is not physically reasonable, since adding a small number
of flavors should not change dramatically the pure gauge dynamics in the large $N_c$
limit.\footnote{Notice
that the backreaction is not problematic if the tachyon itself, and not
just its derivative, diverge: the stress tensor is multiplied by
the tachyon potential, that vanishes exponentially fast as $\tau\to \infty$,
resulting in the recombination of the branes-antibrane pairs in the IR,
which leaves the unperturbed metric and dilaton background.}

Discarding all  but the exponentially divergent solution singles out special initial conditions
in the UV, which correspond to fixing the chiral condensate as a function
of the quark mass \cite{ckp}, i.e. the coefficients of the subleading and leading
terms in the UV expansion of $\tau(r)$.

\subsubsection{Vector mesons}

Once the correct tachyon profile is found from eq. (\ref{tachyoneq}),
this enters the action for the tachyon and  the bulk gauge
fields fluctuations, and determines  their  spectrum. The resulting 4D mass
eigenstates correspond to the various mesons in the dual theory. Here, we only consider
the vector mesons, that  correspond
to the transverse vector components of the 5D gauge fields, $A_i=A^L_{i}+A^R_{i}$.

The quadratic action for the gauge fields is, from eq. (\ref{actiongauge1}):

\be\lab{actiongauge2}
S \sim -{1\over 4}\int dr d^4 x ~ e^{-\Phi}V(\tau) \sqrt{-\hat{g}}
\hat{g}^{\mu\nu}\hat{g}^{\rho\sigma}F_{\mu\rho}F_{\nu\sigma},
\ee
where $\hat{g}$ is the effective (open string) metric felt by the
gauge fields in  the presence of the tachyon:
\be
d\hat{s}^2 = \left(e^{2A_S} +  (\dot{\tau})^2\right) dr^2 + e^{2A_S}\eta_{ij} dx^i dx^j.
\ee
This metric is still asymptotically $AdS$, since $e^{2A_S}$ dominates in the UV,
however, although still conformally flat,  it is not in the conformal frame.
It differs considerably from the bulk background
metric in the IR.

The large $r$ behavior  of $A_S(r)$ and $\tau(r)$ are, from eq. (\ref{strscale}) and (\ref{tachyonIRas2}):
\be\lab{tachyonIRas1}
 A_S(r) \sim {\a -1\over 2} \log r/R,\qquad \tau(r) \sim \tau_0 \exp\left[{2\over\a}
  {R\over\ell^2} \,r\right],  \qquad \a\geq1.
\ee
The second term dominates $\hat{g}_{rr}$ in the infrared. To recast the action
in the form  (\ref{genac}), and read-off the effective Schr\"odinger potential
for the mesons, we change  variables from $r$ to $\tau$. Using
 (\ref{tachyonIRas1})  to express
$A_S$ as a function of $\tau$ in the IR, the effective
metric becomes for large $\tau$:
\be
d\hat{s}^2 \sim d\tau^2 + \left({\a \ell^2 \over 2 R^2} \log \tau/\tau_0 \right)^{\a-1} \eta_{ij} dx^i dx^j,
\ee
where we have neglected the first term in $\hat{g}_{rr}$.  We now pass  to a
new conformal frame,
by  changing  variables from $\tau$ to $\hat{r}$, defined by
\be
d\tau =  \left({\a \ell^2 \over 2 R^2} \log \tau/\tau_0 \right)^{(\a-1)/2}  d\hat{r}+\cdots,
 \ee
which is solved asymptotically for large $\tau$ by:
\be
\hat{r} = \left({2 R^2\over \a \ell^2 }\right)^{(\a-1)/2}
{\tau \over  \left(\log \tau/\tau_0 \right)^{(\a-1)/2}}+\cdots.
\ee
To leading order we can also replace $\log \tau/\tau_0 $ by $\log r /\tau_0 $
in the above relation and the metric reads:
\be
d\hat{s}^2 = e^{2\hat{A}(\hat{r})} \left(d\hat{r}^2 + \eta_{ij} dx^i dx^j\right)
 \sim  \left[{\a \ell^2 \over 2 R^2} \log \hat{r}/\tau_0\right]^{\a-1}
\left(d\hat{r}^2 +  \eta_{ij} dx^i dx^j\right).
\ee
The action for the transverse vector fluctuations becomes:
\be
S =   -{1\over 2}\int d\hat{r} d^4 x  e^{-\Phi}V(\hat{r}) e^{\hat{A}(\hat{r})} \left[(\de_{\hat{r}}A_i)^2  +  (\de_j A_i)^2 \right],
\ee
and has the same form as in (\ref{genac}) with
\be
B(\hat{r}) = {\hat{A}(\hat{r})-\Phi(\hat{r}) \over 2} + {1\over 2} \log V(\tau({\hat{r}}))
\ee
Asymptotically the last term dominates (it behaves like $\tau^2$, which is
exponential in the original $r$ coordinate, while $A_S$ grows logarithmically
and $\Phi$  a power-law of $r$), and we find, using eq. (\ref{tachyonpot}):
\be
B(\hat{r}) \sim - {3\over 4 \ell^2}\left({\a \ell^2 \over 2 R^2}\right)^{\a-1}\, \hat{r}^2 \left(\log \hat{r}/\tau_0\right)^{\a-1}
\ee
>From the general analysis of section \ref{general}, and in particular from eq. (\ref{gluepot}),
the leading behavior of the vector meson  Schr\"odinger potential  is that of
a (logarithmically corrected) harmonic oscillator, therefore it exhibits
an approximately linear mass spectrum\footnote{One can get rid of the extra $\log$ by
a slight modification of the tachyon potential.}. This is a concrete realization of the
general mechanism described in \cite{ckp}.

Notice that the meson spectrum  is generically controlled by a different energy
scale than the one that sets the glueball masses: the two scales are
\be
\Lambda_{glueballs} = {1\over R}, \qquad
\Lambda_{mesons} = {3\over \ell}\left({\a \ell^2 \over 2 R^2}\right)^{(\a-1)/2}
 \propto {1\over R} \left({\ell\over R}\right)^{\a-2}.
\ee

Interestingly, the two scales happen to coincide in the special case $\a=2$, in which
the asymptotic glueball spectrum is \emph{also} linear.

As a final remark we comment on the importance of $1/N_c^2$ corrections due to the bulk (closed sector)
on the meson spectrum.
 It seems that at least for questions of the meson
spectrum, lattice calculations indicate that such corrections are small.
 The errors for the pion and $\rho$-meson masses were
estimated at around 4\% for $N_c=3$ in \cite{bali} working in the quenched approximation.

\section{The parameters of the correspondence}
\lab{pars}

QCD with gauge group $SU(N_c)$ has three parameters: the bare coupling constant $\l_0$,
the theta-angle $\theta$ and the number of colors $N_c$.
On the other hand, through dimensional transmutation the bare coupling constant
is replaced by the dynamically generated strong coupling scale $\Lambda_{QCD}$.
In the large $N_c$ limit therefore, apart from $\theta$, this is the single parameter of the theory.
It sets the scale for  the glueball and meson masses.

\subsection{Parameters in the gravitational action}

On the gravity side, we have a number of parameters entering the
Lagrangian (and the fundamental string action) and a set of
integration constants for the vacuum solutions. The parameters that
appear in the action are the 5D Planck scale $M$, the string scale
$\ell_s$, the AdS radius $\ell$ (via the overall scale of the
potential, $V_0 = 12/\ell^2$). Moreover, the potential as a function
has dimensionless parameters. In its weak coupling expansion, they
are in one to one correspondence with the coefficients of the
$\beta$-function, $b_n$. Among the parameters $b_n$, as we discussed
above, the only scheme independent coefficients are $b_0$ and $b_1$.
For practical reasons we restrict ourselves to potentials
parameterized only by these two. Moreover, the ambiguity in the
normalization of $\l$,  amounts to $b_0$ being a parameter that we
eventually fit. $b_1/b_0^2$ however remains and we take its value
from the QCD $\beta$-function. The strategy however is clear:
although the potential is an a priori arbitrary function, our
eventual choice will have very few adjustable parameters.

The Planck scale governs the strength of bulk interactionss. In the underlying
 string theory, it is determined in terms of the string scale.
However here, this relation is not known and it will have to be taken as an
 extra parameter, that can be fixed by matching to interactions, or
to the finite temperature free energy.

On the other hand, either of the two dimensionfull parameters $\ell$
or $\ell_s$ can be used to set the units. We choose to measure all
the dimensionfull quantities in units of $\ell$.

Apart from the parameters of the action, there are in general three
integration constants that parameterize the solutions to the
Einstein-dilaton system.

Now, we discuss them one by one. In the dual gauge theory the
$\beta$-function  is fixed, and this is  in one-to-one
correspondence with the function $X$ or the superpotential $W$ on
the gravity side. Once the potential $V$ is given, there are
infinite number of $W$ that solve (\ref{VtoW}), parameterized by a
single boundary condition. 
However in appendix (\ref{spotvspot}) we show that the confining asymptotics of the form,
\be\lab{confas1} W\to \f^{\frac{P}{2}} e^{\frac23 \f}, \ee is 
a unique solution to the equation (\ref{VtoW}).
Therefore the requirement of confinement uniquely fixes the
superpotential $W$ and reduces the number of independent integration constants
from three to two. 
 
The IR asymptotics above indeed evolves into an asymptotic AdS space in the UV as,
\be\lab{wuv} W \to
\le(\frac34\ri)^{\frac32}V_0^{\half} + \cO(\l). 
\ee 
Given the superpotential, the equations of motion reduce
to two first-order equations (\ref{einsteinsuper}). 

\subsection{ Reparametrization symmetry and integration constants}

The remaining two initial conditions of the first order system  of motion
 are related  to the ``radial reparametrization symmetry" of the equations
of motion. Indeed from the Einstein's equations (\ref{einsteinsuper}) one learns
that, for any solution $A_*(u)$, $\f_*(u)$ there exist other
solutions parametrized by two numbers:
\begin{equation}\label{rsym}
    A(u) = A_*(u-u_0)+A_0, \qquad \f(u) = \f_*(u-u_0).
\end{equation}
The parameters $A_0$ and $u_0$ are in one-to-one correspondence with the two
integration constants.

\begin{figure}
 \begin{center}
 \leavevmode \epsfxsize=12cm \epsffile{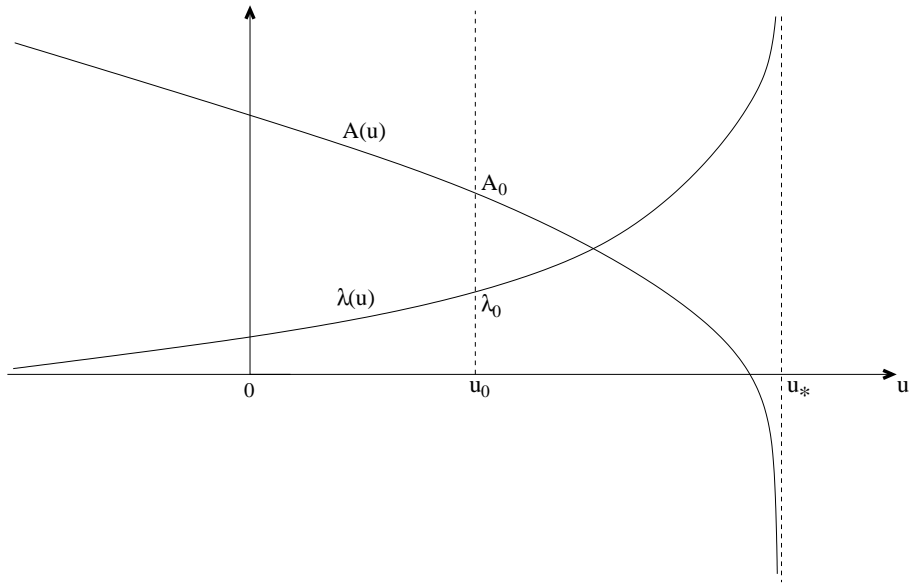}
 \end{center}
 \caption[]{The profile of the scale factor and the dilaton in a typical solution.
$u_*$ is the curvature singularity where the scale factor shrinks to zero
and the dilaton blows up.}
 \label{typical}
\end{figure}

It is useful to describe the symmetries (\ref{rsym}) graphically. In fig.
\ref{typical} we exhibit a typical solution that obeys our UV and IR criteria.
Given the solution in fig. \ref{typical}, one can generate another
by shifting $A(u)$ vertically. This corresponds to the first shift
symmetry $A\to A+A_0$ in (\ref{rsym}). The second shift symmetry
$u\to u+u_0$, corresponds to generating another solution by shifting
both $\l(u)$ and $A(u)$ horizontally in fig. \ref{typical}.
In practice, the solution is fixed by two initial conditions set at
an arbitrary point $u_0$: $\{A(u_0), \l(u_0)\} = \{A_0, \l_0\}$.

The next  observation is that the integration constant $u_0$ is just
a gauge artifact. This is because the original system is translation
invariant in the radial variable. In fact, using this
reparametrization symmetry, one can use $\l$ in place of the radial
variable $u$ and then the physics is completely specified by the
function $A(\l)$. Clearly a horizontal shift in fig. \ref{typical},
although changes $u_0$, leaves $A(\l)$ invariant. In other words,
the only physical  integration constant in the system is the one
that parametrizes the solutions of
\be
dA/d\l = \beta^{-1}(\l).
\ee
The general solution is:
 \be\lab{co1}
  A(\l) = A_0 + \int_{\l_0}^{\l} \frac{d\l}{\b(\l)},
\ee
and it  is invariant under:
\be\lab{eqrep}
\l_0\to\l_0 + \delta, \qquad A_0 \to A_0 +\int_{\l_0}^{\l_0+\delta}  \frac{d\l}{\b(\l)}.
\ee
Therefore, in (\ref{co1}) there is a single combination of $A_0$ and $\l_0$ that specifies the full solution.
This can
be taken as the value of $A$ at some fixed $\l$, e.g. $A_0 = A(\l_0)$.
\footnote{Note that this is not the case for a conformally invariant theory. In that case,
one cannot use the coupling as a new coordinate as it is constant.}

Therefore, all physically distinct solutions only differ by a
constant shift in $A$. Fixing the integration  constant
$A_0=A(\l_0)$ is equivalent to specifying $\Lambda_{QCD}$ in the
gauge theory, because it sets the energy scale through the relation
$E = \exp A$. In fact, this is the only way the integration constant
$A_0$ affects any physical quantity: a change in $A_0$ induces a
constant rescaling of all dimensionfull quantities, such as masses,
confining string tension, etc. In particular, mass ratios are
completely independent of all integration constants, and only depend
on the parameters that appear in the gravity Lagrangian.

Consider for example the scalar or tensor fluctuations. The
corresponding spectral equations follow from an action of the form
(\ref{genac}), with the functions $B$ and $M$ given by  eqs.
(\ref{B0}) and (\ref{B2}). We can change coordinates from $r$  to
$\Phi\equiv \log\l $ in (\ref{genac}), and derive the corresponding
spectral equation for the fluctuations $\zeta(\Phi)$:
\be\lab{co6}
-e^{-3B-A}\de_\Phi \left(e^{3B+A} \de_\Phi W \de_\Phi \zeta \right)
= {e^{-2A}\over \de_\Phi W} m^2 \zeta
\ee
Under  a constant shift  $A\to A+\delta A_0$, and $B$ also shifts by a constant.  The left hand side is therefore
invariant,  and the right hand side is rescaled by $e^{-2\delta A_0}$; Thus,
the only effect on the spectrum is an overall rescaling of all
the mass eigenvalues by     $e^{\delta A_0}$.

The same considerations hold for the confining string tension, eq.
(\ref{qcdstring}): a constant shift in $A$ does not change
the position of the minimum of $A_S(r)$, but it only  rescales
the tension by  $e^{2\delta A_0}$. In particular, the ratios
$m_n^2 /T_s$ are independent of the integration constant.

A change in $A_0$ has the same effect also on
 the scale governing  the perturbative
running of the coupling, that gives rise to  dimensional transmutation:
 if initial conditions
are chosen at  $\l_0 \ll 1$, integration of the $\beta$-function
equation (\ref{co1}) leads to:
\be\lab{lambda0}
{1\over \l} = {1\over \l_0} + b_0 \log {E\over \Lambda_0}, \qquad E =
{e^{A}\over \ell}, \quad \Lambda_0 \equiv    {e^{A_0}\over \ell}.
\ee

We can identify the ``perturbative'' RG-invariant QCD scale as
follows. Integrating eq. (\ref{co1}) for small $\l,\l_0$ up to two
loops, with $\beta(\l)\simeq -b_0 \l^2 - b_1 \l^3 +O(\l^4)$, we
obtain:
\be
\lab{l2l1} A(\l) - {1\over b_0 \l} - {b_1\over b_0^2}
\log (b_0 \l) + O(\l)  = A_0 - {1\over b_0 \l_0} - {b_1\over b_0^2}
\log (b_0\l_0) +  O(\l_0) \simeq constant.
\ee
Therefore, the scale
\be
\label{l2l2} \Lambda_p \equiv {1\over \ell} {\exp \left[A(\l_0) -
{1\over b_0 \l_0}\right]\over (b_0 \l_0)^{b1/b_0^2}}
\ee
 is
approximately independent of $\l_0$ as long as it is small. This
scale appears in the UV expansion of the coupling in the form:
\be\lab{l2l3} {1\over b_0 \l} = \log {E\over \Lambda_p} - {b_1\over
b_0^2}\log \log {E\over \Lambda_p} + \ldots \ee It is the same scale
appearing in the UV expansion  of the solution in conformal
coordinates, (\ref{UVasr}), as one can see by substituting $E\simeq
1/r $ on the l.h.s. of eq. (\ref{l2l3}).

All the different scales we have analyzed above
behave in the same way under a change in the
integration constants, so the relations between them is a property
of the gravity model, not of each  particular solution.

In the explicit examples we present in Section \ref{numerics},
we fix the energy scale to match the lowest glueball mass,
and as we have discussed this fixes unambiguously all
other dimensionfull quantities. In particular  we obtain
a value for $\Lambda_p$ in eq. (\ref{l2l2}):
\be
\Lambda_p = 290\, MeV.
\ee
This value  is larger than the usual QCD  value ($\sim 200\, MeV$). However one  should
keep in mind that the definition of the
strong coupling scale in perturbation theory  is somewhat arbitrary, and moreover we
are not including the effect of quarks in the running
of the coupling.
When comparing  with data (lattice or experiment) it is more
meaningful to  look at unambiguous quantities, e.g. the
value of the strong coupling constant $\a_s$ at a given energy.
For example, we find\footnote{See section \ref{numerics} for a
more detailed discussion, and in particular for the
relation between $\l$ and $\a_s$}.
\be
\alpha_s (1.2 GeV) = 0.34
\ee
which is very close to the experimental value
$\alpha_s^{(exp)} (1.2 GeV) = 0.35\pm 0.02$

In summary, like in QCD,
in the gravity side  solution is specified by the $\beta$-function
plus a single dimensionfull quantity
$e^{A_0}/\ell$,  that parametrizes the different
solutions and sets all the relevant mass scales. It can be related to
 $\Lambda_{QCD}$ of the gauge theory, and it can
 be understood holographically
as the constant of motion that is preserved under the
shift symmetry $u\to u + u_0$. This is precisely $A_0$ in our set-up.

\section{Concrete backgrounds\label{numerics}}

In this section we present explicit  backgrounds that exhibit
all of the features we require (asymptotic freedom,
 confinement, discrete spectrum).  Then we compute
the glueball spectra numerically.

We consider two backgrounds belonging to two distinct classes. The
first is a background with an exponentially decaying scale factor,
and with an infinite range of the conformal coordinate. We focus on
the case $\a=2$ in (see equation (\ref{asymptA})). As shown in section \ref{lco}
this gives an asymptotically  linear glueball spectrum. Secondly, we
analyze an example of a background with finite range of the
conformal coordinate. In both cases we fix the 5D theory by
providing a function $X(\l)$ that interpolates between the required
UV and IR asymptotics. As we discussed, this is equivalent to fixing
the exact $\b$-function. The RG-flow trajectory is further specified
by the UV initial conditions, which we input for the numerical
integration. This fixes the gravity dual completely.

In this paper we only present the glueball spectra. Although
straightforward in principle, the meson spectra require considerably
more complicated numerics. The main obstacle from the numerical
point of view is identifying the correct initial conditions for the
nonlinear tachyon equation, (which is then used as an input in the
computation of the meson spectrum). Therefore we leave the
computation of the meson spectrum for
 future work.

Finally, we compare the glueball spectra with the available lattice
data. For the model with infinite range of $r$ and $\a=2$ we can fix
the parameters in such a way to produce a very good agreement, at a
quantitative level.

\subsection{Background I:  unbounded conformal coordinate}

For an asymptotically free, confining theory, the function $X(\l)$
has the following UV asymptotics (see eq. (\ref{x}))
\be
\lab{uvas}
X(\l) \sim -{b_0\over 3} \l - {b_1\over 3}\l^2 +\ldots \quad \l\to 0
\ee
where $b_k$ are the $k$-th order coefficients of the
perturbative $\b$-function. In the IR we require (see
(\ref{xconf2})):
\be
\lab{iras} X(\l) = -{1\over 2} - {a\over \log \l}
+ \ldots \quad \l\to \infty,
\ee
 where the parameter   $a$
determines the large-$r$ behavior of the scale factor:
\be
\label{Aas} A\sim -C\,  r^\a \qquad a \equiv {3\over 8} {\a -1
\over \a}.
\ee
 We seek for a function of $\l$ that interpolates
between the two asymptotics (\ref{uvas}) and (\ref{iras}). A simple
function that is regular and has this property is,
\be
\lab{xexact} X(\l) = -{b_0 \l
\over 3+ 2 b_0\l} - {(2b_0^2 + 3b_1)\l^2\over 9(1+\l^2)\left(1+
{1\over 9 a}\left(2b_0^2 + 3b_1\right)\log (1 + \l^2)\right)}.
\ee

This expression is motivated by the UV and the IR asymptotics in (\ref{uvas})
and (\ref{iras}) and by the requirement that there are no poles or branch
cut singularities in $\l$. Also, the function $X(\l)$ (hence also $\beta(\l)$)
is strictly negative
for $\l > 0$,  therefore  there are no IR fixed points.

Starting from eq. (\ref{xexact}), we solve for the
metric and dilaton using eqs. (\ref{conf1st}):
\be\lab{eq1r}
\dot{\l} = -{4\over 3\ell} X(\l)W(\l) \l e^A, \qquad \dot{A} = -{4\over 9 \ell} W(\l) e^A.
\ee
The superpotential $W(\l)$, is given in terms of $X$ as in (\ref{spotx}),
\be\lab{superfree}
W = {9\over 4}\left(1+{2\over 3}b_0\l\right)^{2/3}\left[1 + {\left(2b_0^2 +3b_1\right)\over 9a} \log(1+\l^2)\right]^{2a/3},
\ee
 and in writing (\ref{eq1r}) we have explicitly extracted   the overall scale $\ell$.
In the
integration of (\ref{eq1r}), we fix the integration constants as:
\be\lab{init}
A(r_{in})=  A_0, \qquad \l(r_{in})= \l_0.
\ee
for $r_{in}/\ell \ll 1$ and $\l_0\ll 1$, in order to implement the correct UV asymptotics.

\begin{figure}[h]
 \begin{center}
\leavevmode \epsfxsize=7cm \epsffile{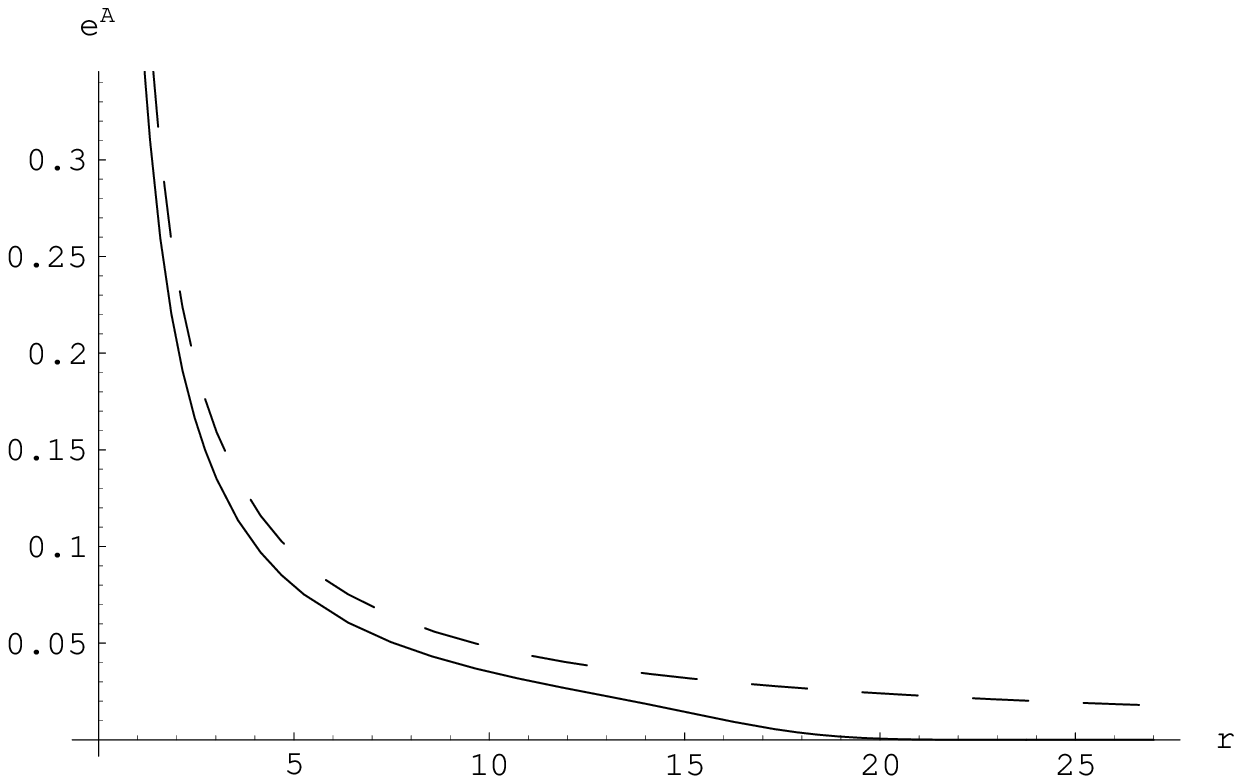} \leavevmode
\epsfxsize=7cm \epsffile{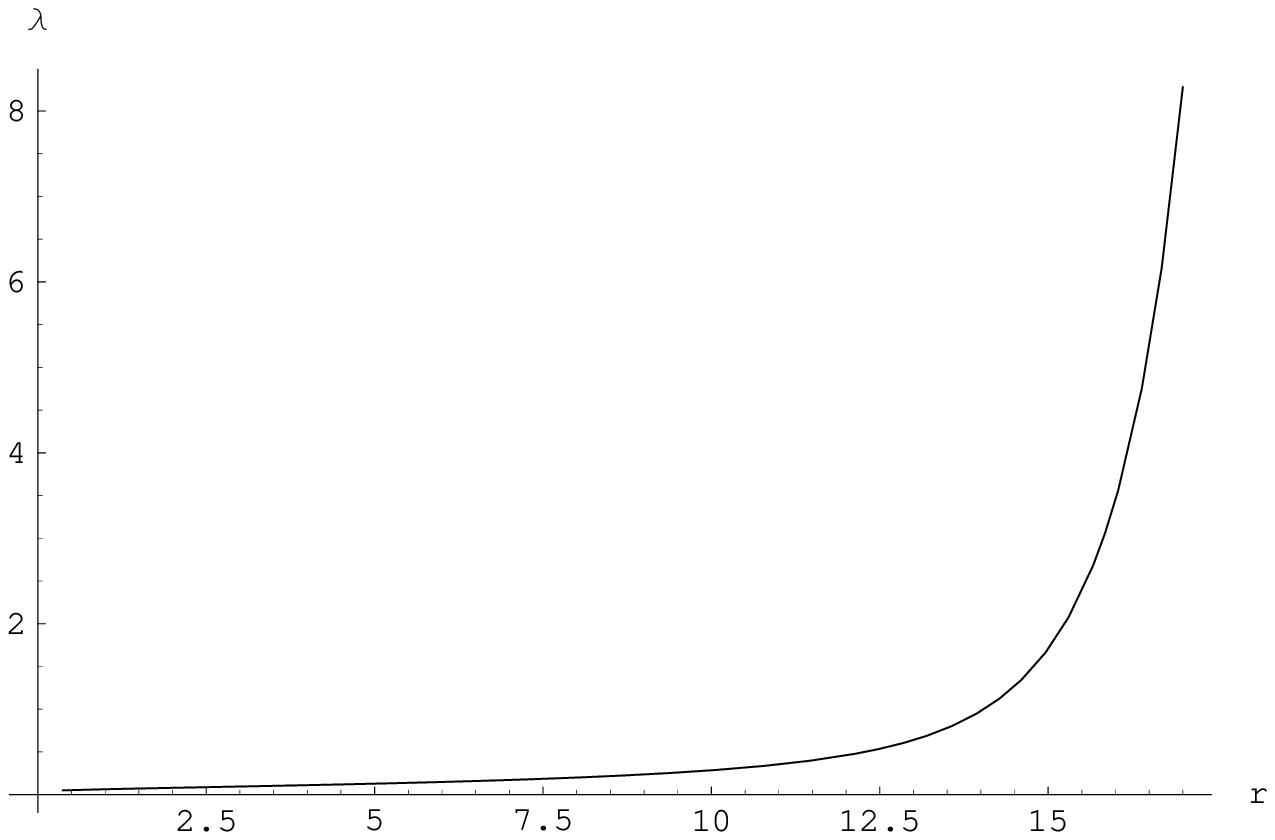}
 \end{center}
 \caption[]{The scale factor and 't Hooft coupling that follow from (\ref{xexact}), $b_0=4.2$, and
initial conditions $A_0=0$, $\l_0=0.05$ at $r=0.36$. The units are such that $\ell=1$. The dashed
line represents the scale factor for  pure $AdS$. }
 \label{geo}
\end{figure}

The scalar and tensor glueball spectra  are completely fixed by the metric
and dilaton background. For the pseudoscalar glueballs, we need to specify
also the axion kinetic function $Z(\Phi)$ appearing in eq. (\ref{axionaction}).
{}From the discussion Sections 5 and 6.2.3, we assume the asymptotic
behaviors:
\be\label{axionasymtotiocs}
Z(\l) \to \left\{ \begin{array}{ll} Z_a &\quad \l \to 0
\\ c_a \l^{4} & \quad  \l \to \infty \end{array} \right. .
\ee We take the function $Z(\l)$ to be the simplest one satisfying
these asymptotics \be\label{zetaa} Z(\l) = 1 + c_a \l^{4} \ee where
we have fixed the  $Z_a=1$ by an overall rescaling of $Z(\l)$, which
does not affect the glueball spectrum. $c_a$ is an extra
dimensionless parameter\footnote{We could take a more general form
that includes the perturbative string theory term $\sim \l^2$, \be
Z(\l) =  1 + b_a \l^2 + c_a \l^{4}, \nonumber \ee however for the
sake of simplicity we set $b_a=0$ in our fits. A non-zero $b_a$
would imply a different preferred value for $c_a$, but  this does
not change the spectrum significatively. However this could have a
non-negligible effect on the axion profile, see Fig. \ref{c}.}.

\subsubsection{The glueball spectra in background I}

We solve the eq. (\ref{schro}) with the Schr\"odinger potential (\ref{gluepot}) numerically.
We compute the spectrum of scalar and tensor glueballs where the function $B$ in (\ref{gluepot})
is given by eqs. (\ref{B0}) and (\ref{B2}) respectively, whereas the 5D mass-term $M$ in (\ref{gluepot})
is zero.

One has to supply the Schr\"odinger equation with the boundary condition in the
UV, (as $r\to 0$),
\be\lab{UVasu} \psi \to C_0 r^{\frac52} + C_1 r^{-\frac32}
\ee
Particle states correspond to
normalizable solutions. Therefore,  normalizability in the UV requires $C_1=0$.
Normalizability in the IR, on the other hand fixes the discrete
values for $m$ in (\ref{schro}).
In practice, we use the shooting method to determine the spectrum:
we scan the values for $m$ and pick the values at which an extra node in the wave function
appears. Precisely at this value of $m$, the wave function becomes
normalizable in the IR.

In principle, the spectrum depends on the parameters of the
background, $b_0$ and $b_1$, the integration constants of the
geometry $\l_0$ and $A_0$ (eqs. (\ref{init})) and the boundary condition of
(\ref{UVasu}), \ie $C_0$. However, not all of these parameters affect
the spectrum nontrivially.
\begin{itemize}
 \item  The constant  $C_0$
is clearly immaterial, due to the linearity of the equation for $\psi(r)$. We
set $C_0=1$ without loss of generality,
\item As we discussed in Section 7, the only physical integration
constant for the background is the choice of
$A_0$ at some value $\l_0$, and it only affects the overall scale
of the masses.   This expectation is
confirmed by the numerical results, as shown explicitly in figure \ref{ScvsL}.
  Thus,  the mass ratios will be  independent of $A_0$ and $\l_0$,
as well as of the $AdS$ scale. The
overall energy scale can then be fixed by matching e.g. the
mass of the lowest state in the spectrum.

\begin{figure}[h]
 \begin{center}
\leavevmode \epsfxsize=7cm \epsffile{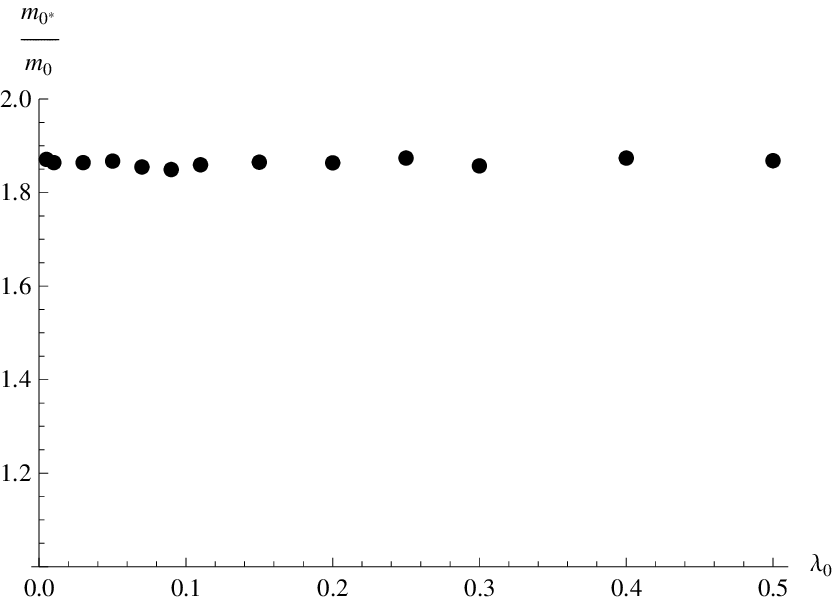}\hspace{0.5cm}
\leavevmode \epsfxsize=7cm \epsffile{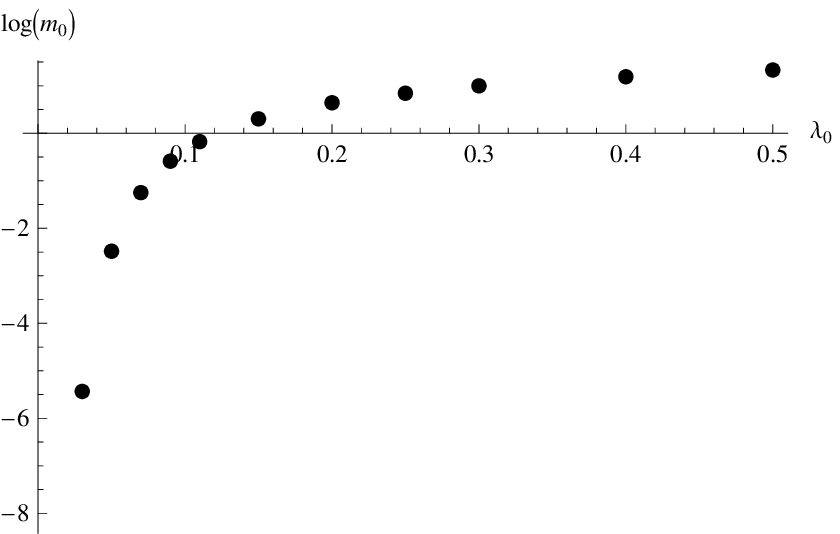}\\
(a)\hspace{6cm}(b)
 \end{center}
 \caption[]{Dependences on initial condition $\l_0$ of (a)  the mass ratios
$R_{00}=m_{0*++}/m_{0++}$ (squares) and $R_{20}=m_{2++}/m_{0++}$ (triangles);
(b)  the absolute scale
of the lowest lying scalar glueball (shown in Logarithmic scale).}
 \label{ScvsL}
\end{figure}

\item As discussed in \cite{part1}, $b_0$  cannot be determined from first
principles in our setup, as the overall coefficient in the relation (\ref{coupling}) between
the dilaton and 't Hooft coupling is not known. We keep $b_0$ as a free parameter.
On the other hand,  the ratio $b_1/b_0^2$ is independent of such normalization. In pure YM this
ratio is given by $51/121$ and this is what we use.
\item The superpotential completely fixes the scalar and tensor glueball
mass ratios; the pseudoscalar glueball masses also depend on the
additional parameter $c_a$ that enter the definition of the axion
kinetic term, eq.  (\ref{zetaa})
\end{itemize}

In light of the above,  we vary
only  $b_0$ 
 for the
purpose of fitting the scalar and tensor glueball lattice data, and $c_a$
to fit the pseudoscalar glueball data.

\begin{figure}[h]
 \begin{center}
\leavevmode \epsfxsize=7cm \epsffile{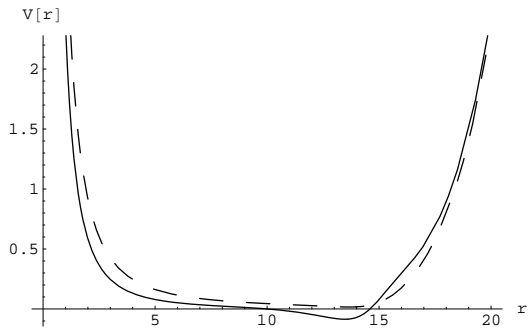}
 \end{center}
 \caption[]{Effective Schr\"odinger potentials for scalar (solid line) and
tensor (dashed line) glueballs. The units are chosen such that $\ell=1$.}
 \label{potentials}
\end{figure}
\begin{figure}[h]
 \begin{center}
\leavevmode \epsfxsize=7cm \epsffile{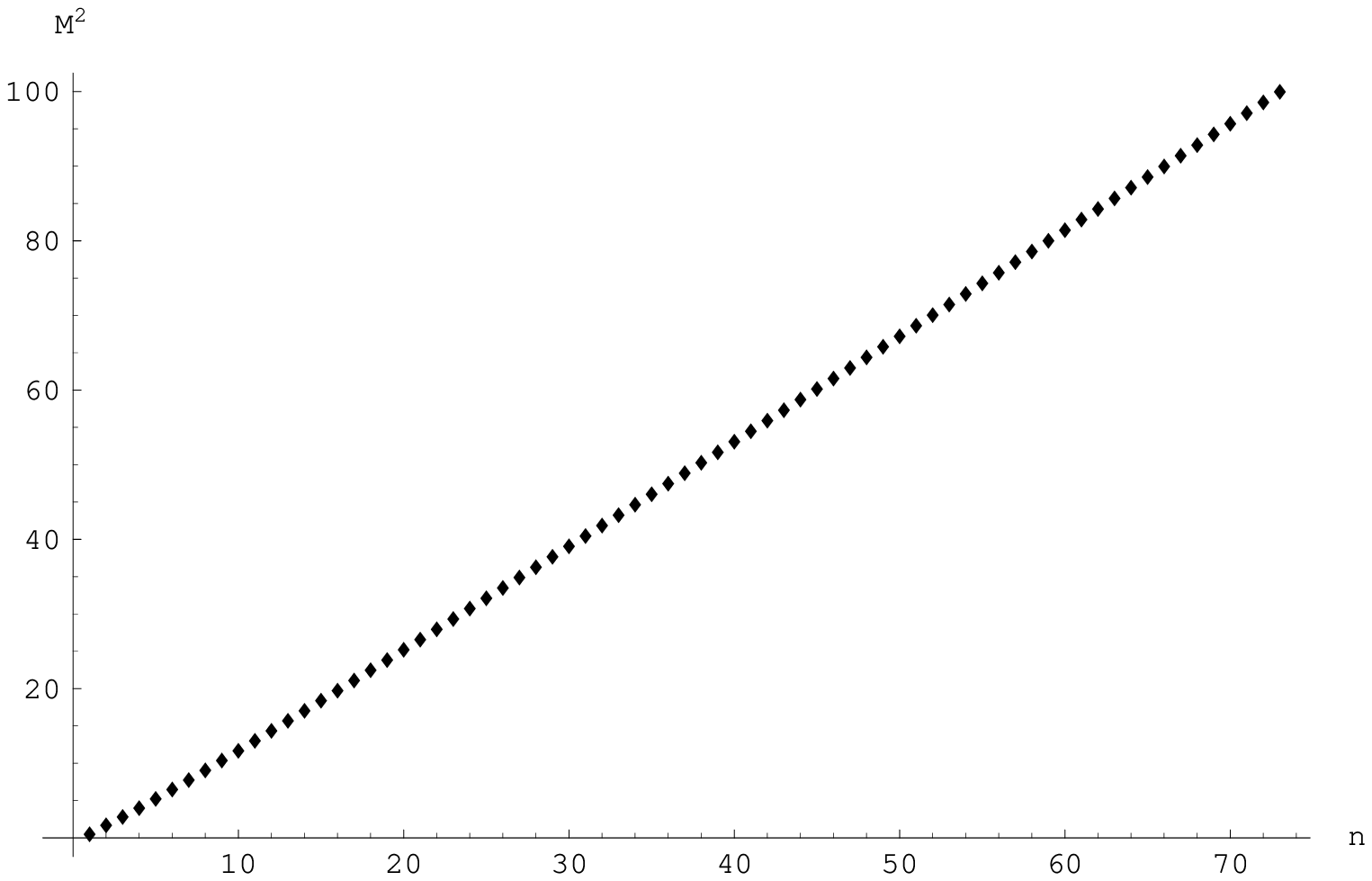}
\leavevmode \epsfxsize=7cm \epsffile{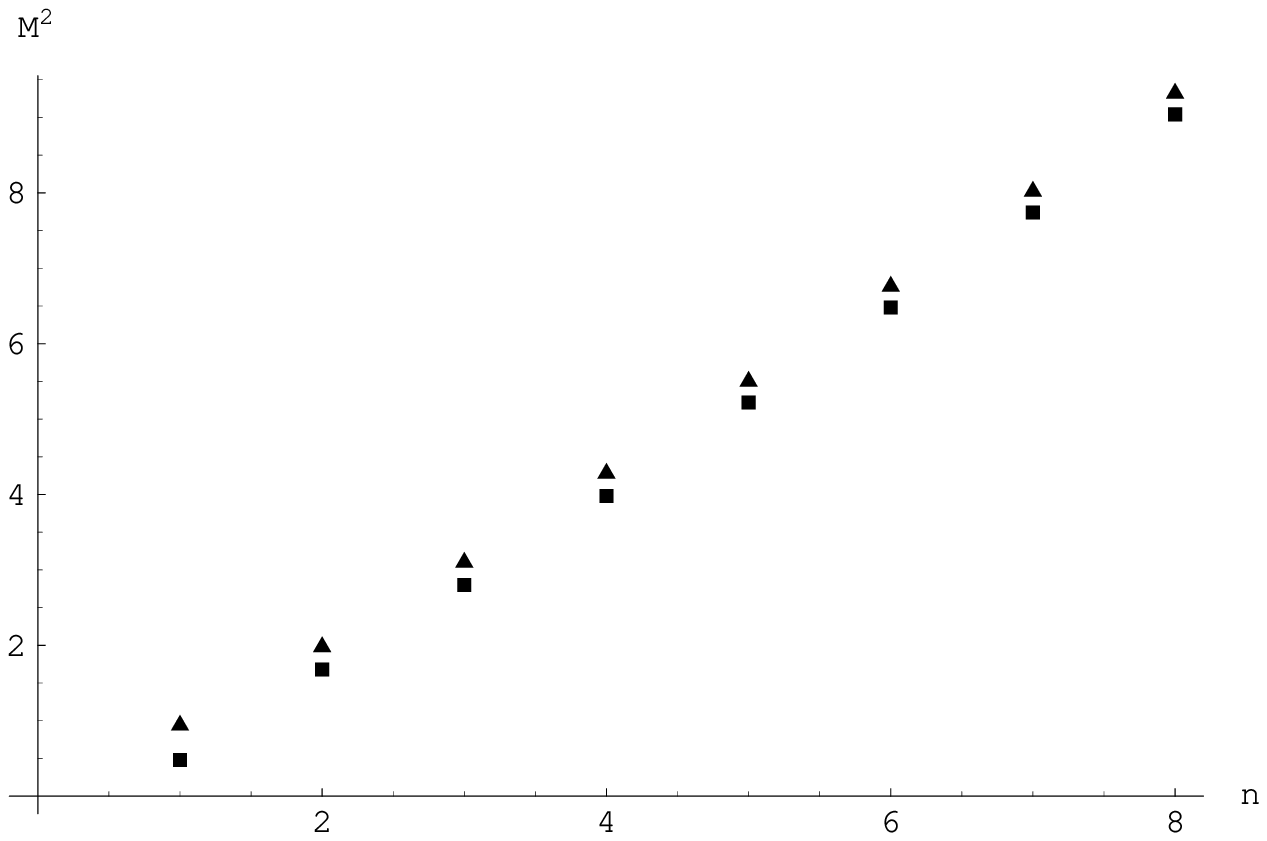}\\
(a)\hspace{6cm}(b)
 \end{center}
 \caption[]{(a) Linear pattern in the spectrum for the
   first 40 $0^{++}$ glueball states. $M^2$ is shown units of $0.007 \ell^{-2}
   $. (b) The first 8 $0^{++}$ (squares) and the $2^{++}$
   (triangles) glueballs. These spectra  are obtained in the background I with
   $b_0=4.2$.}
 \label{linear}
\end{figure}

We perform most of the numerical analysis for the background that
gives linear spectrum, i.e. $\a=2$ (we discuss the dependence of the
spectrum on the parameter $\alpha$ at the end of this section.).
To make the numerics easier, we fix  $\ell=1$ and
work in dimensionless units.  The
geometry looks typically like in Figure \ref{geo}, the effective
Schr\"odinger potentials as in Figure \ref{potentials}, and the
glueball spectrum as in  Figure \ref{linear}.

 We note that, unlike the simple $AdS/QCD$ setup,  the scalar and tensor glueballs are
not degenerate, but the tensor glueballs are generically heavier
than the scalar ones with the same quantum number $n$. The
tensor-scalar mass difference decreases for larger $n$, indicating
that the slopes governing the asymptotics of the two spectra are the
same.
 This is in accord with our discussion in section \ref{spectrum}.

\subsubsection{ Lattice Data}

Available sources for the glueball mass spectra come from
computations on the lattice. Our backgrounds naturally give
predictions for the $N_c=\infty$ theory. Although there are large-$N_c$
extrapolations (see for example \cite{lattice1}), there exist richer
and more precise data for $SU(3)$, especially for the excited glueball
states. Therefore, we choose to fix our
parameters in order to fit the available data for $N=3$. We note
that the error one makes for using $N=3$ data instead of $N=\infty$
is within 5 percent \cite{lattice1}. This is well within the error bars of the
lattice computations for $SU(3)$ (see \cite{lattice2, lattice3, lattice4}).
\begin{table}[h!]
\begin{tabular}{|l|l|l|l|l|l|}
  \hline
  $J^{PC}$ & Ref. I $(m/\sqrt{\sigma})$ & Ref. I (MeV) & Ref. II $(m r_0)$ & Ref. II (MeV) & $N_c\to\infty (m/\sqrt{\sigma})$ \\
\hline
  $0^{++}$ & 3.347(68) & 1475(30)(65) & 4.16(11)(4) & 1710(50)(80) &3.37(15)\\
  $0^{++*}$ & 6.26(16) & 2755(70)(120) & 6.50(44)(7) & 2670(180)(130) & 6.43(50)\\
  $0^{++**}$ & 7.65(23) & 3370(100)(150) & NA & NA & NA\\
  $0^{++***}$ & 9.06(49) & 3990(210)(180) & NA & NA & NA\\
  $2^{++}$ & 4.916(91) & 2150(30)(100) & 5.83(5)(6) & 2390(30)(120) & 4.93(30)\\
  $2^{++*}$ & 6.48(22) & 2880(100)(130) & NA & NA & NA\\
  $0^{-+}$  &  5.11(14) & 2250(60)(100) & 6.25(6)(6) & 2560(35)(120) &  NA  \\
  $0^{-+*}$ &  7.66(35) & 3370(150)(150) & NA   & NA    & NA   \\
\hline
  $R_{20}$ & 1.46(5) & 1.46(5) & 1.40(5) & 1.40(5) & 1.46(11)\\
$R_{00}$ &  1.87(8) & 1.87(8) & 1.56(15) & 1.56(15) & 1.90(17)\\
$R_{A0}$ &  1.52(8) & 1.52(8) &  1.50(5) &   1.50(5) & NA  \\
  \hline
\end{tabular}
\caption{Available lattice data for the scalar and the
  tensor glueballs. Ref. I denotes \cite{lattice4} and
Ref. II denotes \cite{lattice2} and \cite{lattice3}. The first error in the
Ref.I and Ref. II
correspond to the statistical error from the the continuum
extrapolation. The second error in Ref.I is due to the uncertainty in the
string tension $\sqrt{\sigma}$. (Note that this does not affect the mass
ratios). The second error in the Ref. II is the estimated uncertainty from the
anisotropy. In the last column we present the available large $N_c$ estimates according to \cite{lattice1}.
The parenthesis in this column shows the total possible error following by the
estimations in \cite{lattice1}. }\label{latticedata}
\end{table}
\vspace{0.5 cm}

There exist a vast literature on the lattice
computations for the glueball spectra. We take as reference,
the papers \cite{lattice2}, \cite{lattice3} and
\cite{lattice4}.\footnote{We thank H. B. Meyer, C. J Morningstar and
M. Teper for pointing us to these references.} We listed the available
data in table \ref{latticedata}. In that table Ref. I denotes \cite{lattice4} and
Ref. II denotes \cite{lattice2} and \cite{lattice3}. Although we
listed the lattice results also in the units of MeV, it is more
convenient to use the units of $r_0$ (the ``hadronic length scale'') or $\sqrt{\sigma}$ (the confining
string tension). In order to
compare the data according to the two references, one should take
$\sigma r_0^2 \approx 1.36$.\footnote{We thank H.B. Meyer for
explaining this to us.}

In order to avoid the error in the choice of the unit mass scale, we
fit our parameters by the mass ratios ratios, that we denote as:
\begin{equation}\label{rats}
    R_{00} = \frac{m_{0*++}}{m_{0++}},\qquad R_{20}  = \frac{m_{2++}}{m_{0++}}
     ,\qquad R_{A0}  = \frac{m_{0-+}}{m_{0++}}.
\end{equation}
There is a slight mismatch for the values of these ratios in the
refs. \cite{lattice2} and \cite{lattice4}, (see table I). Thus, in
the next section, we shall present our results for fitting our
parameters according to both of these references separately.

Notice that we could have computed the string tension $\sigma$ by
looking at the minimum value of the string frame scale factor,  as
explained in Section  3. To obtain any  numerical information,
however, would further require knowledge of the relation between the
fundamental string tension and the $AdS$ scale. The latter sets the
overall mass unit. Since this relation is not fixed in our model it
does not constitute an independent check.

Finally we should mention that the experimental identification of glueballs in
high energy experiments has a long and not very successful history.
The main problem is to find unambiguous criteria that would distinguish glueballs
from others states (mesons, and hybrids) in the experimental data.
Recent discussions on the status of the experimental glueballs search both for scalar
 and pseudoscalar ones can be found in references \cite{gl1,gl2}.

\subsubsection{Fit for Reference I}

\noindent{\bf $0^{++}$ and $2^{++}$ glueballs}\\

As we discussed above, the
numerical integration of (\ref{schro}) determines the spectrum in
terms of $b_0$, up to a choice of scale. We showed that the mass ratios are
independent of $A_0$ and $\l_0$.

 We fix $\l_0=0.05$, then
vary $b_0$  to obtain the ratios $R_{00}=1.87$ and $R_{20}=1.46$
(table I). We then fix  the overall energy scale  to set $m_{0++} =
1475$. As explained in Section 7, this is equivalent to fixing
$\Lambda_{QCD}$, and completely determines the background solution.
We then compare our results with those in the third column of Table
\ref{latticedata}.

The value of $b_0$ that fits $R_{00}=1.87$ is $b_0=4.2$. Fixing this,
we find $R_{20}=1.40$.
The masses for the lowest lying states are found to be:
\bea
0^{++}&&\quad     m_1,\, m_2, \cdots = 1475,\, 2753,\, 3561 ,\, 4253,\, 4860,\, 5416 \cdots\, MeV. \label{0predI}\\
2^{++}&&\quad     m_1,\, m_2, \cdots = 2055,\, 2991,\, 3739 ,\, 4396,\, 5530, \cdots\, MeV. \label{2predI}
\eea
Notice that the spectrum of  excited states is  in  good agreement with the available data from Ref. I.
We compare our results with the lattice data and the standard
AdS/QCD predictions in Fig. \ref{compareg01}.
The glueball spectrum in the standard AdS/QCD model is worked out
in Appendix \ref{adsqcdglue}\footnote{
There, we fixed $r_0$ by the meson data. If one leaves $r_0$ as a free
parameter in the glueball sector, one can obtain better fits in the
AdS/QCD set-up. For example, \cite{neumann} finds good fit with the
Pomeron trajectory with Neumann boundary conditions.}.
We note that the glueball spectra in the
hard-wall model were first discussed in \cite{BragaI} and \cite{BragaII} with Dirichlet boundary
conditions at the wall.
\begin{figure}[h]
 \begin{center}
\leavevmode \epsfxsize=7cm \epsffile{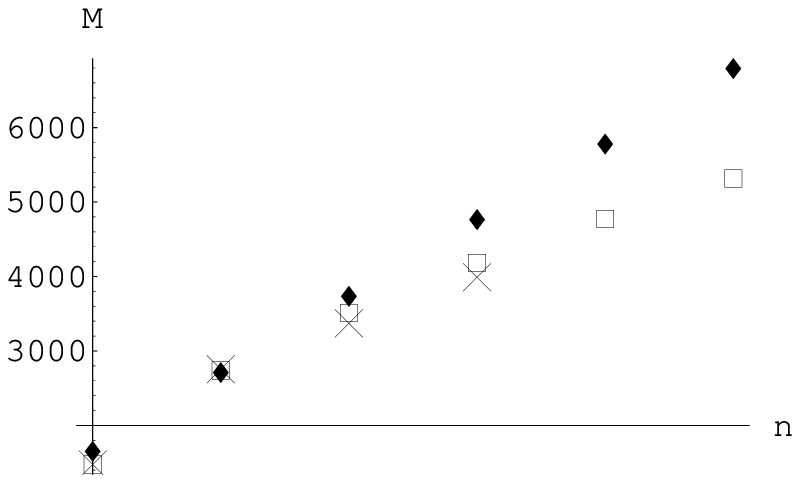}\hspace{0.5cm}
\leavevmode \epsfxsize=7cm \epsffile{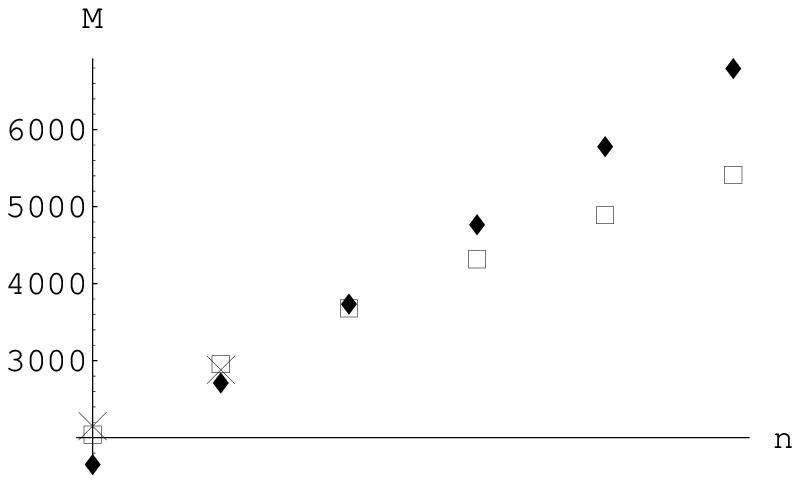}\\
(a)\hspace{6cm}(b)
\end{center}
 \caption[]{Comparison of glueball spectra from our model with $b_0=4.2$ (boxes),
with the lattice QCD data from Ref. I (crosses)  and the AdS/QCD computation (diamonds),
for  (a) $0^{++}$ glueballs; (b) $2^{++}$  glueballs. The masses are in MeV, and
the scale is normalized to match the lowest $0^{++}$ state from Ref. I.}
 \label{compareg01}
\end{figure}

\begin{figure}[h]
 \begin{center}
\leavevmode \epsfxsize=9cm \epsffile{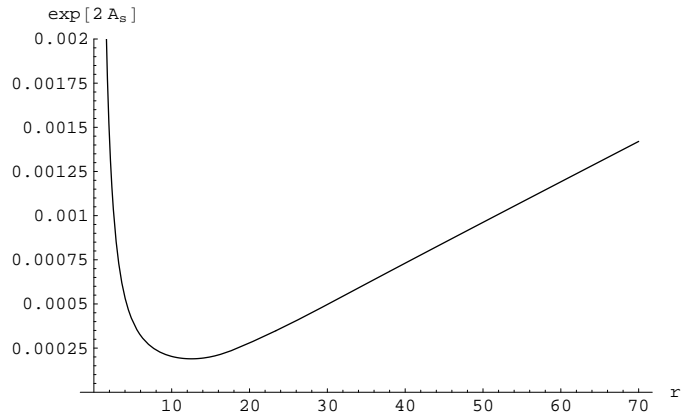}
\end{center}
 \caption[]{The string frame scale factor in background I with $b_0=4.2$. The units on the horizontal axis are such that $\ell=1$.}
 \label{strfigure}
\end{figure}
\vspace{0.5cm}

\noindent{\bf String tension}\\

>From the first column of Table \ref{latticedata} we can estimate
the fundamental string tension $T_f$ in $AdS$ units:
\be
T_f \ell^2 = \sigma \ell^2  e^{-2A_s(r_*)}
= {m^2_{0^{++}}\ell^2 \over (3.347)^2}e^{-2A_s(r_*)}
\ee
The string frame scale factor is shown in Figure \ref{strfigure}, and numerically we find that
 at the minimum $e^{2A_s(r_*)} \simeq 2\times 10^{-4}$. This gives
\be
T_f ~\ell^2 ={\ell^2\over 2\pi \ell_s^2}\simeq 6.24~~~~\to~~~~~{\ell\over \ell_s}\simeq 6.26
\ee
The size of the UV geometry is several times the string length.
This in particular shows  that the dimensionless curvature invariant (in the Einstein frame) near the $AdS_5$ boundary is
\be
\ell_s^2 R\simeq -0.5
\ee
\vspace{0.5cm}

\noindent{\bf Running coupling and QCD scale}\\

Using our choice
of $\l_0$ and $A_0$ we can compute the perturbative QCD scale as defined in eq. (\ref{l2l2}).
We obtain
\be
\Lambda_p \simeq 290\, MeV,
\ee
which is the correct order of magnitude, but not very close to the
generally assumed value of around $200 MeV$. This can be attributed to
the fact that in deriving the latter value, the effect of five flavors
of quarks is assumed to contribute to the running, whereas we are
dealing with  pure Yang-Mills.
Moreover  one should keep
in mind that there is no unambiguous definition of  $\Lambda$ \cite{PDG}.
It is more meaningful to compare with experiment the
value of the strong coupling constant $\alpha_s$ at some energy
scale. In order to do this, we must first identify the relation
 between  our coupling constant $\l$ and the strong coupling constant
$\a_s$. This is fixed once we set the parameter  $b_0$, and can be
obtained by comparing the one-loop beta-functions:
\bea
 && {\de \l \over \de \log E} = - b_0 \l^2 + \ldots \, ,\\
&& {\de \a_s \over \de \log E} = -{\beta_0\over 2\pi} \a_s^2 + \ldots \, ,\qquad  \beta_0  = {11\over 3} N_c.
\eea
>From  the two expressions we find the relation:
\be
\alpha_s = {2 \pi b_0\over 11 N_c/3 }  \l = 2.4 \l
\ee
where we have set $N_c=3$ in the last step\footnote{In ${\cal N}=4$ SYM the
identification is fixed by the D3 brane coupling to the dilaton,
$\a_s  = g_s  = \l/ N_c$
}.
Using our numerical solution we can compute the value of $\a_s$
at some radius $\r$; this  can then be translated into an energy
scale  through the relation $E = E_0 e^{A(r)}$,  in which $E_0$
is fixed by matching the lowest glueball mass (in our case $E_0 = 17630 MeV$).
For example, we find
\be
\a_s (1.2 GeV) = 0.34,
\ee
which is within the error of the quoted experimental value \cite{PDG},
$\a_s^{(exp)} (1.2 GeV ) = 0.35 \pm 0.01$\footnote{The uncertainty on this
value is not reported in \cite{PDG}; rather, it is an estimate obtained from
the corresponding uncertainty in the data $\a_s(M_Z) = 0.1202 \pm 0.005$}.
\vspace{0.5cm}

\noindent {\bf $0^{-+}$ glueballs:}\\

\begin{figure}[h]
 \begin{center}
\leavevmode \epsfxsize=9cm \epsffile{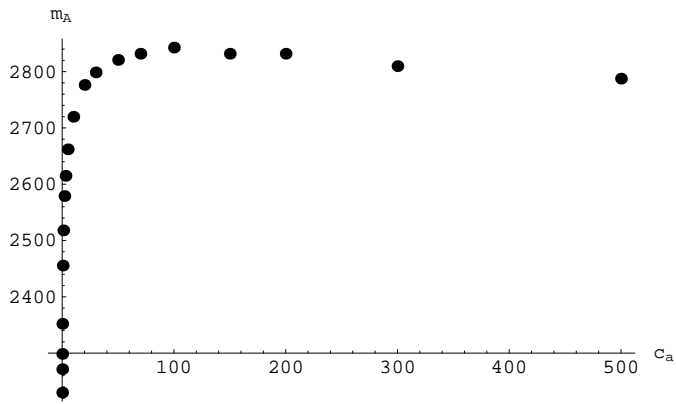}
\end{center}
 \caption[]{Lowest $0^{-+}$ glueball mass in MeV as a function of $c_a$.}
 \label{cmass}
\end{figure}

Having fixed $b_0$,  we  can now vary the parameter $c_a$ to fit
the lowest pseudoscalar glueball mass. First, we notice that
for large values of $c_a$ (greater than $\sim 10$), the spectrum
depends very weekly on this parameter. This is shown in Figure \ref{cmass}

The ref. I value $R_{A0} = 1.52$ is obtained for $c_a=0.05$. For this
value the lowest pseudoscalar glueball masses are found to be:
\be
0^{-+}\quad   m_1,\, m_2,
\cdots =  2243,\,  3436,\, 4396,\, 4911,\, 5541,\, , \cdots\, MeV.
\ee
\begin{table}[ht]
\begin{center}
\begin{tabular}{|c|c|c|c|c|c|}
  \hline
  $J^{PC}$  &  Ref I (MeV) &  Our model (MeV) & Mismatch & $N_c\to \infty$
   \cite{lattice1} & Mismatch to $N_c\to \infty$    \\
\hline
  $0^{++}$  & {\bf 1475 (4\%)} &{\bf 1475} & 0  & {\bf 1475} & 0  \\
  $2^{++}$ &  2150 (5\%) & 2055 & 4\% & 2153 (10\%) & 5\%  \\
  $0^{-+}$  & {\bf 2250 (4\%) } & {\bf 2243 }& 0 &   &  \\
  $0^{++*}$ & {\bf  2755 (4\%) } & {\bf 2753} & 0 & 2814 (12\%) & 2\% \\
  $2^{++*}$ & 2880 (5\%) & 2991 & 4\% &   &  \\
  $0^{-+*}$ & 3370 (4\%) &   3436  & 2\%  &   &  \\
  $0^{++**}$ & 3370 (4\%) & 3561  & 5\% &  &   \\
  $0^{++***}$ & 3990 (5\%) & 4253 & 6\% &  &   \\
\hline
\end{tabular}
\end{center}
\caption{Comparison between the glueball spectra  in Ref. I and in our model. The states
we use as input in our fit are marked in boldface. The parenthesis in the
lattice data indicate the percent accuracy.
 The data in \cite{lattice1} are given
in terms of mass ratios; here they have been rescaled to match the
lowest state mass.}\label{ref1vsus}
\end{table}

\begin{figure}[h]
 \begin{center}
\leavevmode \epsfxsize=18cm \epsffile{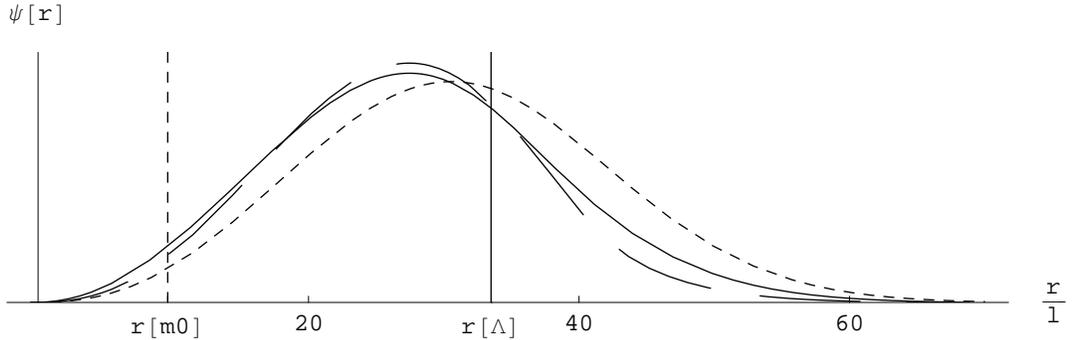}
\end{center}
 \caption[]{Normalized wave-function profiles for the ground states of the $0^{++}$ (solid
line),$0^{-+}$ (dashed line), and $2^{++}$ (dotted  line) towers,
as a function of the radial conformal coordinate. The vertical lines represent the
position  corresponding to $E = m_{0^{++}}$ and $E = \Lambda_p$.}
 \label{wfs}
\end{figure}

Again, after fitting the lowest state, the first  excited state is in good
agreement with the lattice data of Ref. I.
The comparison between our result and the data of ref. I is summarized in table \ref{ref1vsus}.
In figure \ref{wfs} we show the wave-function profiles of the lowest $0^{++}$, $0^{-+}$ and
$2^{++}$ states.
\vspace{1cm}

\subsubsection{Fit to Reference II}

{\bf $0^{++}$ and $2^{++}$ glueballs} \\
We use the data from \cite{lattice2} for our fit,
since that work includes values for some of the  excited states masses.
 Although older, these
data do not differ significantly from the more recent ones reported in
\cite{lattice3}, where however no excited states masses are given.
 We fix $b_0$ to match
$R_{00} = 1.54$. The
preferred value is now $b_0=2.5$. Then,
 we set the energy units so that
$m_{0++} = 1730$. The lowest lying states have masses:
\begin{equation}\label{0predII}
0^{++}:\quad    m_1, m_2, \cdots =  1730,\, 2697,\, 3321,\, 3853 ,\, 4319,\, 4747,\, 5139\,, \cdots\, MeV.
\end{equation}
\begin{equation}\label{2predII}
2^{++}:\quad     m_1,\, m_2, \cdots = 2194,\, 2897,\, 3485,\, 3987,\, 4440,\, 4851, 5229,\,\cdots\, MeV.
\end{equation}

\begin{figure}[h]
 \begin{center}
 \leavevmode \epsfxsize=7cm \epsffile{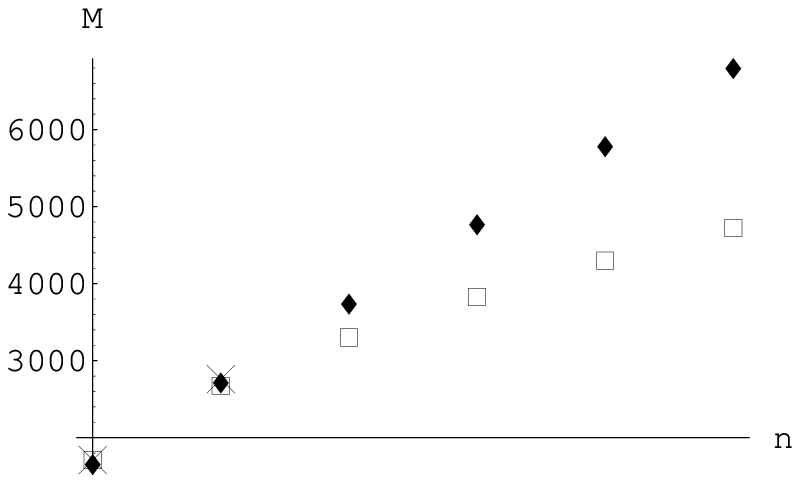}\hspace{0.5cm}
 \leavevmode \epsfxsize=7cm \epsffile{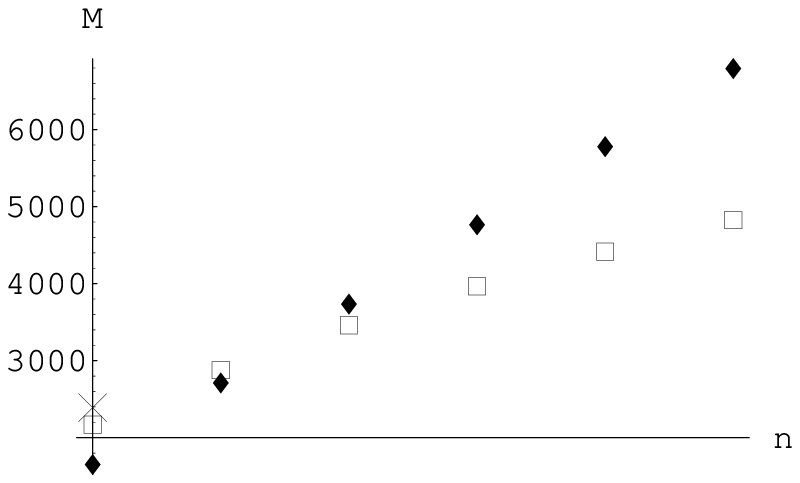}\\
(a)\hspace{6cm}(b)
 \end{center}
 \caption[]{Comparison of glueball spectra from our model with $b_0=2.5$ (boxes),
with the lattice QCD data from Ref. II (crosses)  and the AdS/QCD computation (diamonds),
for  (a) $0^{++}$ glueballs; (b) $2^{++}$  glueballs. The masses are in MeV, and
the scale is normalized to match the lowest $0^{++}$ state from Ref. II.}
 \label{compareg02}
\end{figure}
\vspace{0.5cm}

\noindent {\bf Running coupling and QCD scale}\\
Proceeding like in the previous section, we obtain:
\be
\Lambda_p = 356\, MeV, \qquad \a_s(1.2 GeV) = 0.38
\ee
These values are farther from the observational expectations
compared to the result of the fit to Ref. I.
\vspace{0.5cm}

\noindent {\bf $0^{-+}$ glueballs:}\\
 We  now vary the parameter $c_a$ to fit
the lowest pseudoscalar glueball mass.
\begin{table}[h]
\begin{center}
\begin{tabular}{|c|c|c|c|c|c|}
  \hline
  $J^{PC}$  &  Ref II (MeV) &  Our model (MeV) &  Mismatch & $N_c\to \infty$ \cite{lattice1} & Mismatch to $N_c\to \infty$  \\
\hline
  $0^{++}$  &  {\bf 1710 (5\%)} & {\bf 1710} & 0  & {\bf 1710} & 0\\
  $2^{++}$  &   2390 (5\%)   &   2194  & 8\%  & 2502 (10\%) &  10\% \\
  $0^{-+}$  &   {\bf  2560 (5\%) } & {\bf  2582 } & 0 & &   \\
  $0^{++*}$ &  {\bf 2670 (5\%) }  &{\bf  2697}  & 1\%  & 3262 (12\%) & 18\%  \\
  $0^{-+*}$ &   3640 (5\%)  & 3434  & 6\% & &   \\
\hline
\end{tabular}
\end{center}
\caption{ Comparison between the glueball spectra  in Ref. II (taken from \cite{lattice2}) and in our model. The states
we use as input in our fit are marked in boldface.  The parenthesis in the
lattice data indicate the percent accuracy.
The data in \cite{lattice1} are given
in terms of mass ratios; here they have been rescaled to match the
lowest state mass.}\label{ref2vsus}
\end{table}

The ref. II value $R_{A0} = 1.50$ is obtained for $c=0.023$. For this
value the first pseudoscalar glueball masses are found to be:
\be
0^{-+}\quad   m_1,\, m_2,
\cdots =  2597,\,  3434,\, 3965,\, 4407,\, 4851,\, 5246,\, \cdots\, MeV.
\ee

The comparison between our result and the data of ref. II is summarized in table \ref{ref2vsus}.

\subsubsection{Dependence of the spectrum on the spectral parameter
$\a$}

 Up to now  we have set the spectral
parameter $\a=2$, as it corresponds to linear confinement, $m_n^2
\propto n $ for large $n$. However, unlike in the case of mesons,
there is no direct lattice or experimental evidence for such a behavior
for the glueballs. In particular, the lattice simulations are only
available up to $n=4$ (for  $0^{++}$ only).
Therefore, it is interesting to examine the dependence of the
spectrum on $\a$. We recall that the effective Schr\"odinger
potential in the IR behaves as, \be \lab{schroIR} V(r) \sim r^{2(\a-1)},
\qquad as\,\,\, r\to \infty. \ee
 Hence, one expects
that the mass spectrum will move upwards as one increases $\a$. One
also expects that the hard-wall approximation of AdS/QCD would
correspond to $\a\to\infty$.

\begin{figure}[h]
\begin{center}
\leavevmode \epsfxsize=12cm \epsffile{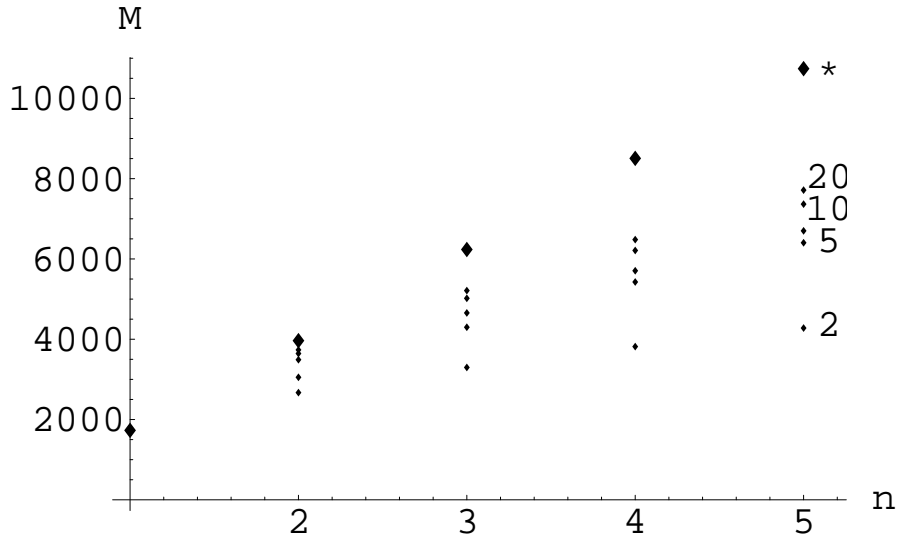}
\end{center}
\caption[]{The $0^{++}$ spectra for varying values of $\a$
that are shown at the right end of the plot. The symbol *
denotes the AdS/QCD result.}
\label{alphadependence}
\end{figure}

We carried out the necessary numerical analysis for the $0^{++}$
glueballs, for fixed values of $\l_0$, $b_0$ and $A_0$ and varying
$\a$. We fix $b_0=4.2$, as in the
fit for Ref. I, so that the mass ratio of $R_{00}$ is 1.87 for $\a=2$.
We normalize the spectra so that the lowest scalar glueball has
the same mass for all $\a$ we consider.
Our results are depicted in fig.\ref{alphadependence}
where we also included the AdS/QCD result for comparison.
\footnote{To compare with AdS/QCD we fixed the value of
$r_0$ of \cite{erlich} such that the first glueball lies at 1475
MeV.} One indeed finds that as $\a$ increases the spectrum of our
background approaches to that of standard AdS/QCD, and the agreement
with Ref. I becomes worse for larger $\a$. However, if we allow to change $b_0$ we
can fit the data equally well for $\a\neq2$ but not too large, so
there is no conclusive evidence that $\a=2$ is preferred.

\subsection{Background II: singularity at finite $r$}

In this section we compute the spectrum in a  5D background with
different IR asymptotics, namely the one in which the IR  singularity is at finite
$r$. We assume a power-law IR singularity,
\be
A(r) \sim \delta \log(r_0-r), \quad r\to r_0.
\ee
For the phase space variable, we take the same UV asymptotics (\ref{uvas}),
whereas in the IR, according to eqs. (\ref{pq4}), one has:
\be\lab{irasfin}
 X(\l) = -{3\over 4}Q +\ldots \qquad Q={2\over 3}\sqrt{1+ \delta^{-1}}
\ee
As interpolating function we choose:
\be\lab{xexactfin}
 X(\l) = -{b_0 \l \over 3+ 2 b_0\l} -
{(2b_0^2 + 3b_1^2)\l^2\over
9 +  {2\over \eta}\left(2b_0^2 + 3b_1^2\right) \l^2}, \qquad \eta \equiv\sqrt{1+\delta^{-1}} -1
\ee

To compute the spectrum we use the same procedure we employed in the
previous example. We first integrate numerically
 the equations for the metric and
dilaton, then we use a shooting method to find the mass eigenstates.
We use $b_0$ and $\delta$ as fitting parameters.

\subsubsection{The glueball spectra in background II}

First, we obtain the spectrum for the same value of
$b_0$ that gives the best fit to the data from Ref. I, namely
$b_0=4.2$, and we vary the parameter $\delta$.\footnote{We always use $\delta>1$ because
of the reasons discussed in Section \ref{beware}}
Varying $\delta$ between $\delta=1.01$ and $\delta=10$ we obtain
the results in Table \ref{tabfinite1} a).
To explore the dependence on $b_0$ we fix $\delta=2$ and vary $b_0$ (see
table \ref{tabfinite1} b). For a wide range of $b_0$  $R_{00}$ and $R_{20}$
are significantly smaller than the lattice values.
\vspace{0.5cm}
\begin{table}
\begin{center}
\begin{tabular}{|l|l|l|}
  \hline
  $\delta $ & $R_{00}$  & $R_{20}$    \\
\hline
  1.01 & 1.50 & 1.20   \\
  1.05 & 1.48  &  1.19  \\
  1.1 & 1.48 & 1.19  \\
  1.5 & 1.41 & 1.16  \\
  2 & 1.37 & 1.13  \\
  3 & 1.27 & 1.09  \\
   4 & 1.27 & 1.08 \\
  5 & 1.24  & 1.07  \\
  7 &  1.20   & 1.05  \\
  10 & 1.16  & 1.04 \\
  \hline
\end{tabular}
\hspace{1cm}
\begin{tabular}{|l|l|l|}
  \hline
  $b_0 $ & $R_{00}$  & $R_{20}$    \\
\hline
 0.5 & 1.47 & 1.17   \\
  0.75 & 1.42  &  1.15  \\
  1 & 1.39 & 1.14 \\
  2 & 1.38 & 1.14  \\
  3 & 1.37 & 1.13  \\
  5 & 1.37  & 1.13  \\
  10 & 1.37  & 1.13  \\
  25 &  1.40   & 1.10  \\
  40 & 1.41  & 1.07 \\
  100 & 1.47  & 1.05 \\
  \hline
\end{tabular}\\
(a) $b_0=4.2$\hspace{2cm} (b) $\delta=2$ \\
$R_{00}^{(II)} = 1.87$, $R_{20}^{(II)} = 1.46$
\end{center}
\caption{ Lowest glueball mass ratios for a) $b_0=4.2$., $l_0=0.05$, for
  varying $\delta$; b) $\delta=2$., $l_0=0.05$, for
  varying $b_0$  }\label{tabfinite1}
\end{table}
\subsubsection*{$0^{++}$ and $2^{++}$ glueballs: Fit
for Reference I}

 To fit the data in Ref. I we use the following procedure: for
different  values of $\delta$, we fix $b_0$ to obtain the mass ratio
$R_{00}=1.87$ as close as possible. Then we compare our finding for $R_{20}$
with the lattice value. Since the dependence
on $b_0$ for a given $\delta$ does not follow a clear
pattern, it is very hard to fit exactly any particular value
of $R_{00}$. It turns out that we were not able, with this
ansatz for $X(\l)$,  to obtain an $R_{00}$
larger than 1.65,  for which $R_{20}=1.3$.

\subsubsection*{$0^{++}$ and $2^{++}$ glueballs: Fit
for Reference II}

Contrary to the case of Ref. I above, one can fit the value $R_{00}=1.56$ in
Ref. II (table \ref{latticedata}), by choosing $b_0=0.96$ and $\delta=1.01$.
However, we cannot find a set of parameters which also gives a good result for
$R_{20}$. For the aforementioned values of $b_0$ and $\delta$, one obtains
$R_{20}=1.25$.

\subsection{Estimating the effect of the UV running}

In this subsection we investigate how the logarithmic running of the
coupling in the UV affects the IR properties, such as the glueball
mass spectrum. To address this issue, we compare the spectrum
of background I ($\a=2$, $b_0=4.2$) with another background obtained by
keeping the same IR properties, but with a conformal fixed point in the UV. In
the latter background, the geometry is asymptotically $AdS_5$ up to power-law
corrections, and the  't Hooft  coupling flows to a non-zero  value $\l_*$, which
can be chosen to be small. Such a
geometry has the following asymptotics for the superpotential and $\b$-function in the UV (i.e. for $\l\sim\l_*$):
\bea
W_{conf} = W_0 +  W_1 (\l-\l_*)^2 + \ldots, \qquad W_0  = {9\over 4 \ell} \lab{confas}\\
\beta_{conf}(\l) \sim - \tilde{b}_0\l_*(\l_*-\l) \qquad
\tilde{b}_0>0,\l_*\ll1 \eea In the IR, we take the new background to
have the same large $\l$ asymptotics as background I, as in
(\ref{iras}) and  (\ref{Aas}) with $\a=2$. Moreover, we fix the
initial conditions and the parameter $\l_*$ such that the
strong-coupling scale of the two backgrounds are the same. As a
definition of the strong coupling scale  we take the slope of the
scalar glueball mass spectrum:  $m_n^2 = \Lambda^2 n$ for large $n$.

As a simple  example of an asymptotically conformal background with the desired IR properties
we can take:
\be\lab{exactmetric}
e^A(r) = {\ell\over r} e^{-(r/R)^2}, \qquad \Phi(r)= \Phi_0 + {3\over 2}{r^2\over R^2}
\sqrt{1+3{R^2\over r^2}}+ {9\over 4}\log {2 {r\over R} + 2\sqrt{{r^2\over R^2} + {3\over2}} \over \sqrt{6}}.
\ee
One can easily check that the above solves Einstein's for an appropriate choice of superpotential, that is detailed in appendix \ref{analyt}.
This is an example of an
explicit   ``soft wall'' background, in which both the metric and the dilaton are known exactly, and
 that can be derived as a consistent solution of Einstein's equations.

 We use the same shooting method
as before to compute the mass eigenvalues. We can fix $\Phi_0\equiv \Phi(0)$
and $R$ in (\ref{exactmetric})
to match the slope of the glueball masses found in the asymptotically free background.

As an alternative background, we start with the exact superpotential:
\be\lab{superconf}
W_{conf}=W_0 \left(1+{4\over 9}b_0^2 (\l -\l_*)^2)^{1/3}\right)\left(9 a+(2b_0^2 + 3b_1)\log\left[1+(\l-\l_*^2)\right]\right)^{2a/3}.
\ee
This amounts to a small modification of the superpotential (\ref{superfree}), but it behaves asymptotically as (\ref{confas})
in the UV.

\begin{figure}
\begin{center}
\leavevmode \epsfxsize=12cm \epsffile{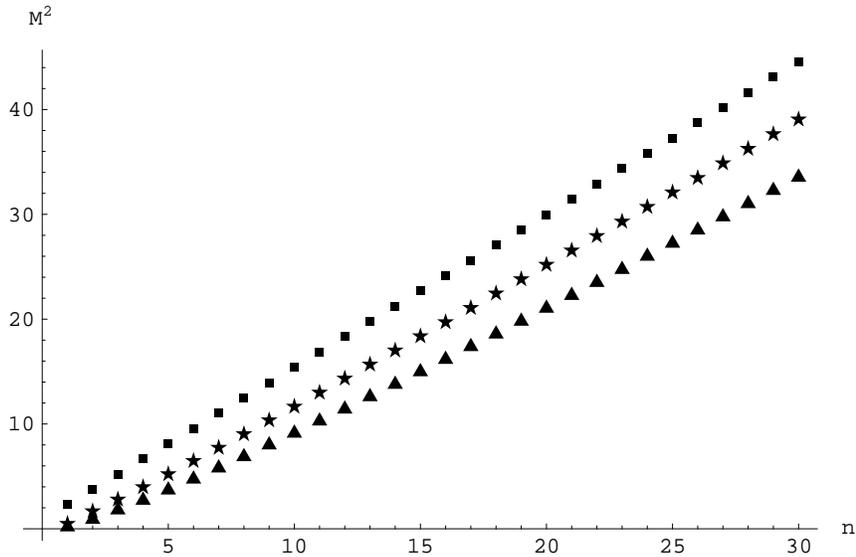}
\end{center}
\caption[]{The comparison of the scalar glueball masses for the asymptotically
free and the two conformal backgrounds: the stars correspond to the
asymptotically free background (\ref{xexact}) with $b_0=4.2$ and $\l_0=0.05$; the squares correspond the results
obtained in the background (\ref{exactmetric}) with $R=11.4\ell$; the
triangles denote the spectrum in the background given the
superpotential (\ref{confas}) with $b0=4.2$, $\l_0=0.071$ and $\l_* =
0.01$. These values are chosen so that the slopes coincide asymptotically for
large $n$. }
\label{confvsfree}
\end{figure}

The results are  shown in figure \ref{confvsfree}.

\subsection{Discussion}

Here we summarize the results of our numerical analysis. From the
qualitative point of view, our general setup can reproduce the known
features of the scalar and tensor glueball spectra. For example, as
in the lattice studies, the $0^{++}$ states are lighter than the
$2^{++}$ states, contrary to the $AdS$/QCD models of
\cite{erlich,pomarol}, in which the two towers are exactly
degenerate. The pattern $m^{(0)}_n < m^{(2)}_n$  seems to be a
generic feature of the dual backgrounds in which the dilaton is
taken to be non-trivial. We see numerically that this behavior is
realized in all the backgrounds we considered, and it was also
observed in \cite{cr}. Moreover, we always observe $R_{00}>
R_{20}$, which is common to all lattice results.

The ``linear'' model with $\a=2$ seems to reproduce the pattern of
excited spin-0 glueballs found in the lattice study \cite{lattice4}
which to our knowledge is the only work that computes the masses of
such states.
 From the quantitative point of view, we can make the following comments. We remind
the reader that our fits refer to mass ratios, as we can always choose
arbitrarily the absolute energy scale.

\begin{itemize}
\item For the infinite range background (background I) one can fit both sets of
  the available lattice data, Ref I. and Ref II, by fixing the parameter
  $b_0$. To check agreement with the lattice, one
should look at the last column of Table \ref{latticedata}, as our
setup is supposed to describe 4D YM at large $N_c$. Notice that the
large $N_c$
mass ratios $R_{00}$ and $R_{20}$ are very close to the ones of Ref.
I. Moreover, the uncertainties in  $R_{00}$ and $R_{20}$ for large
$N_c$ are  larger than the ones reported for the glueball masses in
both Refs. I and II. Our best fit for Ref. I is well within the
large $N_c$ error-bars.
\item The value of the spectral parameter $\a$ affects the
  results. We fix it to $\a=2$ in order to obtain a linear Regge
  trajectory. We note however that it is possible to fit the lattice data for a different
 set of values for $b_0$ and $\a\ne 2$. In this case the large $n$ asymptotics
 in the spectrum will not be linear.
\item As a general conclusion for the finite range background (background II),
  we can say that we could not find a range of parameters that yield good fits
  for both the scalar \emph{and} tensor glueball masses. In particular, if one adjusts
  the parameters in order to fit the scalar ratio $R_{00}$, then the tensor
  glueball masses turn out to be significantly lower than the lattice results,
  and outside the large $N_c$ error bars.
\item We analyzed the dependence of the spectrum on the logarithmic running of
  the coupling in the UV, by comparing our results with a background where one
  has the same IR but a conformal fixed point in the UV. This background has power law
  running for the coupling. One finds that for a fixed slope of the glueball
  spectrum, the overall scale of the masses do change. However it is possible
  to fit the lattice data by a choice of different parameters. Therefore, one
  can obtain in principle the same spectrum (at least for small $n$)
in a theory where the UV is a conformal fixed point.
\item A final word on fitting the lattice data:
  our strategy is to fit $R_{00}$ by fixing the parameter $b_0$ in our backgrounds
and then obtain a prediction for the ratio $R_{20}$. As we
mentioned, this prediction falls into the  error bars in the
  references I and II that account for
the uncertainty in the large $N_c$ limit  (see
table(\ref{latticedata})). Furthermore our
 predictions for the higher excited states also turn out within
 those
error bars, if we assume the same large $N_c$ uncertainty as for the
lowest states\footnote{There is no large $N_c$ extrapolation
available for the third and fourth excited $0^{++}$ states.}
 This is despite the fact that
our method of fitting the data is somewhat crude. A better method
would be to apply a global fit both for $R_{00}$ and $R_{20}$. One
expects from this method to produce better results for the higher
excited states as well.
\end{itemize}

{\bf Note added:} while this paper was being completed we became aware of
work along similar directions, which appeared in \cite{aldo}

 \addcontentsline{toc}{section}{Acknowledgments}
\acknowledgments

\noindent It is a pleasure to thank  B. Bringoltz, R. Casero,  L. Giusti, D. K. Hong,
K. Intriligator, M. Luscher, J. Mas, H.B. Meyer,  C. Morningstar, V. Niarchos,
C. Nunez, H. Panagopoulos, I. Papadimitriou, S. Pal,  A. Paredes, G. Policastro, F. Sannino,
C. Skenderis, M. Shifman, E. Shuryak, S. J. Sin, J. Sonnenschein, M. Teper, J. Troost,
A. Vainshtein,
G. Veneziano, A. Vladikas for useful discussions and correspondence.
We would also like to thank A. Schwimmer for reminding one of us the dictum of St John the Chrysostom.

UG and FN are supported by European Commission Marie Curie
Postdoctoral Fellowships, under contract number MEIF-CT-2006-039962 and MEIF-CT-2006-039369.
This work was also partially supported by
INTAS grant, 03-51-6346, RTN contracts MRTN-CT-2004-005104 and
MRTN-CT-2004-503369, CNRS PICS \#~2530,  3059 and 3747,
 and by a European Union Excellence Grant,
MEXT-CT-2003-509661.

\newpage

\appendix
\renewcommand{\theequation}{\thesection.\arabic{equation}}
\addcontentsline{toc}{section}{Appendices}
\section*{APPENDIX}

\section{Characterization of confining  backgrounds} \label{confinement}

We consider the  Einstein frame  metric in the conformal
coordinates,
\be
\label{metricf1} ds^2 = e^{2A(r)}\left(dr^2 +
\eta_{ij}dx^i dx^j\right), \qquad 0<r<r_0,
\ee
 where  $r=0$ is the  AdS boundary.
The corresponding string frame metric is
\be\label{metricf2} ds^2 =
e^{2A_s(r)}\left(dr^2 + \eta_{ij}dx^i dx^j\right),\qquad
A_S(r) = A(r) + {2\over 3} \Phi(r).
\ee
Given the behavior of the
scale factor close to the singularity, the asymptotic behavior of the
dilaton is uniquely fixed by the first of  eqs. (\ref{einsteqconf}),
\be
\label{dilr} \dot{\Phi}^{2}(r) = - {9\over
4}\left(\ddot{A}(r)-\dot{A}^2(r)\right).
\ee
 Knowledge of  $A(r)$
and $\Phi(r)$ uniquely determines the asymptotics of the phase space
variable $X$, therefore those of the $\beta$-function from eqs.
(\ref{x}). Asymptotics of the superpotential $W$ can be determined
from eqs. (\ref{x}), or from the second eq. in (\ref{conf1st}). $X$,
$\b$ and $W$ can then be expressed as functions of $\Phi$ by
inverting asymptotically the relation between $\Phi$ and
$r$\footnote{This can be done in backgrounds where the NEC is satisfied,
see Section 2.}.

Therefore, we can parametrize different backgrounds by the
asymptotics of the scale factor alone, since this completely
determines the asymptotics of all other quantities. The singularity
can be at a finite or an infinite value in the conformal coordinate.
We discuss these two cases separately. For all cases analyzed below,
we give the IR asymptotics of the following quantities, found by the
following equations:
\begin{itemize}
\item Einstein frame scale factor $A(r)$,
\item Dilaton and  't Hooft coupling  $\Phi =\log \l$.
\item String frame scale factor :
\be \label{a1}A_S = A +{2\over 3} \Phi \ee
\item  Einstein frame and string frame curvatures\footnote{In the
Einstein  frame there are two independent curvature invariants,
$(\de_r \Phi)^2$ and the Ricci scalar. They both behave
asymptotically as $e^{-2A}\dot{A}^2$, and will be denoted
collectively by $R$. The same holds for the string frame.}
: \be
\label{a2} R \sim e^{-2A}  \dot{A}^2, \quad  R_S \sim e^{-2A_S}
\dot{A}_S^2 \ee
\item Phase space variable and $\beta$-function,
\be \label{a3} X(\Phi) = {\dot{\Phi}\over 3 \dot{A}}, \quad  \b(\l)=3\l X(\l) \ee
\item Superpotential
\be \label{a4} W(\l) \sim \exp\left[-{4\over 3}\int {d\l\over \l} X(\l)\right] \ee
\item Dilaton potential (in Einstein and string frame actions):
\be \label{a5} V(\Phi) = -{4\over 3}\left({d W\over d\Phi}\right)^2 + {64\over 27} W^2, \quad V_S(\Phi) = e^{-4\Phi/3} V(\Phi) \ee
\item Metric and dilaton asymptotics  in the domain-wall coordinate $u$:
\be \label{a6} u = \int dr e^A(r) \ee
\end{itemize}

\subsection{Unbounded  conformal coordinate}

If the space-time extends over an infinite range of the $r$
coordinate, the Einstein frame scale factor $e^A(r)$ necessarily
vanishes as $r\to \infty$, as a consequence of eq.
(\ref{derboundr}). Therefore, $A(r)\to -\infty$ as $r\to \infty$. We
analyze two possible types of behavior for $A(r)$, logarithmic and
power law (the latter was also discussed in Section \ref{conf}). In
both cases the singularity is at a finite value $u_0$ in ``domain
wall'' coordinates.

\subsubsection{Logarithmic divergence}

Consider backgrounds such that, for large $r$:

\be \label{metric-asy-log} A(r) \sim -\gamma\log r + \ldots\qquad
\gamma\geq1. \ee The constraint $\g\geq1$ comes from the Null Energy
Condition discussed in Section 2. $\g=1$ corresponds to AdS
asymptotics in the IR, which does not lead to confinement. For any
$\g>1$, there is no confinement either,  as we show below. We have,
asymptotically: \be \dot{A} \sim -{\gamma\over r}, \quad \ddot{A}
\sim {\gamma\over r^2}. \ee
 From (\ref{dilr}) we obtain:
\be\label{dil2}
\dot{\Phi}^{2} \sim {9\over 4} {\gamma^2 - \gamma \over r^2}.
\ee
Eq. (\ref{dil2}) integrates to:
\be
\Phi \sim {3\over 2} \sqrt{ \gamma^2 - \gamma} \log r.
\ee
>From eq. (\ref{metricf2}), the string frame scale factor behaves asymptotically as:
\be
A_S (r) \sim - \gamma\left(1 - \sqrt{1 - \gamma^{-1}}\right) \log r,
\ee
Since the overall coefficient is negative
($\gamma\geq 1$), $A_S(r) \to -\infty$ as $r\to \infty$. Therefore
the string tension vanishes and there is no area law in this case.
These are the asymptotics of the relevant quantities:
\bea
&& \textrm{as} \quad r\to \infty:\nn \\
&& A \sim -\gamma \log r, \qquad \gamma >  1; \quad Q\equiv {2\over3}\sqrt{1 - {1\over \gamma}} < {2\over 3} \\
&& \textrm{color confinement:}\,\, \textrm{NO}\\
&& \Phi \sim {3\over2} \sqrt{\gamma^2 -\gamma} \log r = {9\over 4} \gamma Q \log r, \\
&& A_S \sim  - \gamma\left(1 - {3\over2}Q\right) \log r, \\
&& R \sim r^{2(\gamma-1)}\to \infty, \\
   && R_S \sim r^{2(\gamma-1) - 3 \gamma Q}
\to \left\{\begin{array}{ll} 0 &\quad  1<\gamma < {1\over 2}(1+\sqrt{5}) \\
                             \infty & \quad \gamma >{1\over 2}(1+\sqrt{5})\end{array}
\right.  \\
&&
X(\l) \sim -{1\over 2} {3Q\over 2} , \quad W(\l)\sim \l^Q \label{pq1}\\
&& V \sim \l^{2Q} ,\quad
 V_S = \l^{-{4 \over 3}} V
\sim  \l^{2Q-{4\over 3}}\\
&& u \sim u_0 - O\left({1 \over r^{\g-1}}\right)\\
&& A(u) \sim -{\g\over \g-1}\log (u_0-u).
\eea

\subsubsection{Power-law divergence}
Next we consider the following large $r$ behavior:
\be \label{metric-asy-power}
A(r) \sim -C r^\a + \ldots, \qquad C>0, \a>0,
\ee
where the precise nature of the subleading terms is immaterial. This
case was discussed in Section (\ref{conf}). It leads to confinement
if and only if $\a\geq 1$.
We have:
\be
\dot{A} \sim -C \a r^{\a-1},  \quad \ddot{A} \sim -C \a (\a-1)r^{\a-2}
\ee
Notice that $\ddot{A}/\dot{A} \sim r^{-1}$, therefore
  eq. (\ref{dilr}) is solved, asymptotically, by:
\be
\Phi = -{3\over 2} A + {3\over 4} \log |\dot{A}| + \Phi_0 + O\left({1\over r}\right)
\ee
where we have kept the first subleading term, which is universal
and independent of the subleading terms in (\ref{metric-asy-power}).
The string frame metric, from eq. (\ref{metricf2}), is:
\be   \label{strframef}
A_S \sim {(\a -1 )\over 2}\log r + {2\over 3}\Phi_0+  O\left({1\over r}\right)
\ee
Notice that the leading terms cancel. (\ref{strframef}). Therefore:
\be
A_S\to \left\{\begin{array}{ll}-\infty, & 0<\a<1  \\
                                const,  &  \a=1 \\
                               +\infty, & \a> 1
\end{array}\right.
\ee
and we have confinement if and only if $\a\geq 1$.
In the borderline case $\a=1$,  $A_S$ asymptotes to a finite constant as
$r\to \infty$. The
string frame metric is asymptotically Minkowski, and the dilaton is linear
in $r$, up to subleading corrections. The string of minimal world-sheet area
stretches all the way to $r=\infty$, but the confining
string tension is nevertheless finite.

We list below various relevant quantities:
\bea
&& \textrm{as} \quad r\to \infty:\nn \\
&& A \sim -C r^\a , \qquad \a>0, C>0 ; \qquad P\equiv {\a-1\over \a}<1  \\
&& \textrm{color confinement:} \,\, \textrm{if}\,\, \a\geq 1\\
&& \Phi \sim {3\over2} C r^\a + {3\over 4}(\a-1)\log r, \\
&& A_S \sim  {(\a -1 )\over 2}\log r, \\
&& R \sim e^{2Cr^\a} r^{2(\a-1)}\to \infty, \\
   && R_S \sim {1\over r^{a+1}}\to 0  \\
&&
X(\l) \sim -{1\over 2} \left(1+ {3 P\over 2}{1\over \log\l}\right) , \qquad W(\l)\sim (\log\l)^{P\over2} \l^{2\over 3} \label{pq2}\\
&& V \sim  (\log\l)^{P}\l^{4\over3} ,\qquad
 V_S = \l^{-{4 \over 3}} V \sim  (\log\l)^{P}
\eea
The domain wall coordinate $u$ terminates at a finite value $u_0$, as
the integral in eq. (\ref{a6}) converges as $r \to \infty$. The metric
and dilaton in this frame are, close to the singularity:
\bea
&& u\to u_0, \qquad \log (u_0 - u) \sim -Cr^\a \\
&& A(u)\sim  \log(u_0-u) +  P \log \left[-\log (u_0 -u)\right] + \ldots, \\
&&\Phi(u)\sim -{3\over2}\log(u_0-u) -{3\over4}P \log \left[-\log (u_0 -u)\right] \eea

\subsection{Finite range of the conformal coordinate}

Now suppose that the singularity is at a finite value of the
conformal coordinate, $r=r_0$. By monotonicity of $A(r)$, the scale
factor at the singularity  either vanishes, or stays finite.
\subsubsection{Finite $A(r_0)$}
If $A(r_0)$ is
finite,  the singularity must be caused by
non-analyticity in $A$.
The dilaton may stay finite at $r_0$, or asymptote to $+\infty$
(we are assuming strong coupling in the IR, so we exclude the case
$\Phi(r_0) = -\infty$). In any case, the string frame scale factor,
$A + 2\Phi/3$, is either finite at $r_0$ or asymptotes to $+\infty$, therefore
it must have a minimum for some $r_*$ in the range $(0,r_0]$. The value
at the minimum must be finite (otherwise there would be a singularity
at $r_*<r_0$), leading to a confining  string with non-zero tension.

According to the identification (\ref{energy}),  the fact that the
Einstein frame scale factor is nowhere vanishing means that the dual
4D theory is defined only above a certain energy $E_{min}\sim
e^{A_{min}}$. We will discard this case for a different reason:
there is no screening of the magnetic color charge.
\subsubsection{Power-law divergence}

Next, we consider the case when the Einstein metric scale factor
vanishes at some $r=r_0$ as a power-law: \be\label{metric-asy2} A(r)
\sim  -{C\over(r_0-r)^{\tilde{\a}}}, \qquad \tilde{\a}>0, C>0. \ee
Below we show that the string has a finite tension for all
acceptable values of $\tilde{\a}$ and $C$.  The argument we present
holds for  any generic subleading behavior. One can easily check
that the solution of (\ref{dilr}) close to $r_0$  is given by \be
\Phi(r) \sim -{3\over 2} A(r) + {3\over 4} \log |\dot A(r)| +
\Phi_0. \ee This ansatz solves (\ref{dilr}) up to a term
proportional to $(\ddot{A}/\dot{A})^2 \sim (r_0 -r)^{-2}$, which for
$\a>0$ is subleading w.r.t the term $\dot{A}^2 \sim
(r_0-r)^{2\tilde{\a}+2}$ in eq. (\ref{dilr}) . The string frame
metric asymptotes as: \be   \label{strframefin} A_S \sim  {1\over 2}
\log \dot{A}\sim -{(\tilde{\a} +1 )\over 2}\log (r_0-r). \ee
The leading
terms cancel, and the first subleading term is universal. Eq.
(\ref{strframefin}) shows that $A_s \to +\infty$ as $r\to r_0$ for
any positive $\tilde{\a}$, and we always obtain a confining string.

We list below various relevant quantities:
\bea
&& \textrm{as} \quad r\to r_0:\nn \\
&& A \sim - {C \over (r_0-r)^{\tilde{\a}}} , \qquad \tilde{\a}>0, C>0 ;
\qquad P\equiv {\tilde{\a}+1\over \tilde{\a}}> 1  \\
&& \textrm{color confinement} = \textrm{YES} \\
&& \Phi \sim {3\over2} {C\over(r_0-r)^{\tilde{\a}}} - {3\over 4}(\tilde{\a}+1)\log( r_0-r), \\
&& A_S \sim  -{(\tilde{\a} +1 )\over 2}\log (r_0-r), \\
&& R \sim {1\over (r_0-r)^{2(\tilde{\a}+1)}}e^{{2C\over (r_0-r)^{\tilde{\a}}}}\to \infty, \\
   && R_S \sim (r_0-r)^{\tilde{\a}-1}\to \left\{\begin{array}{ll}  \infty & \quad 0\leq \tilde{\a}<1\\
                                               const &  \quad \tilde{\a} =1 \\
                                                0 & \quad \tilde{\a}>1\end{array}
                     \right.  \\
&&
X(\l) \sim -{1\over 2} \left(1+ {3 P\over 2}{1\over \log\l}\right) , \qquad W(\l)\sim (\log\l)^{P\over2} \l^{2\over 3} \label{pq3}\\
&& V \sim  (\log\l)^{P}\l^{4\over3} ,\qquad
 V_S = \l^{-{4 \over 3}} V
\sim   (\log\l)^{P}, \\
&& u\sim  u_0 - e^{-C/(r_0-r)^{\tilde{\a}}}, \nn\\
&& A(u)\sim  \log(u_0-u) + P \log \left[-\log (u_0 -u)\right] + \ldots, \\
&&\Phi(u)\sim -{3\over2}\log(u_0-u) -{3\over4}P \log \left[-\log (u_0 -u)\right]. \eea

\subsubsection{Logarithmic divergence}

In this case we have, asymptotically:
\be \label{scalefin}
A \sim \delta \log (r_0-r), \qquad \delta >0,
\ee
and
\be  \dot{A} \sim -{\delta\over (r_0-r)}, \quad \ddot{A} \sim -{\delta\over (r_0-r)^2}.
\ee
 From (\ref{dilr}) we obtain:
\be\label{dil2fin} \dot{\Phi}^{2} \sim {9\over 4} {\delta^2 + \delta
\over (r_0-r)^2}, \ee Eq. (\ref{dil2fin}) integrates to:
\be\label{dil3fin} \Phi \sim -{3\over 2} \sqrt{ \delta^2+\delta}
\log (r_0-r). \ee where we chose the branch ($\Phi>0$). The string
frame scale factor behaves asymptotically as: \be A_S (r) \sim
\delta\left(1 - \sqrt{1 + \delta^{-1}}\right) \log (r_0-r). \ee For
large $r$ it asymptotes  to $+\infty$, as the overall coefficient is
negative for positive $\delta$. Thus, the fundamental string
confines.

In this case we have:
\bea
&& \textrm{as} \quad r\to r_0:\nn \\
&& A \sim \delta \log (r_0-r), \qquad \delta >  0; \quad Q\equiv {2\over3}\sqrt{1 + {1\over \delta}} > {2\over 3} \\
&& \textrm{color confinement} = \textrm{YES}\\
&& \Phi \sim {3\over2} \sqrt{\delta^2 +\delta} \log r = {9\over 4} \delta Q \log r, \\
&& A_S \sim  - \delta\left(1 - {3\over2}Q\right) \log r = - {1\over1+3Q/2} \log(r_0-r), \\
&& R \sim {1\over (r_0-r)^{2(\delta+1)}}\to \infty, \\
   && R_S \sim  (r_0-r)^{-{3Q\over 3Q/2 +1}}\to \infty  \\
&&
X(\l) \sim -{1\over 2} {3Q\over 2} , \quad W(\l)\sim \l^Q \label{pq4}\\
&& V \sim \l^{2Q} ,\quad
 V_S = \l^{-{4 \over 3}} V
\sim  \l^{2Q-{4\over 3}}\\
&& u\sim u_0 - O\left((r_0-r)^{\delta+1}\right)\\
&& A(u) \sim {\delta\over \delta+1}\log(u_0 -u)
\eea

\section{Magnetic charge screening: the finite range\label{monopoleapp}}

Here we want to determine the potential between two magnetic charges
at large separation, for the type of backgrounds with $r_0<+\infty$.
The case $r_0=+\infty$ was treated in section (\ref{monopole}).

\subsection{$A(r_0)$ finite}
If the Einstein frame scale factor does not vanish at
the IR singularity, the D-string frame scale factor cannot vanish either,
and   there is no difference between the calculation of the 't Hooft loop
and that of the Wilson loop on the same background. As explained in
Appendix (\ref{conf}), Section A.2.1, in this case the electric
string confines. Therefore  the magnetic string confines too.
These kinds of background fail to satisfy an important
test for a candidate holographic dual of QCD. The same consideration
applies to all theories where the 5th dimension terminates
at a regular  IR boundary.

\subsection{$A(r_0) \to  -\infty$}
We treat the case of power-law decay of the scale factor $e^A$. The
exponential case can be discussed along the same lines. We take \be
A \sim \delta \log(r_0-r), \qquad \delta>0. \ee From eqs.
(\ref{scalefin}) and (\ref{dil3fin}) we see that the D-string scale
factor is asymptotically (as $r\sim r_0$) \be A_D = A + {\Phi\over
6} \sim \delta\left(1 - {1\over 4}\sqrt{1+ \delta^{-1}}\right) \log
(r_0-r). \ee For $\delta < 1/15$,   $A_D \to +\infty$, the scale
factor diverges at the singularity and  the magnetic string
confines. For $\delta>1/15$ the scale factor vanishes as a
power-law: \be e^{2A_D} \sim (r_0 - r)^\g, \qquad \g =
2\delta\left(1 - {1\over 4}\sqrt{1+ \delta^{-1}}\right) >0. \ee In
this case the magnetic string tension vanishes. To investigate the
potential between two monopoles at large $L$, it is sufficient to
translate our setup into the notation of \cite{cobi} and use their
results: defining $s = r_0 - r$, we are in the situation described
in  \cite{cobi}, with $f(s) = g(s) \sim s^\g$ as $s\to 0$. In their
notation, this is the case $f(0)=0$ and $k= < j+1 $ (since
$k=j=\g$). From their general analysis it follows that,  for small
$s_F$ (the turning point of the world-sheet), \be L(s_F) \sim s_F^{k}
\ee i.e. $L(s_F)$ vanishes as $s_F$ approaches the singularity.  The
same is true in the UV:  $L(s_F)$ always vanishes close to an
asymptotic $AdS$ region.  Therefore, it must be that $L$ has a
maximum value $L_{max}$ for some  $r_{max}<r_0$, and there is no
smooth solution of the geodesic equation for $L>L_{max}$. As we
argued earlier in the case of infinite $r_0$,   the magnetic charges
are free for $L>L_{max}$.

The behavior of $L(r_F)$ in the case with exponential fall-off
close to $r_0$ cannot be deduced directly from the results of \cite{cobi},
but it can be  addressed by adapting the discussion
in  Section \ref{monopole},  and the result is the same, i.e. $L(r_F)$ cannot
diverge.

\section{Fundamental string world-sheet embeddings in the presence of a non-trivial dilaton}

The relevant world-sheet action is
\be S={1\over
 4\pi\ls^2}\int d^2\xi~\sqrt{g}g^{\a\b}G_{\mu\nu}(X)\partial_{\a} X^{\mu}\partial_{\b} X^{\nu}
 +{1\over 4\pi}\int d^2\xi\sqrt{g}R^{(2)}\Phi(X) \,,\label{wsac}
\ee

Instead of solving the equations  we will do a simpler test.
We will show that
 the contribution of the dilaton coupling to the full energy of the
string is negligible in the limit where the distance between the endpoints of the string becomes large.

We evaluate the action in the vicinity of the point $r=r_*$
at which the scale factor of the target space metric has a minimum.
We use the conformal coordinate system:
\be
\lab{conmet} ds^2 =
e^{2A_s(r)}\le(dx^2 + dr^2 \ri),
\ee
 where $A_s$ is the string frame
scale factor, $A_s = A + \frac23 \f$. We assume that the
contribution of the second term in (\ref{wsac})  is
small with respect to the first term and we confirm our assumption,
a posteriori. Then the leading term in the solution to the equation of motion that follows from (\ref{wsac})
is,
\be
\lab{wssol} g_{ab} = \hat{g}_{ab} = G_{\m\n}\6_a X^{\m}\6_b
X^{\n}.
\ee
We fix the diffeomorphism invariance on the world-sheet
by choosing $\tau = X^0$, $\s = Y$. Here, $Y$ is the direction in
the Minkowski space on which the quark pair lies. Using
(\ref{conmet}) and (\ref{wssol}), it is straightforward to compute
the Ricci scalar on the world-sheet. One finds, \be\lab{wsric}
\sqrt{g}R =
\frac{-2}{\le(1+B(r)^2\ri)^{\frac32}}\le((1+B^2)B^2A_s''+BB'A_s'\ri),
\ee where we defined $B(r) = (dy(r)/dr)^{-1}$ and the primes denote
derivatives w.r.t. $r$. Notice that $B=0$ at the world-sheet turning
point.
 The second term in (\ref{wsac})
becomes, \be\lab{wsscd} S_{(2)} = T \int
\frac{dy}{(1+B^2)^{\frac32}}\le((1+B^2)B^2A_s''+BB'A_s'\ri)\f(r).
\ee We assume that the scale factor $A_s$ has a minimum at a point
$r_*$. When the world-sheet turning point reaches $r_0$,
 $A'$ and $B$ in (\ref{wsscd}) both  vanish and $A''$ and $\f$ are some
positive constants and the quark pair distance  $L = \int dy$
diverges. Then it is clear from above that,
$$S_{(2)} \to const,$$
therefore it is bounded in $L$, whereas the Polyakov term in (\ref{wsac}) diverges linearly in $L$
(under the aforementioned assumptions). Hence we can ignore the dilaton coupling in (\ref{wsac}) consistently.
However, one has to be careful about the situations in which the integrand in (\ref{wsscd}) asymptotes to a constant.
In these cases, $S_{(2)} \propto L$ and one cannot ignore the dilaton corrections to the induced metric.

The picture we assume is as follows:  the string world-sheet is
smooth with a single turning point at $r_t$. The geometry of the
string is determined by a single boundary condition that we can take
as the length between the end-points of the string on the boundary,
$L$. As $L$ is made larger the turning point $r_t$ approaches the
minimum of $A_s$ that we call $r_*$. In particular we are assuming
that there is a single minimum for $A_s$. As $L$ approaches
infinity, the greater part of the world-sheet falls into the minimum
$r_*$. This picture is valid for all of the backgrounds that we
analyzed in this paper. Thus, indeed the only term that has a
potential divergence is the first term in (\ref{wsac}).

\section{Singularities of the tachyon}\label{tachyonapp}

Here we analyse the properties of eq. (\ref{tachyoneq}) in
 the case with the following asymptotics
\be\lab{tachyonIRas1app} A(r) \sim - \left({r\over R}\right)^\a,
\quad A_S(r) \sim {\a -1\over 2} \log r/R, \qquad \Phi(r) \sim
{3\over 2} \left({r\over R}\right)^\a, \quad \a\geq1 \ee

First, assume $\tau(r)$ is nonsingular for any finite $r$. We want to analyse the behavior
near $r=\infty$.
Asymptotically, (\ref{tachyoneq})  becomes:
\be\label{tacIRas2app}
  \ddot{\tau} -  {3\a \over 2 R}\left( {r\over R}\right)^{\a-1}\dot{\tau} + {3\over \ell^2}
\left({r\over R}\right)^{\a-1} \tau -
{3\a \over 2 R}(\dot{\tau})^3 +  {3\over \ell^2} \tau (\dot{\tau})^2=0.
\ee
We are interested in solutions that diverge as $r\to\infty$. First, suppose
that  $\tau\to \infty$, but $\dot{\tau}$ stays finite. In this case,
the third term in eq (\ref{tacIRas2app}) would
be much larger than all others, and the equation would not
be solved asymptotically. Then we conclude that as $\tau\to \infty$,
$\dot{\tau}\to \infty$ as well. Then the
last two terms dominate  eq. (\ref{tacIRas2app}), and the solution
behaves as:
\be\lab{tacIRas3app}
\tau(r) \sim \tau_0 \exp\left[{2\over\a} {R\over\ell^2} \,r\right]  \qquad r\to \infty.
\ee
where $\tau_0$ is an integration constant.

Now we want to check if it is possible for the tachyon to diverge at some
finite value $r_*$, where the metric and the dilaton are non-singular. Then, close
to $r_*$ :
\bea\lab{taylor}
&& A_S =  A_0 + (r-r_*)A_1 + {A_2\over 2}(r-r_*)^2 + \ldots \\
&&\Phi =  \Phi_0 +\Phi_1  (r-r_*) + {\Phi_2\over 2}(r-r_*)^2 + \ldots
\eea
and we can approximate  eq. (\ref{tachyoneq}) by:
\be\lab{tachyoneq2}
\ddot{\tau} + \left(3A_1 -  \Phi_1\right)\dot{\tau} +  e^{2A_0} \mu^2\tau +
e^{-2A_0}\left[4A_1 -\Phi_1\right](\dot{\tau})^3 + \mu^2 \tau (\dot{\tau})^2 =0.
\ee
 If $\tau\to \infty$ at $r_*$, then the ratios $\dot{\tau}/\tau$,  $\ddot{\tau}/\dot{\tau}$ and
 $\ddot{\tau}/\dot{\tau}$ all diverge at $r_*$, implying that the terms in
eq. (\ref{tachyoneq2}) proportional  to $\ddot{\tau}$ and $(\dot{\tau})^3$ diverge faster
than all other terms. Therefore close to $r_*$ we can further approximate eq.  (\ref{tachyoneq2}) by:
\be\lab{tachyoneq3}
\ddot{\tau}  +
e^{-2A_0}\left[4A_1 -\Phi_1\right](\dot{\tau})^3 = 0.
\ee
This equation is  solved by:
\be\lab{tachyonsqrt}
\tau \simeq \tau_* + c \sqrt{r-r_*}, \qquad c = {1\over 2}{e^{2A_0}\over 4A_1 -\Phi_1},
\ee
which is not consistent with the assumption  that $\tau$ diverges at $r_*$. Notice
however that the approximation we made in writing eq. (\ref{tachyoneq3}) still
holds if we make the weaker assumption that $\dot{\tau}$, and not necessarily $\tau$,
diverges at $r_*$. Then, eq. (\ref{tachyonsqrt}) correctly describes the
asymptotics near $r_*$. This is, in fact, the generic behavior  for
arbitrary boundary conditions, the point $r_*$ being determined by initial
conditions.

There is one situation when the above argument breaks down, i.e. when there exists
a point $r_{**}$ at which $4A_1 - \Phi_1=0$. In this case the term in (\ref{tachyoneq2})   proportional
to $(\dot{\tau})^3 $ acquires an extra $(r-r_{**})$ factor, and it is possible to solve
the equation asymptotically with the last two terms:
\be
(r-r_{**})e^{-2A_0}\left[4A_2 -\Phi_2\right](\dot{\tau})^3 + \mu^2 \tau (\dot{\tau})^2
\simeq0\quad \Rightarrow \quad \tau \sim (r-r_{**})^{1+h}, \quad h = {e^{2A_0}\mu^2 \over 4A_2 -\Phi_2}.
\ee
If  $1+h<0$, this is  consistent with $\tau(r) $ diverging at $r_{**}$.

\section{The superpotential versus the potential\lab{spotvspot}}

In this appendix we analyze the (important) details of passing from the potential to the superpotential.
The first order non-linear differential equation that relates the two is
\be\lab{potspot}
V(\l) = -{4\over3}\l^2 \left({dW\over d\l}\right)^2 + {64\over 27}W^2,
\ee
For a given choice of $V(\l)$ there is in principle a one-parameter family
of solutions to (\ref{potspot}). This extra parameter that corresponds to the initial condition
of the first order differential equation is compensating from the fact that the equations of motion, written in terms of the superpotential are
first order and therefore have one less initial condition.

According to our previous discussion, the different solutions to (\ref{potspot}) correspond to different $\beta$-functions,
and hence to different 4D gauge theories.
However, here we show that the requirement of color confinement in the IR
singles out a unique superpotential among these.

As described in detail in \cite{part1} asymptotic freedom implies that for small $\l$:
\be\lab{UV3}
V =\sum_{n=0}^{\infty}V_n~\l^n= V_0 + V_1\l + O(\l^2).
\ee
A straightforward calculation shows that the one-parameter family of solutions to the non-linear equation (\ref{potspot})
has the structure
\be
W_C={C\over \l^{4\over 3}}+{1\over C}\sum_{n=0}^{\infty}C_n~\lambda^{{4\over 3}+n}+{1\over C^3}\sum_{n=0}^{\infty}D_n~\lambda^{5+n}
+{1\over C^5}\sum_{n=0}^{\infty}E_n~\lambda^{{23\over 3}+n}+\cdots
\label{e1}\ee
where $C\not=0$ is the one free parameter, while all other coefficients are uniquely fixed by the expansion of the potential:
\be
C_0={27\over 256}V_0\sp C_1={27\over 352}V_1 \sp C_2={27\over 448}V_2\sp C_3={27\over 544}V_3
\ee
\be
C_4={27\over 640}V_4\sp C_5={27\over 736}V_5 \sp C_6={27\over 832}V_6\sp C_7={27\over 928}V_7
\sp C_8={27\over 1024}V_8
\ee
\be
D_0={3\over 19}C_0C_1\sp D_1={33C_1^2+48C_0C_2\over 176}\sp D_2={27C_1C_2+18C_0C_3\over 50}
\ee
\be
D_3={42C_2^2+75C_1C_3+48C_0C_4\over 112}\sp D_4={57C_2C_3+48C_1C_4+30 C_0C_5\over 62}
\ee
\be
E_0={11\over 171}C_0^2C_1
\sp
E_1={3421 C_0C_1^2+2128 C_0^2C_2\over 16720}
\ee

There exists however a second branch of solutions again parametrized by a single integration constant 
$\tilde{C}$, disconnected from the family $W_C(\l)$:
\be\label{UV5}
W_{\tilde{C}}(\l) = W_0 + W_1 \l+{\cal O}(\l^2) + \tilde{C} \l^{\frac{16}{9}} e^{-\frac{1}{b_0 \l}} + \cdots, 
\quad W_0 = \sqrt{{27\over 64} V_0}, \; W_1 = \sqrt{{27\over64 V_0 }} {V_1\over2},
\ee
where the dots denote terms that vanish faster than $\exp(-4/b_0\l)$. 
(\ref{e1}) together
with (\ref{UV5}) provide the general solution of (\ref{potspot}). All
the solutions of the continuous family are singular for $\l\to 0$,
and the corresponding metric is not asymptotically $AdS_5$. AdS asymptotics
require that $W\to const $ in the UV and therefore single out the $W_{\tilde{C}}$ branch. 

It is the
one that leads to the correct $\beta$-function, at least perturbatively. 
However one wonders how the value of $\tilde{C}$ is fixed among the solutions of this branch. 
For this purpose we turn to the analysis of the IR asymptotics below. 

We investigate the same problem, namely the solution of equation (\ref{potspot}) but in the IR region.
We have argued that confinement requires that  the  IR asymptotics of the superpotential must be of the form,
\be\lab{confas4}
V\to A\f^{P} e^{2Q \f},   \qquad \f \to \infty.
\ee
with  $1\leq Q<4/3$. Solving the equation for the superpotential we find the one-parameter family of solutions
\be\lab{genIRspot}
W_C = \le(\frac34\ri)^{\frac32}\sqrt{A}\le\{Ce^{\frac43\f}
+\frac{1}{C}(4-3Q)^{-1}e^{(2Q-\frac43)\f}\f^P\ri\}+\cdots,
\ee
that is parametrized by one integration constant $C$.
In addition there is
the ``singular" solution,
\be\lab{sinIRspot}
W_* =\sqrt{27A\over 4(16-9Q^2)}~e^{Q \f}\f^{\frac{P}{2}}+\cdots.
\ee
Consider now the IR asymptotics of the potential that we expect to give confinement in the case where the range of the
radial coordinate is infinite.
\be
V=A~e^{{4\over 3}\f}~\f^P
\ee
From (\ref{genIRspot}) and (\ref{sinIRspot}) it is the singular solution $W\sim e^{{2\over 3}\f}~\f^{P/2}$ that we need, 
rather than the continuous family.

Now, one can show by scanning the space of solutions with all possible asymptotic behaviors in the IR, that the $W_*$ above leads in the UV to $W_{\tilde{C}}$ with $\tilde{C}=0$. This is done by solving (\ref{potspot}) numerically. 
%
%
We observe that the requirement of confinement singles-out the unique singular
solution out of an infinite set.

In view of this phenomenon, the practical strategy for choosing
the correct potential is to first choose the appropriate superpotential (or the function $X$) 
and then use the differential equation (\ref{potspot}) to compute the relevant potential.

\section{Standard AdS/QCD Glueball spectrum\label{adsqcdglue}}

In this appendix we consider the standard AdS/QCD model
\cite{erlich} where the background geometry is $AdS_5$ with an IR
cut-off at $r=r_0$. The dilaton is constant. The glueball spectra in this model were first
computed by employing Dirichlet boundary conditions in the papers \cite{BragaI} and \cite{BragaII}.
However, here we shall allow for more general boundary conditions.

In this geometry, both the scalar and spin-two glueballs spectra are
determined by the following equation:
\begin{equation}\lab{adsfluc} \ddot{\xi}-\frac{3}{r}\dot{\xi}+ m^2 \xi = 0,
\end{equation}
The corresponding effective Schr\"odinger potential is,
\begin{equation}\lab{adsschro}
V_s = \frac{15}{4}\frac{1}{r^2}, \qquad r<r_0
\end{equation}
and there is an infinite wall at $r=r_0$.

The solution to (\ref{adsfluc}) that is normalizable in the UV is,
\be\lab{adsol} \xi = r^2 J_2(k r) \approx r^4\quad as\quad r\to 0.
\ee

The important difference between our backgrounds and AdS/QCD is that
in AdS/QCD the normalizability condition in the IR does not restrict
the spectrum. Indeed all the solutions of (\ref{adsfluc}) with the
UV asymptotics (\ref{adsol}), are normalizable in the IR. What
discretizes the spectrum is the boundary condition at $r=r_0$. In
general this can be a mixed boundary condition that may be written
as, \be \lab{adsbcIR} \dot{\xi}(r_0)-C_i~{\xi}(r_0)=0 \ee
 Here $C_i$ are real numbers and
one can have different $C_i$ for different particle species $i$.
Therefore, the free parameters to fit the data are $r_0$, $C_{0++}$ and
$C_{2++}$. In the standard AdS/QCD model, the value of $r_0$ is
determined by fitting the pion mass which yields $r_0 = 1/322\,\,
MeV^{-1}$.

\begin{figure}[h]
 \begin{center}
 \leavevmode \epsfxsize=12cm \epsffile{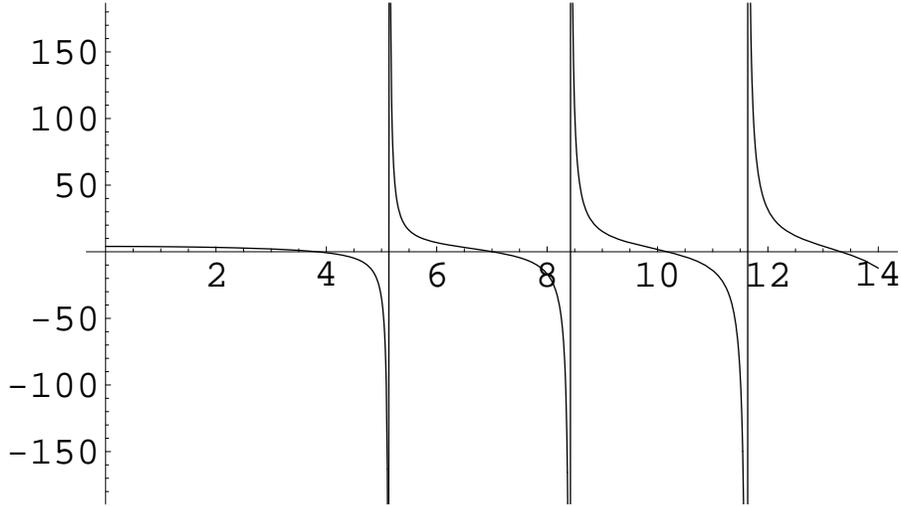}
 \end{center}
 \caption[]{Plot of $\dot{\xi}/\xi$ as a function of $m r_0$. The spectrum is
   determined by the points that correspond to the intersection of this plot
   and the horizontal line $\dot{\xi}/\xi=const $.  }
 \label{adg}
\end{figure}

We want to determine $C_{0++}$ and $C_{2++}$ to obtain a best fit to
the lattice data. To obtain a best fit
to the first $0^{++}$ glueball (1730 MeV), one has to
avoid the first solution that is shown in fig. \ref{adg}. We note
that $\lim_{m\to 0} \dot{\xi}/\xi(m r_0)=2$. Hence one needs
$C_{0++}>2$. A quick glance at the fig.\ref{adg} shows that the best
fit (highest possible mass) for the first $0^{++}$ mass is obtained
by setting $C_{0++} = 2+\epsilon$ in the limit $\epsilon \to 0^+$.

Then one determines the $0^{++}$ masses as, \be\lab{adspred} m_1,\,
m_2, \cdots = 1651,\, 2710,\, 3734,\, 4764 \, 5778,\, 6792\cdots\,
MeV. \ee

However now the best fit for the $2^{++}$ masses is given by the
same IR boundary condition in the IR, \ie $C_{2++} = C_{0++}$
\footnote{There is of course the possibility of choosing $C_{0++}$
bigger than $C_{2++}>2$. However then the first scalar glueball
masses becomes smaller than 1651 Mev and the discrepancy with the
lattice data increases. }. This gives the same mass spectrum for the
spin-2 glueballs as in (\ref{adspred}).

\section{A simple analytic solution with AdS and confining asymptotics\label{analyt}}

We present here a simple analytic solution for the metric and the dilaton that interpolates
between $AdS_5$ and a constant dilaton near the UV boundary and an Gaussian confining behavior
and strong coupling in the IR.
It is given by
\be\lab{exactmetric11}
e^A(r) = {\ell\over r} e^{-(r/R)^2}, \qquad \Phi(r)= \Phi_0 + {3\over 2}{r^2\over R^2}
\sqrt{1+3{R^2\over r^2}}+ {9\over 4}\log {2 {r\over R} + 2\sqrt{{r^2\over R^2} + {3\over2}} \over \sqrt{6}}.
\ee

There are two dimensionless parameters that characterize this solution. The first is the asymptotic value
of dilaton $\Phi_0$ in the UV parametrizing the UV gauge coupling. To mimic YM, it should be taken large and negative.
The other is the ratio $R/\ell$, that characterizes the confinement scale $R$.

The associated superpotential can be given in implicit form (as a function of $r$):
\be
W(r) = {9\over 4 \ell}e^{r^2/R^2}\left(1 -2 r^2/R^2\right)
\ee
Derivatives with respect to $\Phi$ can be computed using the chain rule, $W'={\dot W/\dot \Phi}$ etc.
Note that this solution may look similar but is very   different from the so called soft-wall AdS/QCD model.
Although the dilaton has the same behavior, the metric here in the IR is very different from AdS.
On the other hand the metric in the soft-wall AdS/QCD model is AdS everywhere.

\end{document}